\documentclass[12pt, twocolumn]{aastex701}

\usepackage{multirow}

\begin{document}

\title{Rapid bulge assembly in young galaxy disks at Cosmic Dawn}
\shorttitle{Bulge assembly at Cosmic Dawn}
\shortauthors{Borgohain \& Saha}

\author[orcid=0000-0002-2870-7716]{Anshuman Borgohain}
\email[show]{anshuman@iucaa.in}
\affiliation{Inter-University Centre for Astronomy and Astrophysics, Pune, Maharashtra - 411007, India
}

\author[orcid=0000-0002-8768-9298]{Kanak Saha}
\email[show]{kanak@iucaa.in}
\affiliation{Inter-University Centre for Astronomy and Astrophysics, Pune, Maharashtra - 411007, India
}

\begin{abstract}
Recent observations with the \textit{James Webb Space Telescope (JWST)} have begun to reveal a surprising morphological diversity in galaxies within the first billion years after the Big Bang, including indications of structural maturity previously thought to arise much later. These findings raise fundamental questions about when and how well-known structural components of galaxy morphology, such as bulges and disks, first emerged. However, directly identifying and resolving such structures at $z > 6$ remains challenging due to limited spatial resolution and sensitivity. In this work, we present a clear and robust morphological analysis of a sample of 187 galaxies at $z \ge 6$, demonstrating that distinct bulge and disk components were already beginning to emerge during this early epoch. Using multi-component light profile fitting, we identify a subset of 20 galaxies whose surface brightness distributions can be modeled with an inner spheroidal (Sersic) component and an underlying exponential disk component. These systems exhibit high bulge-to-total (B/T) light ratios ($\sim 0.47$) and central stellar mass surface densities, $\sim$ 7.82$\times$10$^{8}$ M$_{\odot}$kpc$^{-2}$ -- close to those of nearby quiescent galaxies. Combined with their intense central star formation rate surface densities, $\sim$1.21$\times$10$^{1}$ M$_{\odot}$yr$^{-1}$kpc$^{-2}$, our results indicate a rapid building of inner stellar mass and bulge assembly within these young systems. We propose that these early bulge–disk galaxies represent progenitors of massive star-forming and quiescent systems observed at lower redshifts. Their subsequent evolution may proceed through physical processes such as disk growth, compaction, quenching, or bulge-disk co-evolution, driven by both internal dynamics and external interactions.
\end{abstract}


\section{Introduction} 
\label{sec:intro}

The origin of the rich morphological diversity observed in present-day galaxies remains a central challenge in galaxy formation and evolution \citep{blanton_moustakas2009, conselice2014, SomervilleDave2015, naab_ostriker2017}. Observations from the \textit{Hubble Space Telescope} (HST), \textit{Atacama Large Millimeter/submillimeter Array} (ALMA), and now the \textit{James Webb Space Telescope} (\textit{JWST}) have pushed the redshift frontier, revealing galaxies actively assembling their key components. Deep surveys such as JADES, CEERS, REBELS, and PRIMER programs, have begun to resolve internal galaxy structures at $z \gtrsim 6$, revealing mature morphologies within the first billion years of cosmic history \citep[e.g.,][]{ferreira-etal2023, kartaltepe-etal2023, huertas-company-etal2024, baker-etal2025, fujimoto-etal2025}. Studies have revealed rotating disk systems at very early cosmic epochs, including dynamically cold disks at $z \sim 7$ \citep{rowland-etal2024} and clumpy rotating disks without massive bulges at $z \sim 6$ \citep{fujimoto-etal2025}, demonstrating the diversity of early galaxy structures during this period. In fact, a significant fraction of galaxies ($\gtrsim45\%$) up to $z \sim 6$ and beyond already exhibit regular, disk-like morphologies, and nearly 20\% of the massive galaxies show prominent central bulges \citep{huertas-company-etal2024, sun-et-al2024}. However, the $\Lambda$CDM framework tends to over-predict both the abundance and mass of bulges \citep{brooks_christensen2016}, a discrepancy often referred to as the ``over-cooling problem''. Conversely, the discovery of bulgeless, pure-disk galaxies at $z \sim 1$ \citep{sachdeva_saha2016} raises questions about the processes that suppress bulge formation while preserving disk structure. Feedback mechanisms are thought to play a key role: stellar feedback can mitigate over-cooling, while AGN feedback at high redshift is a likely driver of bulgeless disks in simulations \citep{governato-etal2007, okamoto-etal2008, choi-etal2014}. Still, the presence of evolved features - such as bulges and spiral arms - within the first Gyr after the Big Bang challenges the standard $\Lambda$CDM paradigm and highlights the need to better understand the formation and evolution of these structures \citep[e.g.,][]{de_Lucia-etal2011}.

Stellar bulges are spheroidal components located at the centers of both early- and late-type galaxies. More than half of present-day spiral galaxies host a central bulge \citep{fisher_drory2008, gadotti2009, kormendy2016}. Although they resemble scaled-down elliptical systems, bulges differ significantly from ellipticals when kinematic properties are considered \citep[e.g.,][]{dressler_sandage1983, davies_illingworth1983, kormendy_kennicutt2004, fisher_drory2008, gadotti2009}. Photometrically identified bulges are broadly classified into two types -- \textit{classical} and \textit{pseudo} -- based on their evolutionary characteristics \citep{drory_fisher2007, kormendy_kennicutt2004}. \textit{Classical} bulges, more commonly found in early-type galaxies, share several properties with elliptical galaxies: redder colors, older stellar populations, dynamically hot and dispersion-supported kinematics, and high Sérsic indices ($n > 2$, typically $\sim$4) \citep{kormendy_illingworth1982, zoccali-etal2006, lecureur-etal2007, kormendy_kennicutt2004, fisher_drory2008, gadotti2009}. They also follow the same scaling relations and lie on the fundamental plane of ellipticals \citep{barroso-etal2002, driver-etal2007, macarthur-etal2008, fisher_drory2011}. In contrast, \textit{pseudo} bulges are more prevalent in late-type galaxies \citep{thomas_davies2006}, and exhibit disk-like properties such as bluer colors (indicating ongoing star formation and younger stellar populations), rotational support, and lower central light concentrations ($n < 2$) \citep{drory_fisher2007, fisher_drory2008, gadotti2009, fisher_drory2011}.

The formation of stellar bulges is a key process in galaxy evolution, closely linked to the quenching of star formation in galactic disks \citep[e.g.,][]{Sachdevaetal2019,lang-etal2014}. Observationally, about $\sim$60\% of massive galaxies ($M_{\star} > 10^9~M_{\odot}$) in the Local Volume host a stellar bulge, while the remainder are either bulgeless or in the process of forming one \citep{fisher_drory2011}. Although the fraction of bulge-hosting galaxies declines with increasing redshift, evolved bulges and signs of quenching have been detected as early as $z \sim 6$ \citep[e.g.,][]{whitaker-etal2012b, barro-etal2017, Sachdevaetal2019,lelli-etal2021, huertas-company-etal2024, jain_wadadekar2024,xiao-etal2025, genin-etal2025}, suggesting that bulge formation and the associated suppression of star formation can occur rapidly, and more efficiently than predicted by standard models for young galaxies.

Current theories identify two broad formation channels: Classical bulges are thought to have formed through rapid, violent processes such as monolithic collapse \citep{larson1974} or major mergers within a Gyr \citep{hernquist_barnes1991, aguerri-etal2001, mihos2004, springel-etal2005b}, often accompanied by intense starbursts or AGN activity \citep{mihos2004, okamoto2013}. These systems can later rebuild their disks via gas accretion \citep{brook-etal2004}. In contrast, pseudo bulges form gradually over longer timescales ($\gtrsim$ Gyr) through secular processes such as bar-driven gas inflows, intermittent central star formation, and internal disk instabilities \citep{combes-etal1990,courteau1996, kormendy_kennicutt2004, SahaCortesi2018}. Additional mechanisms like radial migration, minor mergers, or dynamical rearrangement of the disk can also contribute to the bulge growth \citep{springel_hernquist2005, rokar-etal2012, johnston-etal2012, guedes-etal2013, morelli-etal2016}. Faster bulge formation is also possible via the coalescence of massive star-forming clumps -- driven by dynamical friction or through strong, centrally concentrated gas inflows, both acting over a few hundred Myr \citep{noguchi1999, bournaud-etal2007, elmegreen-etal2008, mandelker-etal2017,SahaCortesi2018}. These rapidly evolving systems may appear as compact, star-forming ``blue nuggets'', which have been observed across cosmic time \citep{dimauro-etal2022, lapiner-etal2023}.

It is challenging to explain the assembly of massive bulges within a short span of time (less than a few hundred Myr) after the Big Bang. A complete census of bulges at the earliest epochs will help constrain the interplay of mechanisms that can act within a Gyr such as rapid gas inflows, violent disk instabilities and massive clump migration \citep{dekel-etal2009,bournaud-etal2014, conselice2014, SahaCortesi2018}. This has long been limited by the small sizes and low surface brightness of bulges at these redshifts and insufficient spatial resolution. With the unprecedented resolving power of \textit{JWST}, we can now probe such structures in detail in the primordial galaxies. Also, \textit{JWST}'s ability to probe rest-frame optical and bluer wavelengths in high redshift galaxies enables us to trace the ongoing stellar mass assembly. In this study, we aim to develop a comprehensive picture of when and how quickly galactic bulges emerge and evolve. 

Throughout this paper, we adopt a standard flat $\Lambda$ CDM cosmology with Hubble constant, H$_{0}$ = 70 km s$^{-1}$ Mpc$^{-1}$ , $\Omega_{m}$ = 0.3 and $\Omega_{\Lambda}$ = 0.7; consistent with WMAP nine-year results \citep{Hinshaw-etal2013}. We quote all magnitudes in the AB magnitude system \citep{oke1974}.

\section{Data and Sample selection}
\label{sec:sample}

We use publicly available \textit{JWST} imaging observations and associated photometric and spectroscopic catalogs for the GOODS North and South fields -- JADES \citep{eisenstein-etal2023,eisenstein-etal2023b,rieke-etal2023, hainline-etal2024, bunker-etal2024, deugenio-etal2025, https://doi.org/10.17909/8tdj-8n28} and FRESCO \citep{oesch-etal2023, meyer-etal2024, covelo-paz-etal2025, https://doi.org/10.17909/gdyc-7g80}. In this work, we focus on galaxies that assembled within the first billion years after the Big Bang, and therefore select objects with secure spectroscopic redshifts, $z_{\mathrm{spec}} \geq 6$. As emphasized in previous studies, accurate broadband colors are crucial for estimating unbiased stellar masses and star-formation rates (SFRs) \citep[e.g.,][]{sawicki2012, merlin-etal2024}. Therefore, we cross-match this sample with the point-spread function (PSF)-matched photometric catalog of \citet{merlin-etal2024} to obtain consistent multiband photometry for 301 galaxies spanning $z_{\mathrm{spec}} = 6.00-13.35$. The photometric measurements cover 16 broadband filters, ranging from HST/F435W to F160W and \textit{JWST}/F090W to F444W. We then construct and model the spectral energy distributions (SEDs) as described in the following section.

\section{Spectral energy distribution modeling}
\label{sec:sed}

%-----new cigale table--------
\begin{table*}[htp]
\centering
\caption{The CIGALE input parameters used for the SED modeling.}

\begin{tabular}{l c}
\toprule
CIGALE parameter & Values \\
\hline

\multicolumn{2}{c}{Star Formation History (SFH): \texttt{sfhdelayed}} \\
\hline

$\tau_{\rm main}$ (Myr) & 10, 50, 100, 200, 300, 500, 750, 1000  \\
Age$_{\text{main}}$ (Myr) & 50, 100, 150, 200, 250, 300, 400, 500, 600, 700, 800 \\

$\tau_{\rm burst}$ (Myr) & 10.0, 20.0, 50.0 \\
Age$_{\text{burst}}$ (Myr) & 10.0, 20.0, 50.0 \\
$f_{\rm burst}$ & 0.0, 0.01, 0.02, 0.05, 0.07, 0.1 \\

\hline
\multicolumn{2}{c}{Stellar Population: \texttt{Bruzual \& Charlot 2003}} \\
\hline
Z$_{\star}$ & 0.0001, 0.0004, 0.004, 0.008 \\

\hline
\multicolumn{2}{c}{Nebular Emission: \texttt{nebular}} \\
\hline
$\log$ U  & -3.0, -2.5, -2.0, -1.5, -1.0\\
Z$_{\rm gas}$ & 0.0001, 0.0004, 0.004, 0.008 \\

\hline
\multicolumn{2}{c}{Dust Attenuation: \texttt{dustatt\_modified\_starburst}} \\
\hline
E(B-V)$_{\rm lines}$ & 0.02, 0.05, 0.1, 0.2   \\
E(B-V)$_{\rm factor}$ & 0.44 \\
$\delta$ & -0.5, -0.1, 0.0, 0.1 \\

\hline
\end{tabular}

    \label{tab:cigale_config}
\end{table*}

We use the Python-based SED modeling code CIGALE \citep{boquien-etal2019, burgarella-etal2025} to derive the stellar masses of our galaxy sample using broadband photometric fluxes in 16 filters as mentioned above. As described above, we use the PSF-matched photometric fluxes from \citet{merlin-etal2024}, which would provide unbiased colors over multiband observations. We correct all the fluxes for foreground Galactic extinction using dust maps from the NED database \citep{schlafly-etal2011}. For the SED modeling, we use a Chabrier \citep{chabrier2003} initial mass function (IMF) with BC03 \citep{bruzual_charlot2003} stellar population models and assume a delayed star-formation history with a recent burst of star formation. To model the attenuation, we use a modified Calzetti extinction law \citep{calzetti-etal2000} and dust models from \citep{draine_li2007, draine-etal2014}. The final CIGALE configuration generates 6,082,560 models per object. Since we have the $z_{\text{spec}}$ values for each object, we do not fit the redshifts. We present a detailed information on the configuration file in Table \ref{tab:cigale_config}.

\section{2D modeling}

In this work, we are interested to study the morphology of the galaxies as traced by their young stellar population -- to track the ongoing assembly. In that, we utilize those broadband \textit{JWST} images that probe the rest-frame U ($\sim$3200-4000$\AA$) and B ($\sim$4000-5000 $\AA$) emission. The rest-frame $B-$band is an appropriate tracer of the underlying stellar mass distribution in galaxies less than a Gyr old. We adopt this wavelength range for the primary morphological analysis of the galaxies. This allows us to maximize the redshift coverage in the optical regime. In addition, we consider the rest-frame U band, which traces a comparatively younger stellar population and is sensitive to the recent star-formation. A comparison of the structure in these two rest-frame bands can provide insight into how the recent star-formation activity is distributed relative to the underlying stellar structure. As we discuss systems requiring two-component fits in later sections, we use the above comparison to explore whether any of the disk components exhibit extended UV (XUV) disk-like phases. In cases where multiple \textit{JWST} filters probe these wavelength ranges, we use the one with which there is a higher overlap. We prepare the galaxy sample for modeling their 2D light distributions using \textsc{GALFIT} \citep{peng-etal2002, peng-etal2010} by obtaining cutouts of 2.5x2.5 arcsec. We mask all neighboring objects prior to modeling, by using the segmentation image obtained from the JADES archive. We note that full coverage of all sources across all the \textit{JWST} filters is not available and thus not every galaxy in the sample has imaging in the desired filters. For example, an object may be present in the FRESCO or JADES catalog, but no imaging in let's say, the F356W filter. Additionally, we do a visual inspection and discard further objects from the sample because they are evidently too faint and/or compact to model. The resulting sample consists of 222 galaxies for which we perform the morphological analysis.

\subsection{Single component fit}

We begin with a single Sersic function to model the galaxy sample. Here, the intensity at a distance R is given by

\begin{equation}
\label{eq:sersic_function}
I(R) = I_e \exp\left\{ -b_n \left[ \left( \frac{R}{R_e} \right)^{1/n} - 1 \right] \right\}
\end{equation}

where, I$_{e}$ is the intensity at the half-light or effective radius R$_{e}$, $n$ is the Sersic index and b$_{n}$ is such that $\Gamma$(2n) = 2$\gamma$(2n, b$_{n}$). During the modeling, we take the PSF of the appropriate \textit{JWST} filter into account, so that we obtain the intrinsic structural parameters. We apply conservative bounds on the parameter ranges for stability and to avoid unphysical results as follows: 0.2 $\leq$ $n$ $\leq$ 5.0 and 1.5 $\leq$ R$_{e}$ $\leq$ 10 pixels. The lower limit for R$_{e}$ is chosen so that it is not smaller than the PSF of the image ($\sim$ 3 pixel FWHM). An upper limit on R$_{e}$ =10 pixels (0.3 arcsec) is used which is equivalent to $\sim$1.7 kpc at $z\sim$6; and 1 kpc at $z\sim$ 13.

%----new figure set for 2-component objects----

\figsetstart
\figsetnum{1}
\figsettitle{Morphological decomposition carried out for two-component galaxies in the sample}

\begin{figure*}[htp]
\figurenum{1}
\centering
\plotone{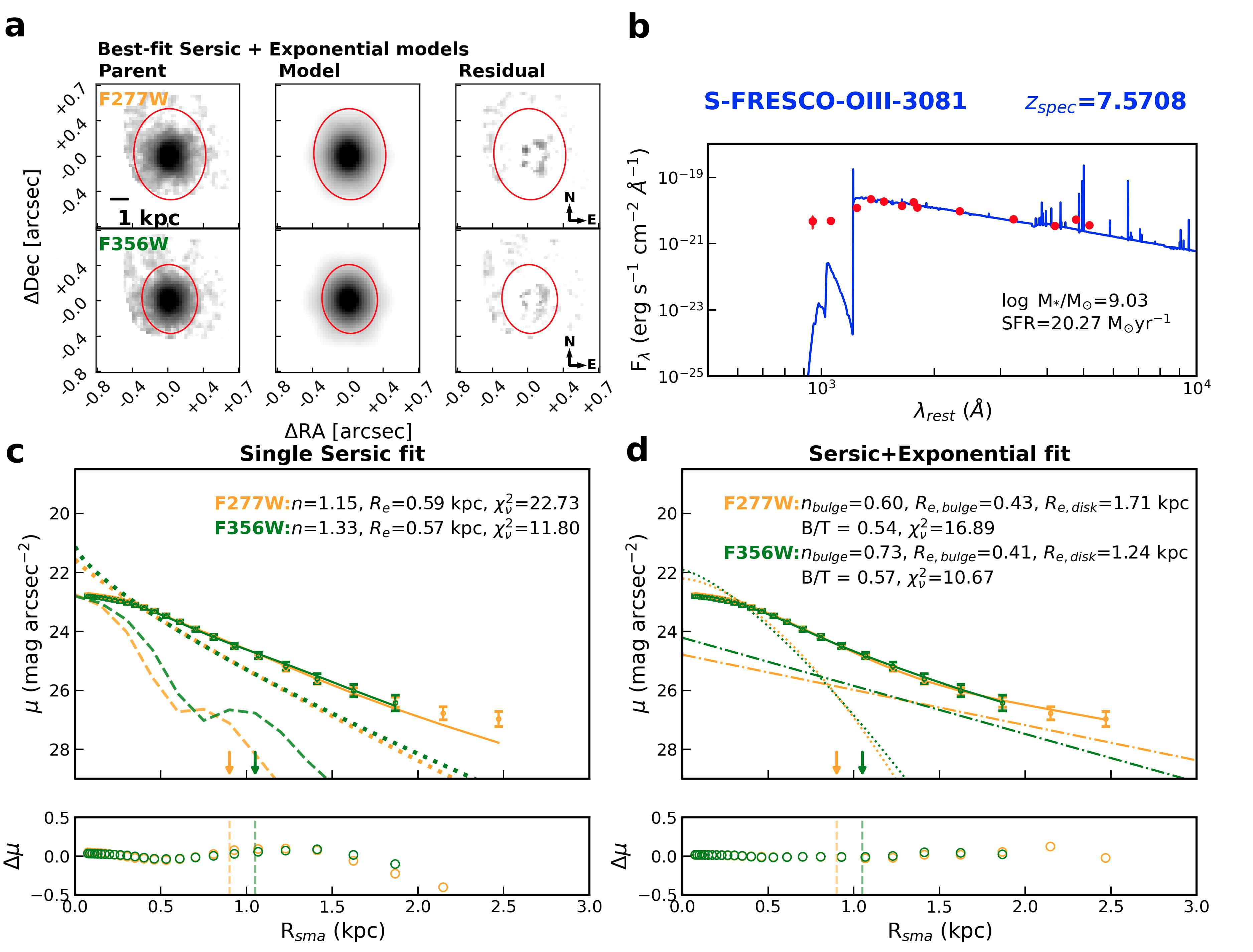}
\caption{\textbf{Morphological decomposition carried out for two-component galaxies in the sample. a:} The grayscale images, models and residuals of the galaxy in the appropriate \textit{JWST} filter probing rest-frame U and B band emission. The red ellipses mark the S/N$\sim$3 extent of the galaxy in the respective filters. \textbf{b:} The modeled rest-frame SED of the galaxy. \textbf{c:} The observed and modeled surface brightness profiles using a single Single Sersic model. The solid curves represent the modeled profiles (PSF convolved model) fitted to the observed datapoints. The dotted curves represent the intrinsic models of the galaxy in the respective filter. The dashed curve represents the PSF profile in the appropriate \textit{JWST} filter. The upside down arrows represent the extent that encloses 80\% of the PSF flux. These are represented below by vertical lines in the panel for the residual profiles, $\Delta\mu$. \textbf{d:} Same as panel c, but using an inner Sersic + Exponential Disk model for the galaxy. The errorbars represent 1-$\sigma$ errors. The complete figure set (20 images) is available in the online journal.}   
\label{fig:representative}
\end{figure*}

\figsetgrpstart
\figsetgrpnum{1.1}
\figsetgrptitle{N-FRESCO-Ha-1324}
\figsetplot{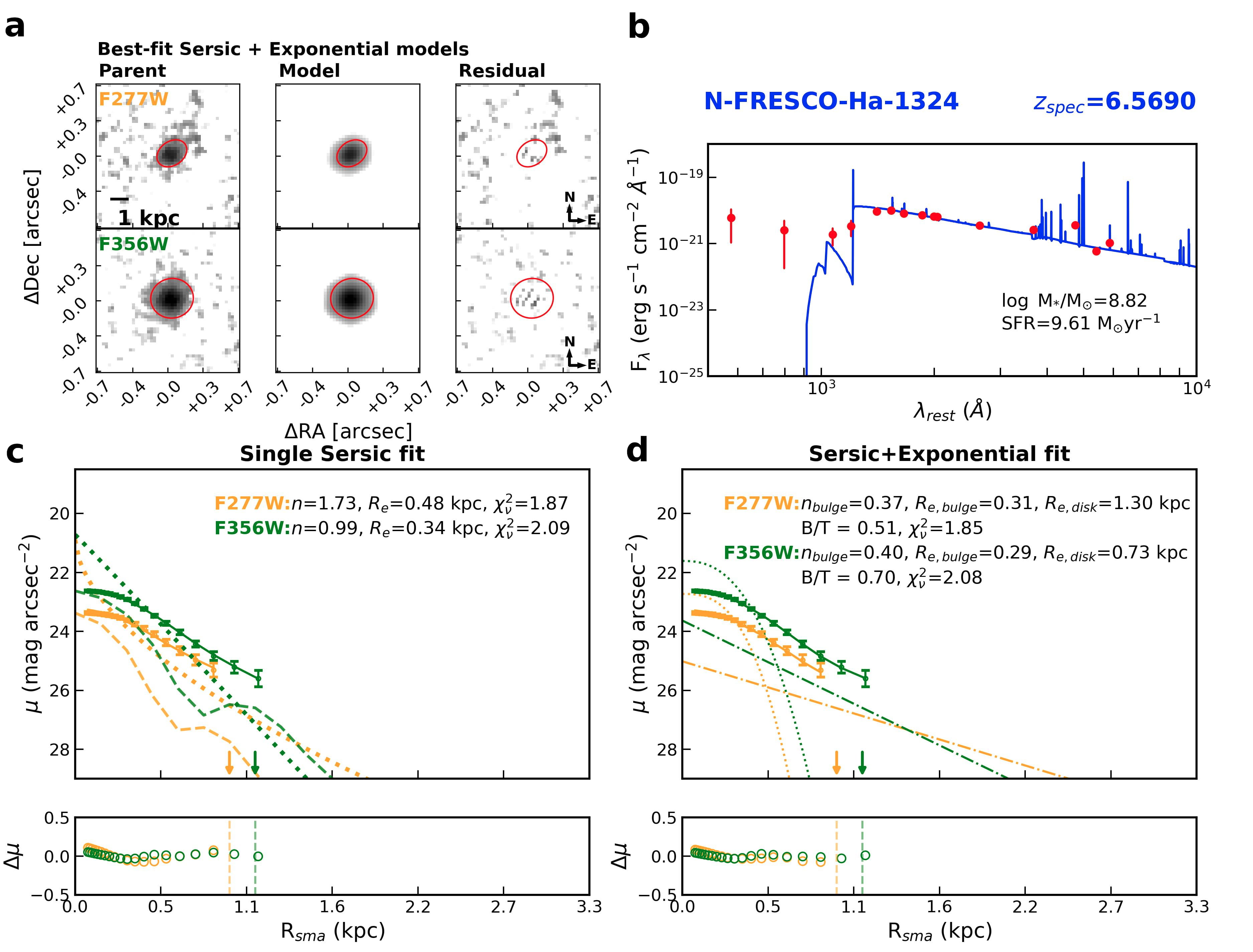}
\figsetgrpnote{See Figure \ref{fig:representative} for description.}
\figsetgrpend

\figsetgrpstart
\figsetgrpnum{1.2}
\figsetgrptitle{N-FRESCO-OIII-2203}
\figsetplot{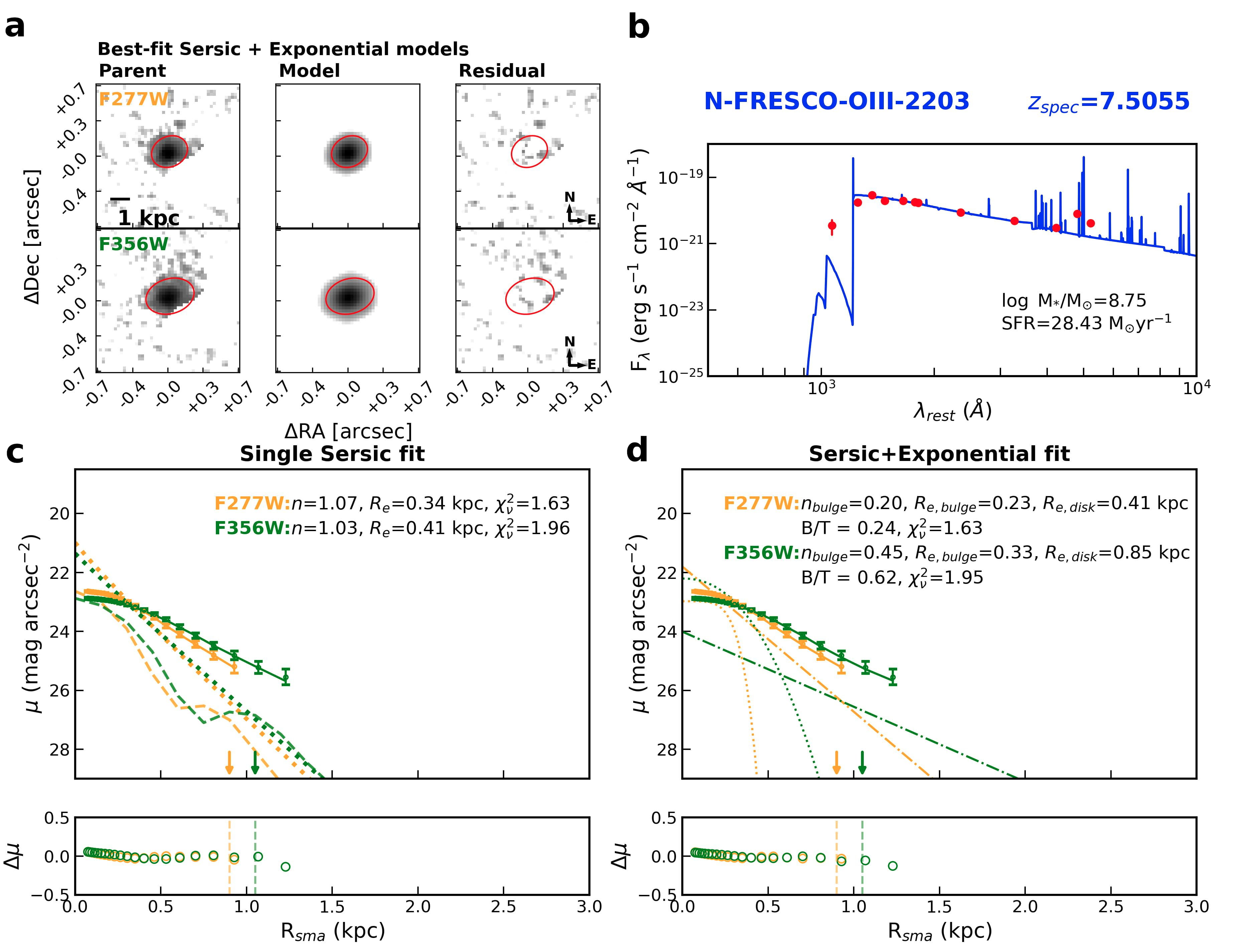}
\figsetgrpnote{See Figure \ref{fig:representative} for description.}
\figsetgrpend

\figsetgrpstart
\figsetgrpnum{1.3}
\figsetgrptitle{N-JADES-1009077}
\figsetplot{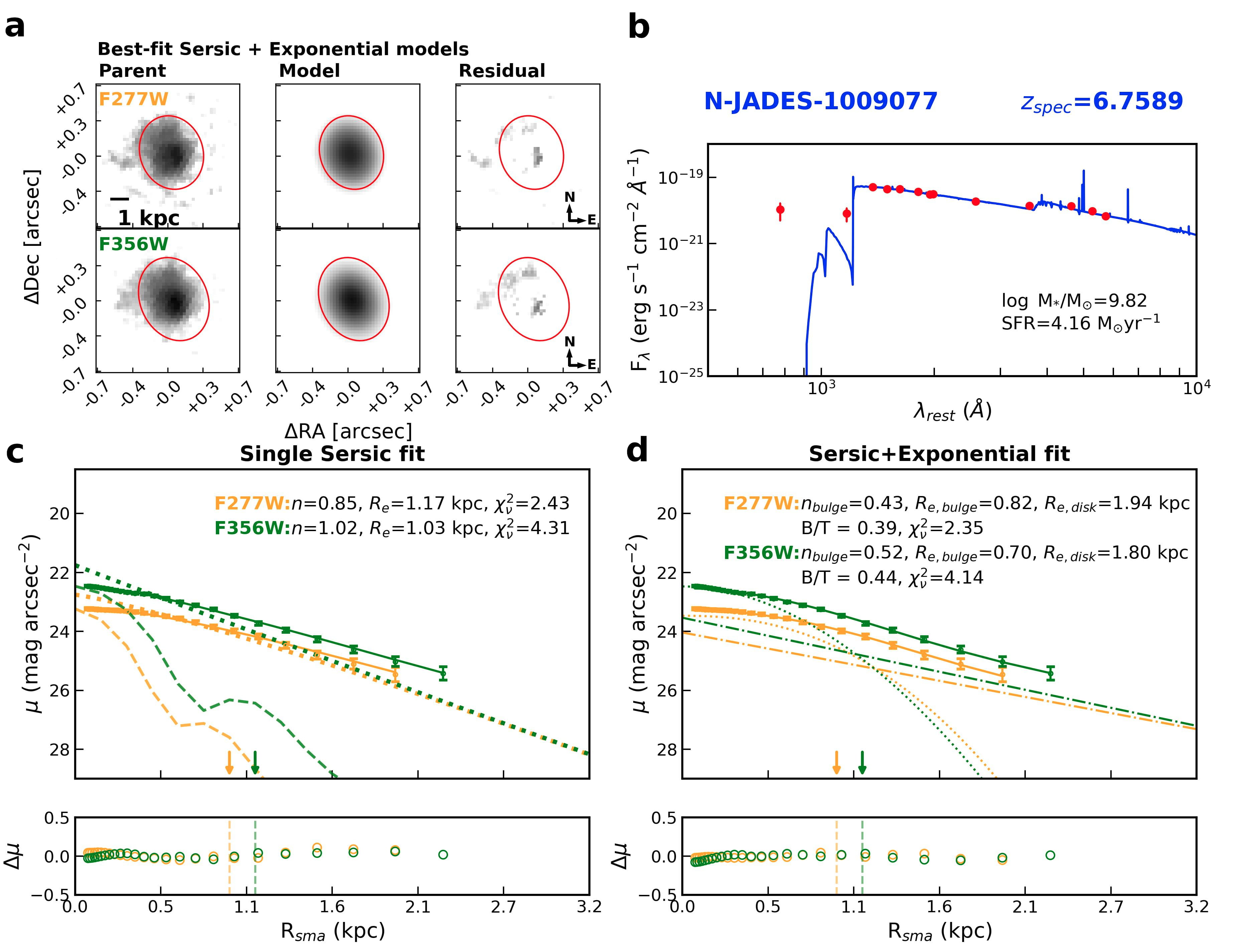}
\figsetgrpnote{See Figure \ref{fig:representative} for description.}
\figsetgrpend

\figsetgrpstart
\figsetgrpnum{1.4}
\figsetgrptitle{N-JADES-1010816}
\figsetplot{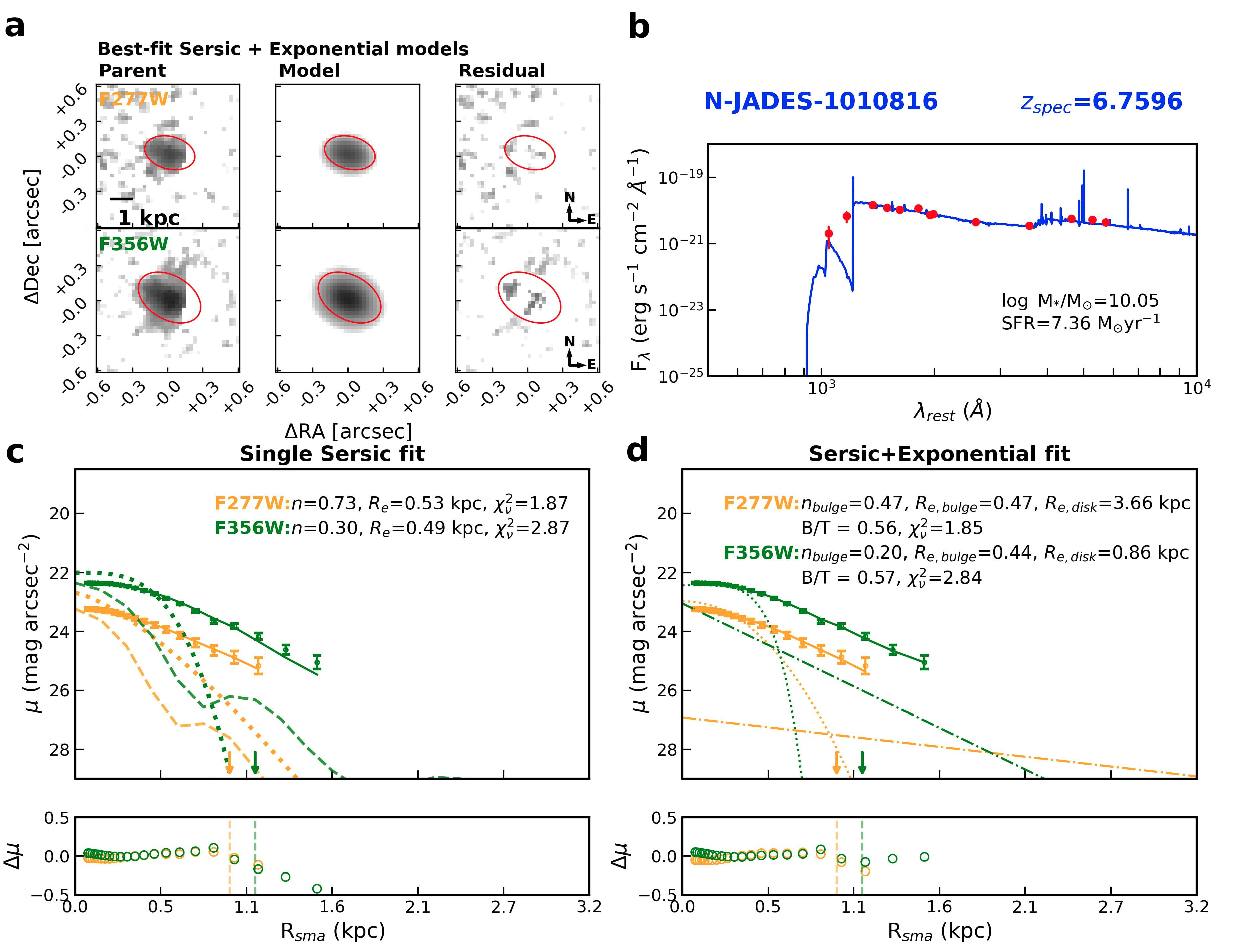}
\figsetgrpnote{See Figure \ref{fig:representative} for description.}
\figsetgrpend

\figsetgrpstart
\figsetgrpnum{1.5}
\figsetgrptitle{N-JADES-1022943}
\figsetplot{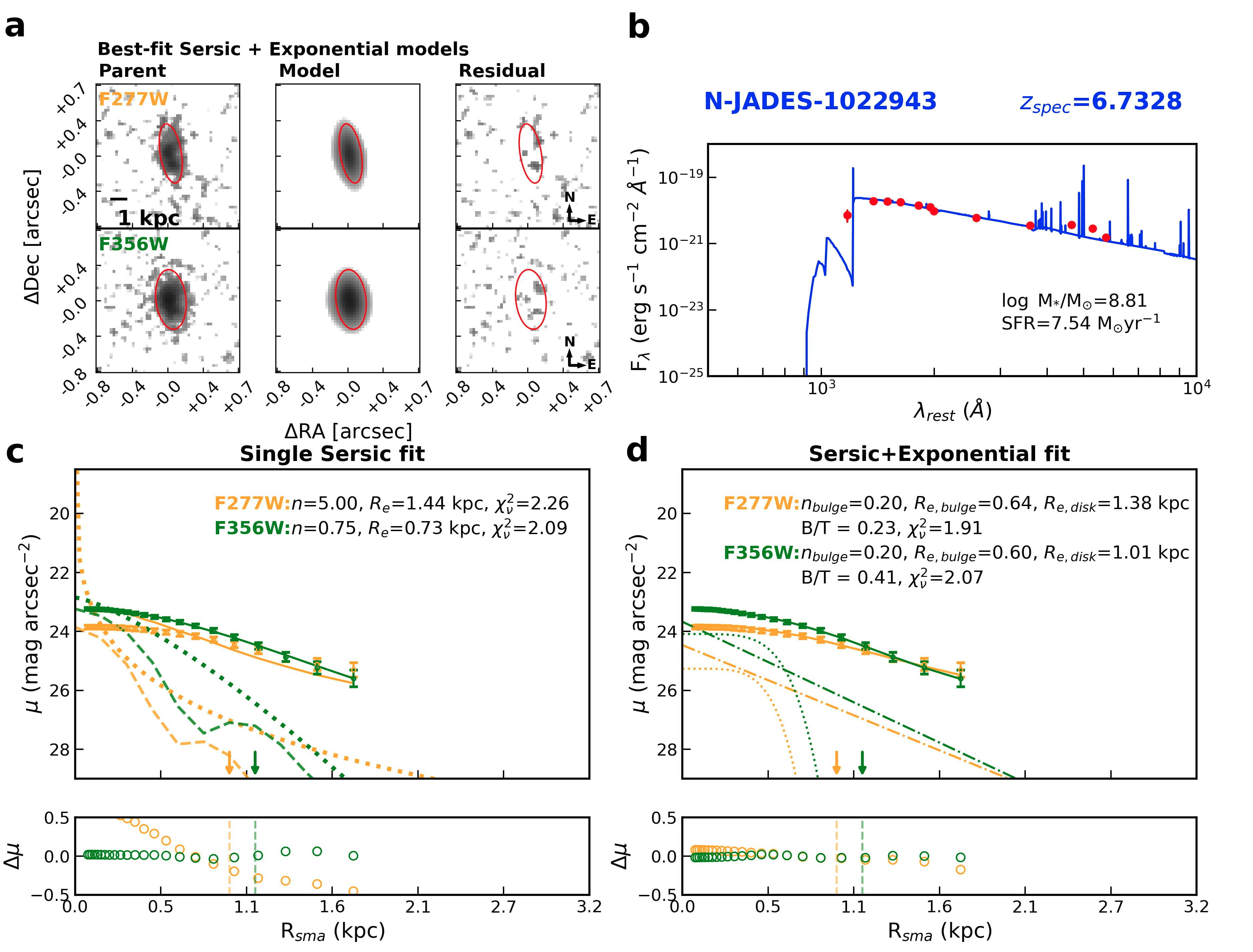}
\figsetgrpnote{See Figure \ref{fig:representative} for description.}
\figsetgrpend

\figsetgrpstart
\figsetgrpnum{1.6}
\figsetgrptitle{N-JADES-1078455}
\figsetplot{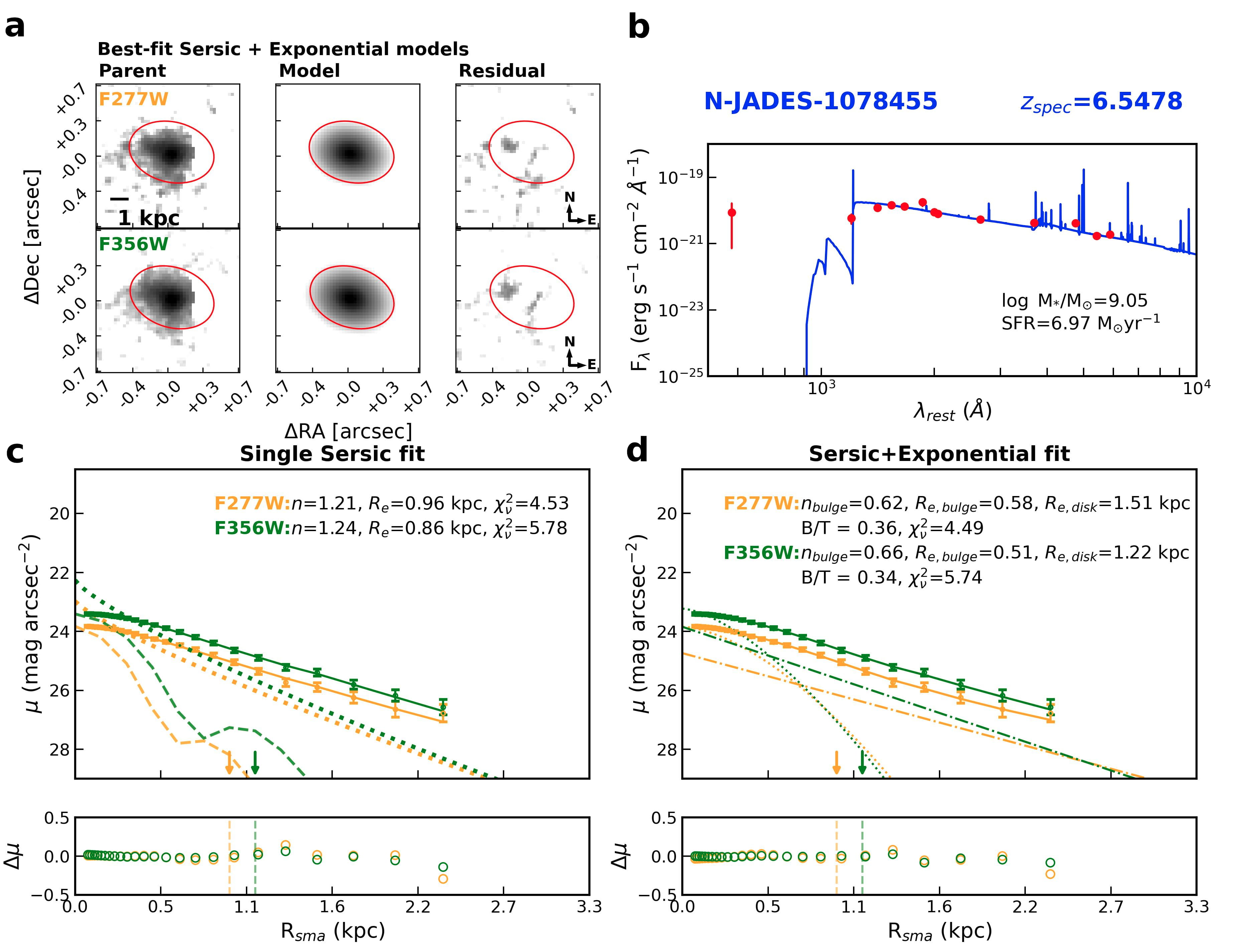}
\figsetgrpnote{See Figure \ref{fig:representative} for description.}
\figsetgrpend

\figsetgrpstart
\figsetgrpnum{1.7}
\figsetgrptitle{N-JADES-1078891}
\figsetplot{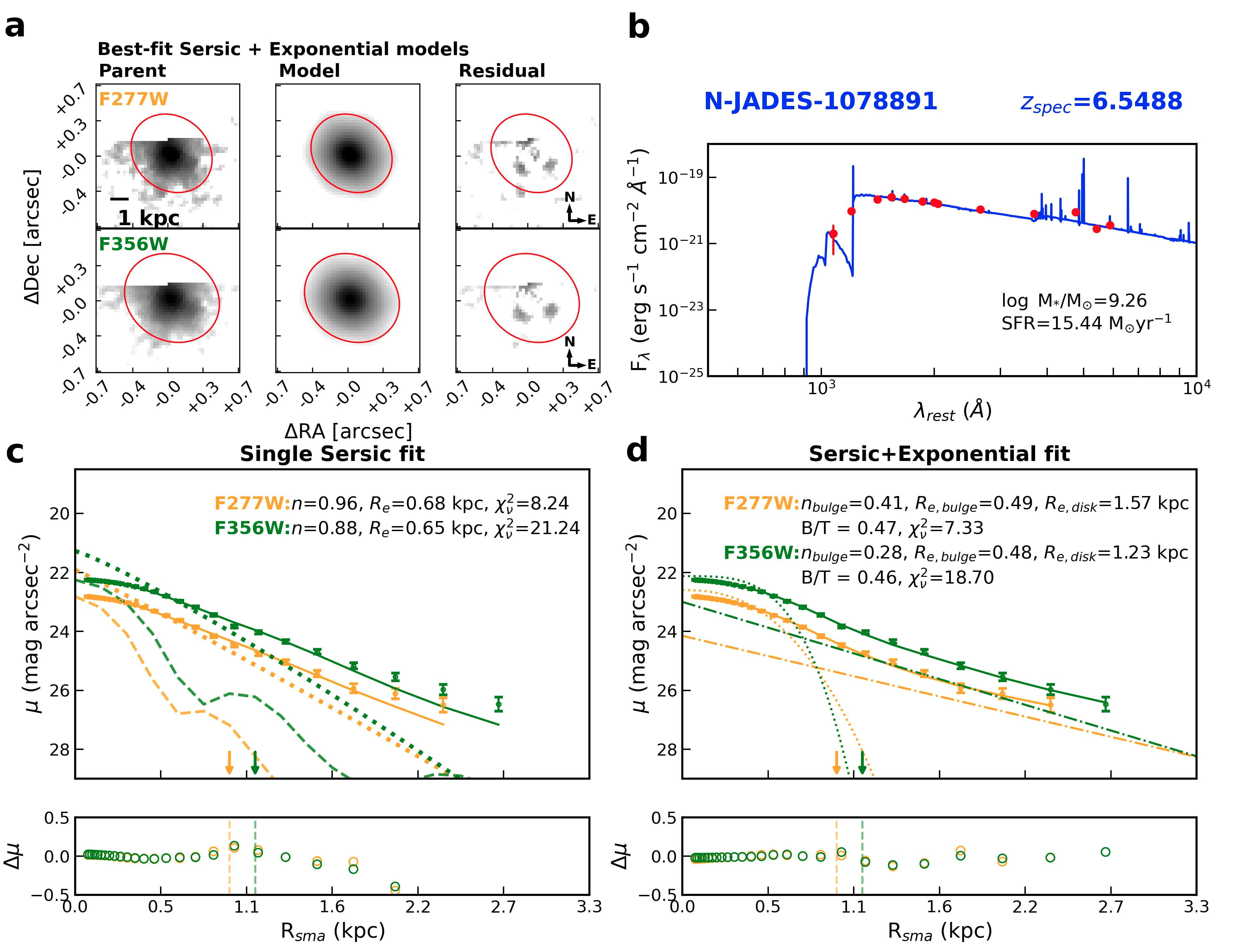}
\figsetgrpnote{See Figure \ref{fig:representative} for description.}
\figsetgrpend

\figsetgrpstart
\figsetgrpnum{1.8}
\figsetgrptitle{S-FRESCO-Ha-18617}
\figsetplot{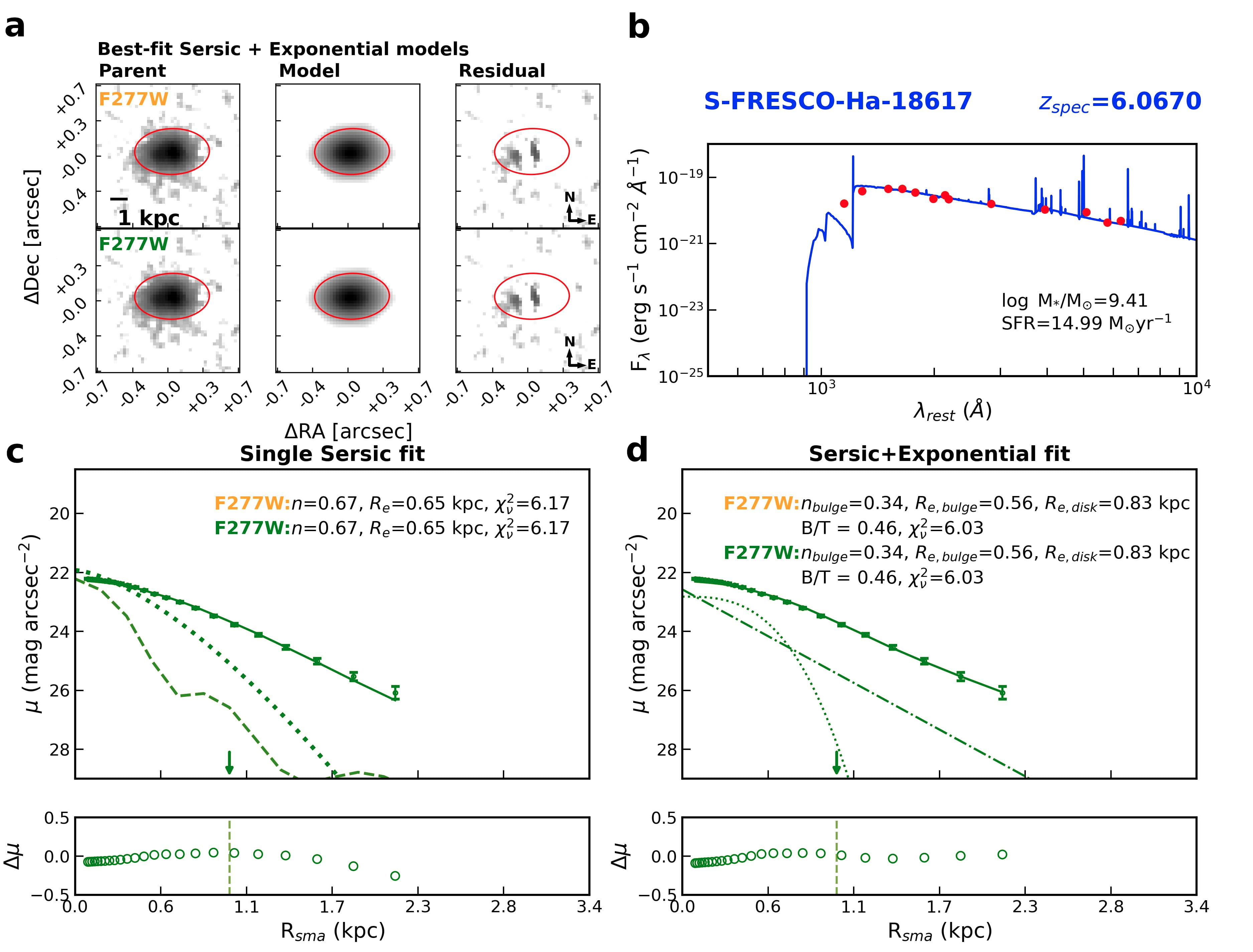}
\figsetgrpnote{See Figure \ref{fig:representative} for description.}
\figsetgrpend

\figsetgrpstart
\figsetgrpnum{1.9}
\figsetgrptitle{S-FRESCO-OIII-3081}
\figsetplot{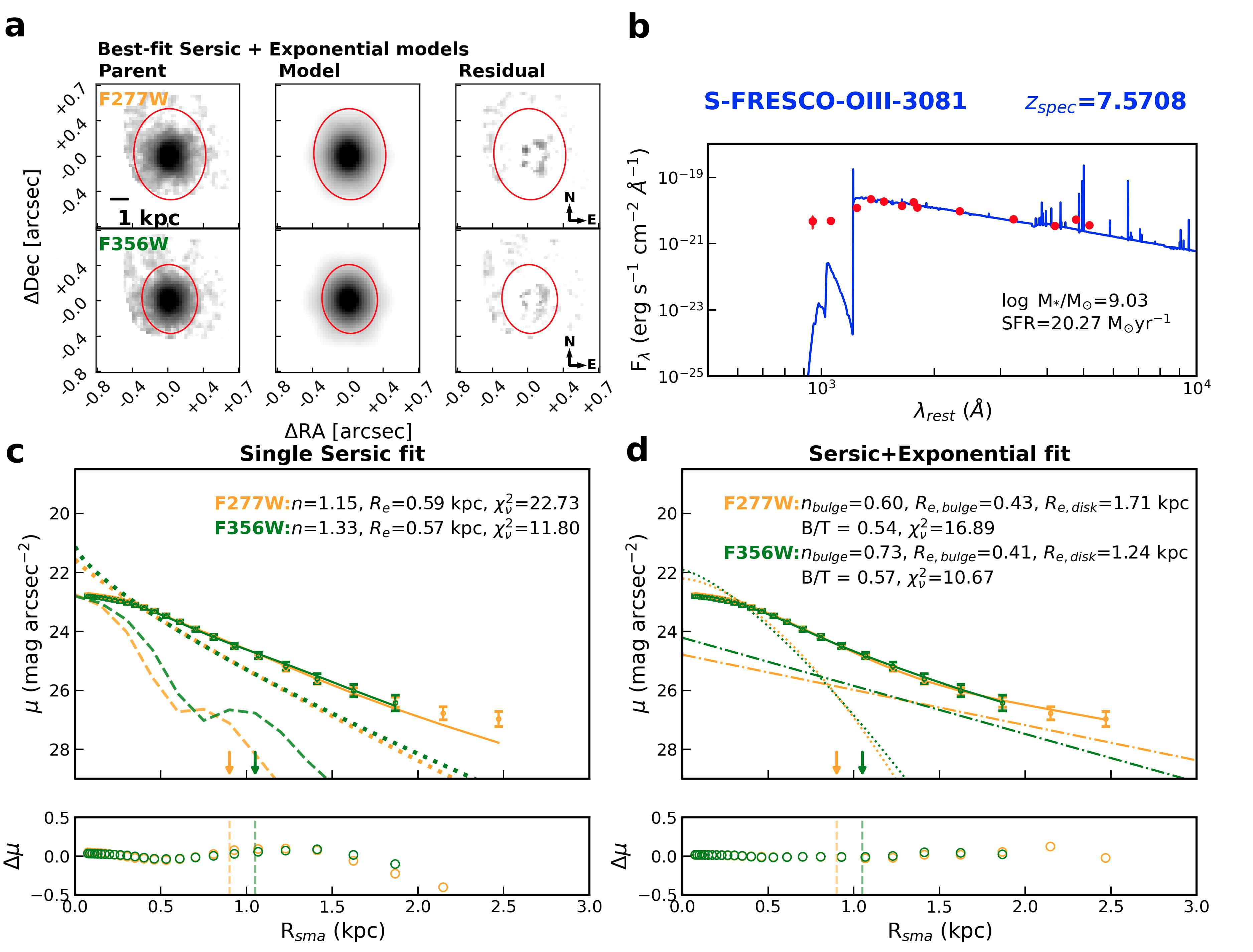}
\figsetgrpnote{See Figure \ref{fig:representative} for description.}
\figsetgrpend

\figsetgrpstart
\figsetgrpnum{1.10}
\figsetgrptitle{S-FRESCO-OIII-23077}
\figsetplot{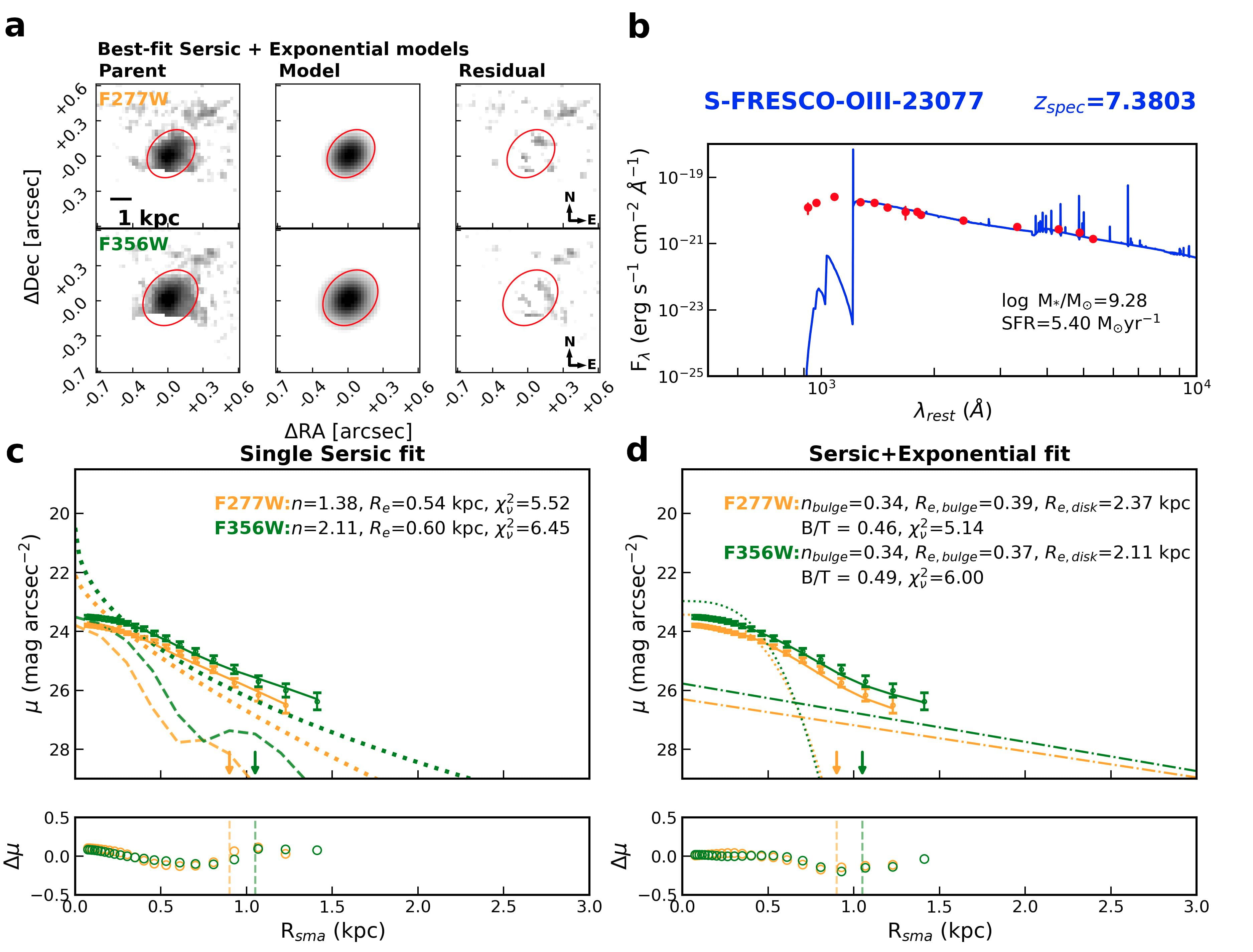}
\figsetgrpnote{See Figure \ref{fig:representative} for description.}
\figsetgrpend

\figsetgrpstart
\figsetgrpnum{1.11}
\figsetgrptitle{S-FRESCO-OIII-26147}
\figsetplot{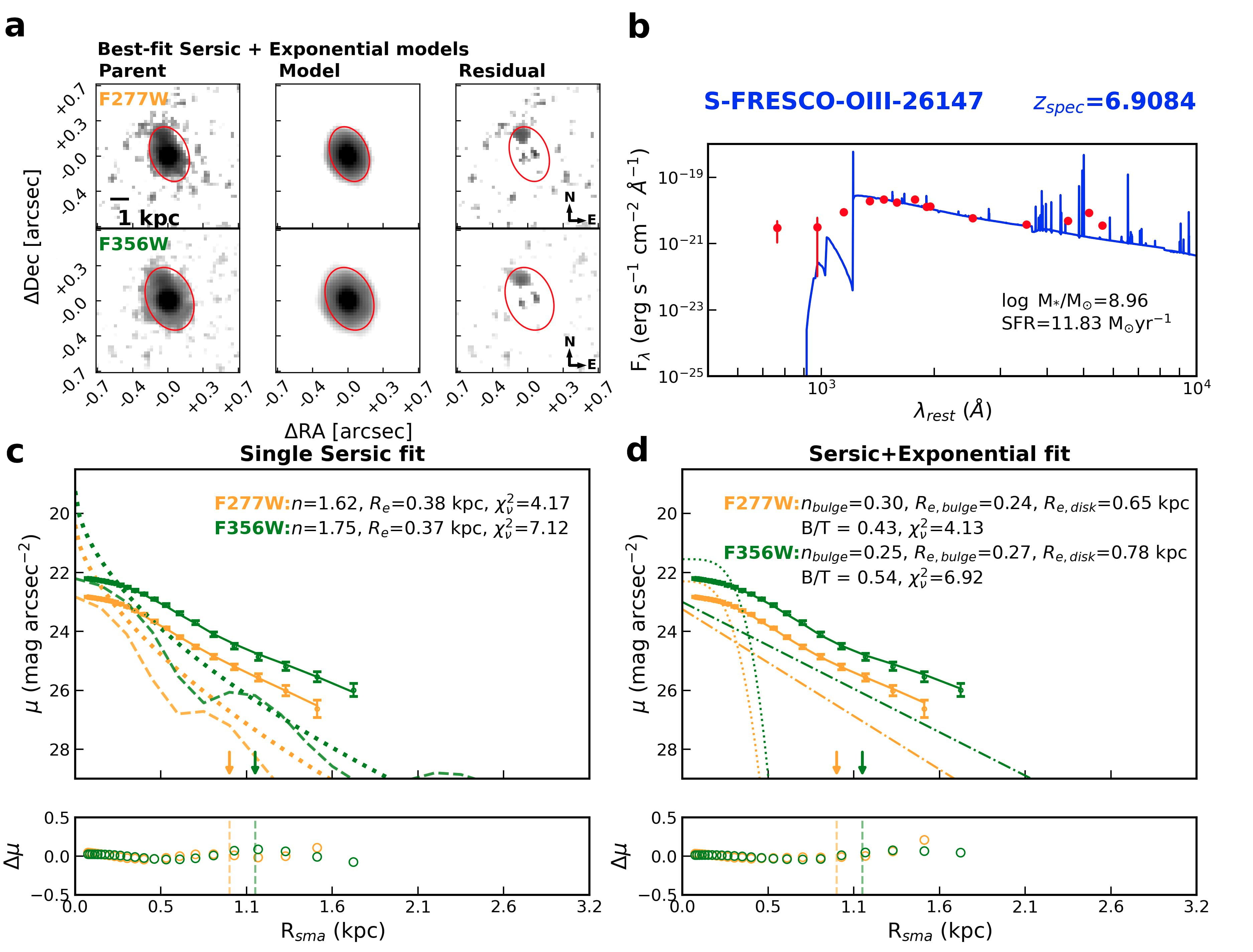}
\figsetgrpnote{See Figure \ref{fig:representative} for description.}
\figsetgrpend

\figsetgrpstart
\figsetgrpnum{1.12}
\figsetgrptitle{S-FRESCO-OIII-28631}
\figsetplot{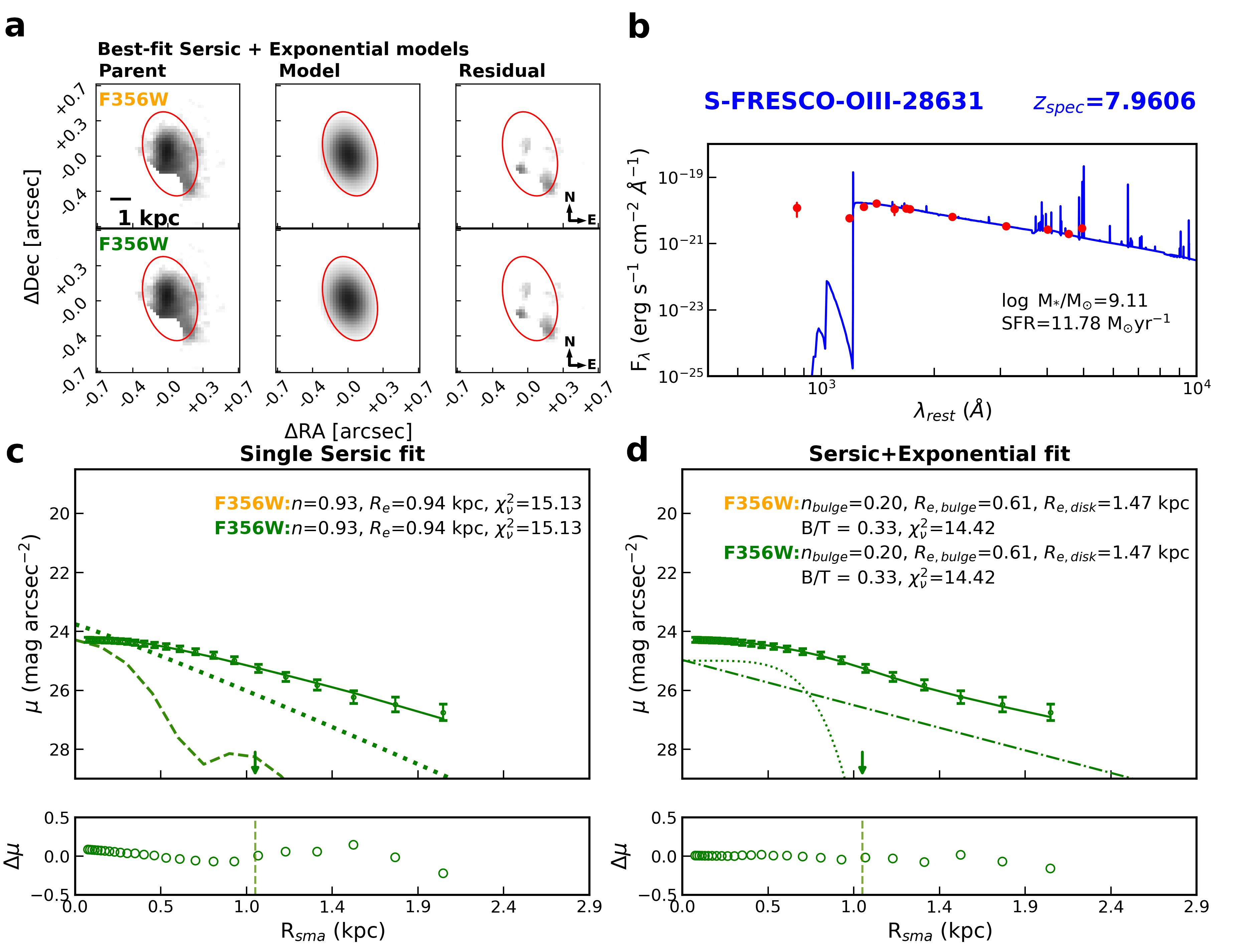}
\figsetgrpnote{See Figure \ref{fig:representative} for description.}
\figsetgrpend

\figsetgrpstart
\figsetgrpnum{1.13}
\figsetgrptitle{S-JADES-107324}
\figsetplot{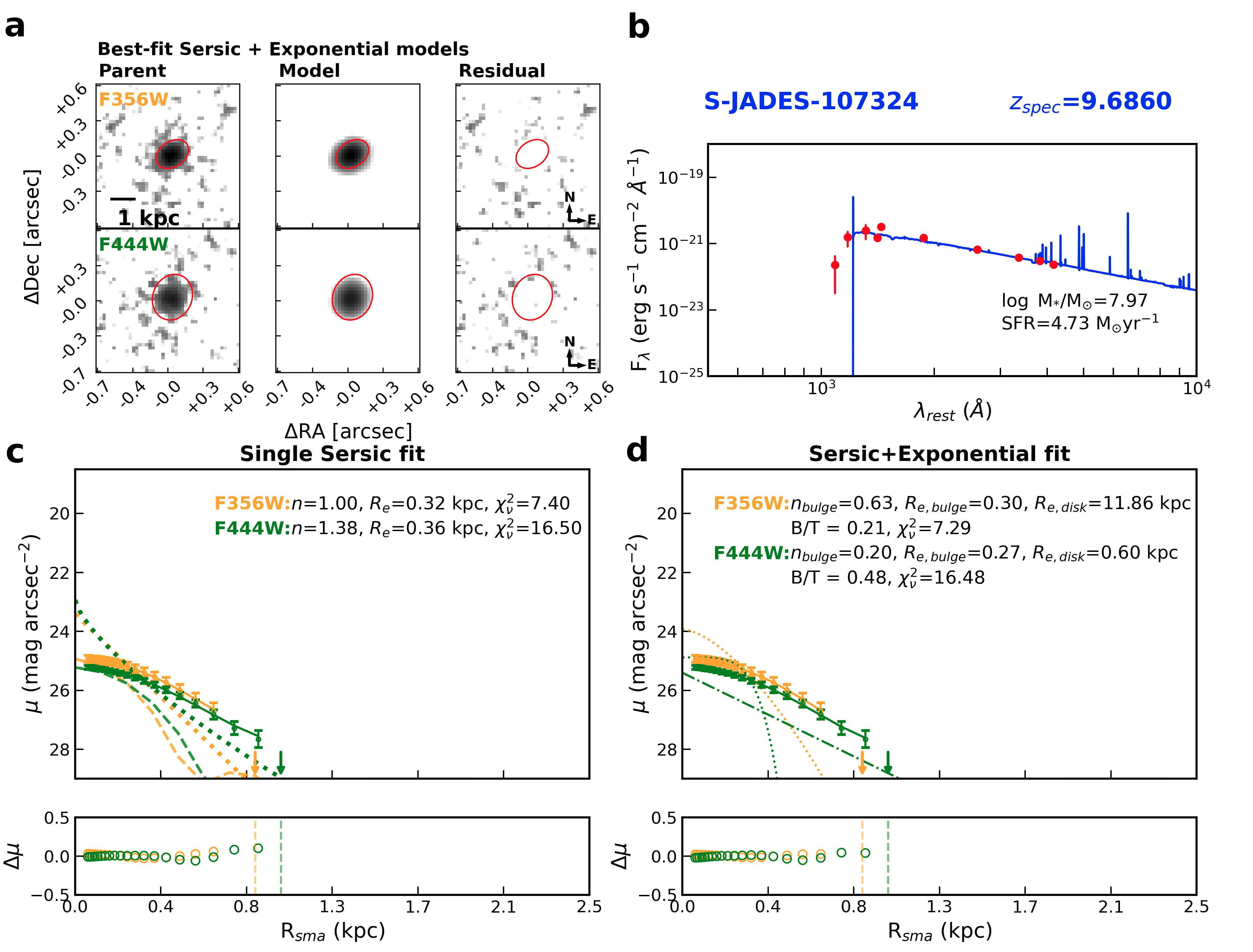}
\figsetgrpnote{See Figure \ref{fig:representative} for description.}
\figsetgrpend

\figsetgrpstart
\figsetgrpnum{1.14}
\figsetgrptitle{S-JADES-116352}
\figsetplot{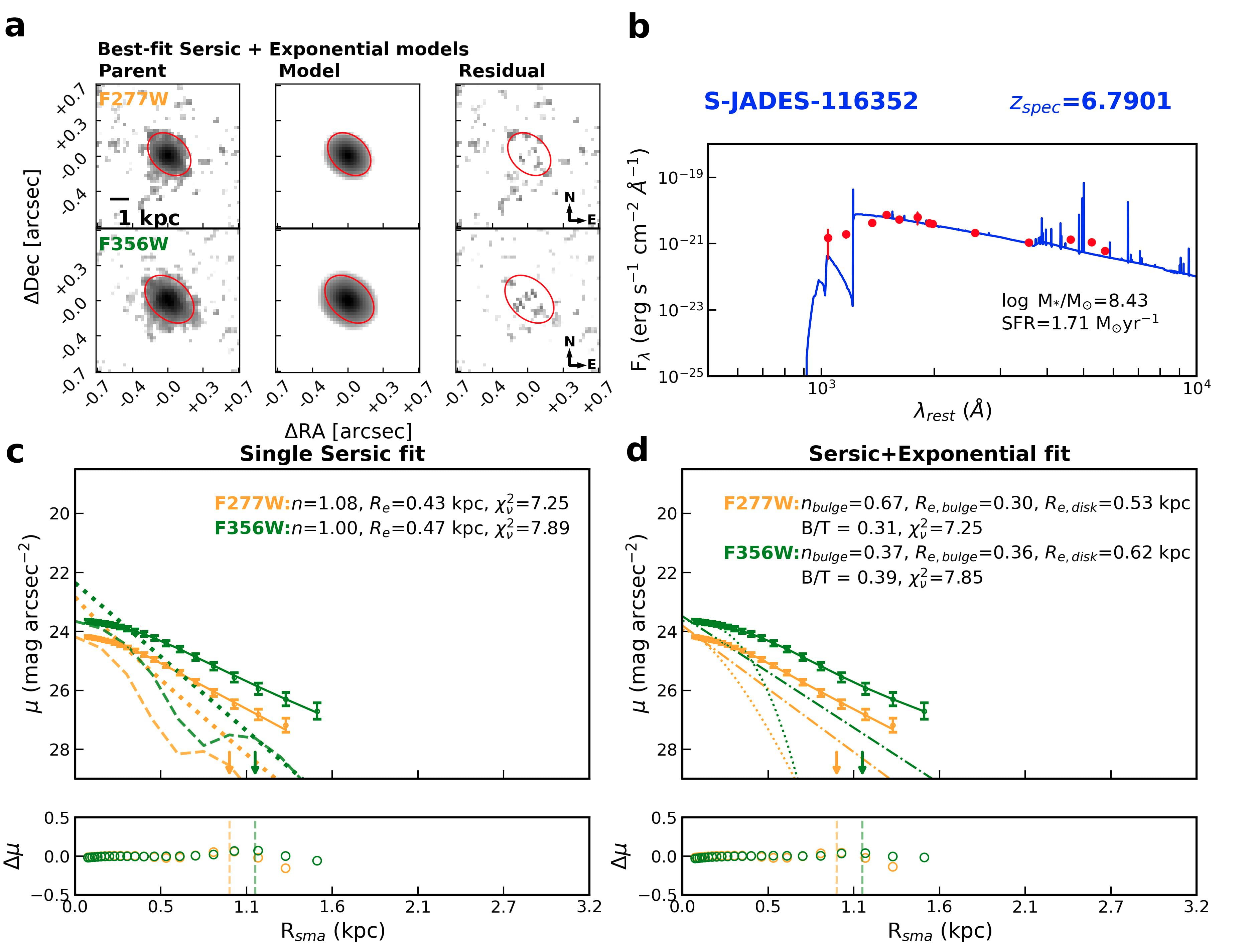}
\figsetgrpnote{See Figure \ref{fig:representative} for description.}
\figsetgrpend

\figsetgrpstart
\figsetgrpnum{1.15}
\figsetgrptitle{S-JADES-117602}
\figsetplot{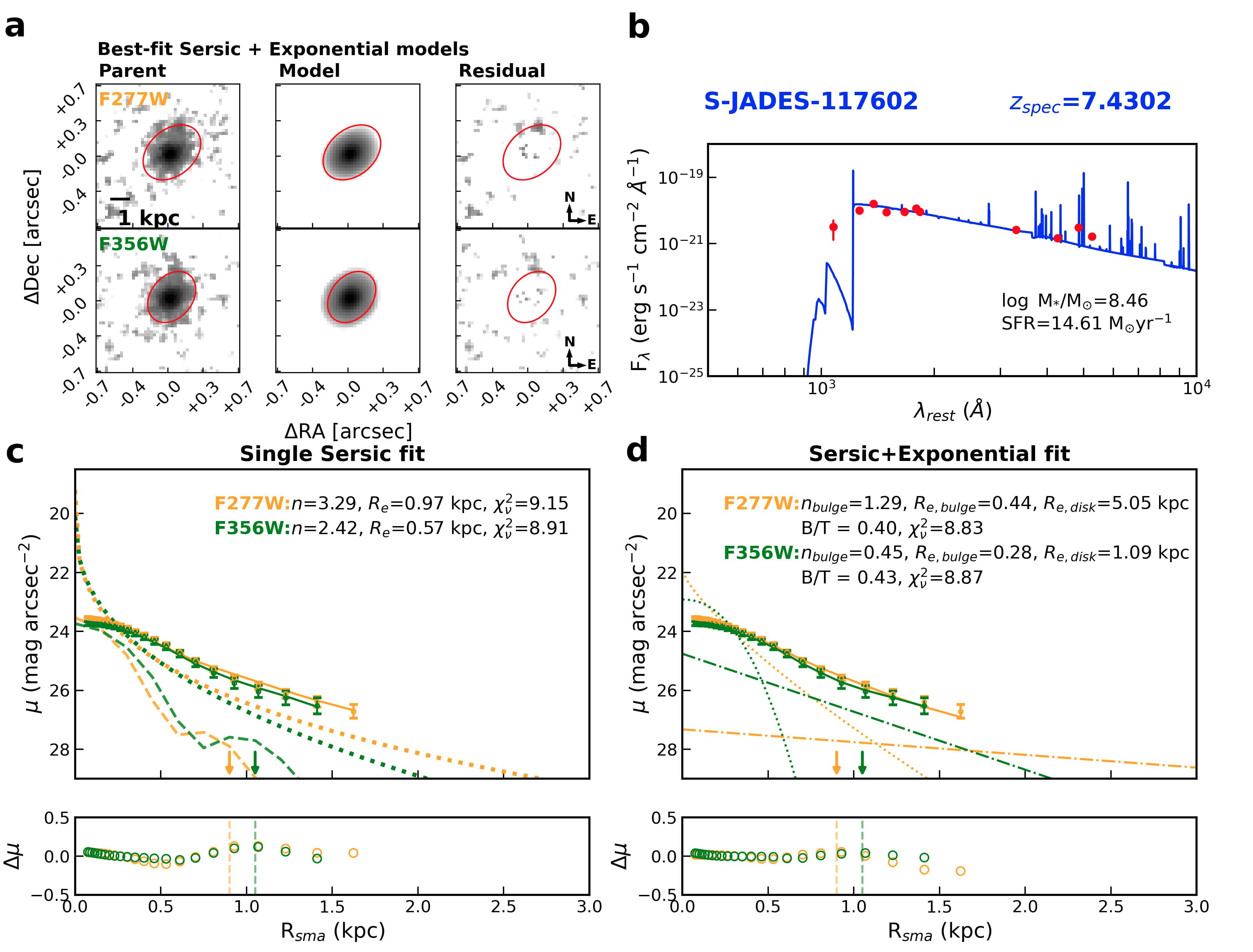}
\figsetgrpnote{See Figure \ref{fig:representative} for description.}
\figsetgrpend

\figsetgrpstart
\figsetgrpnum{1.16}
\figsetgrptitle{S-JADES-122436}
\figsetplot{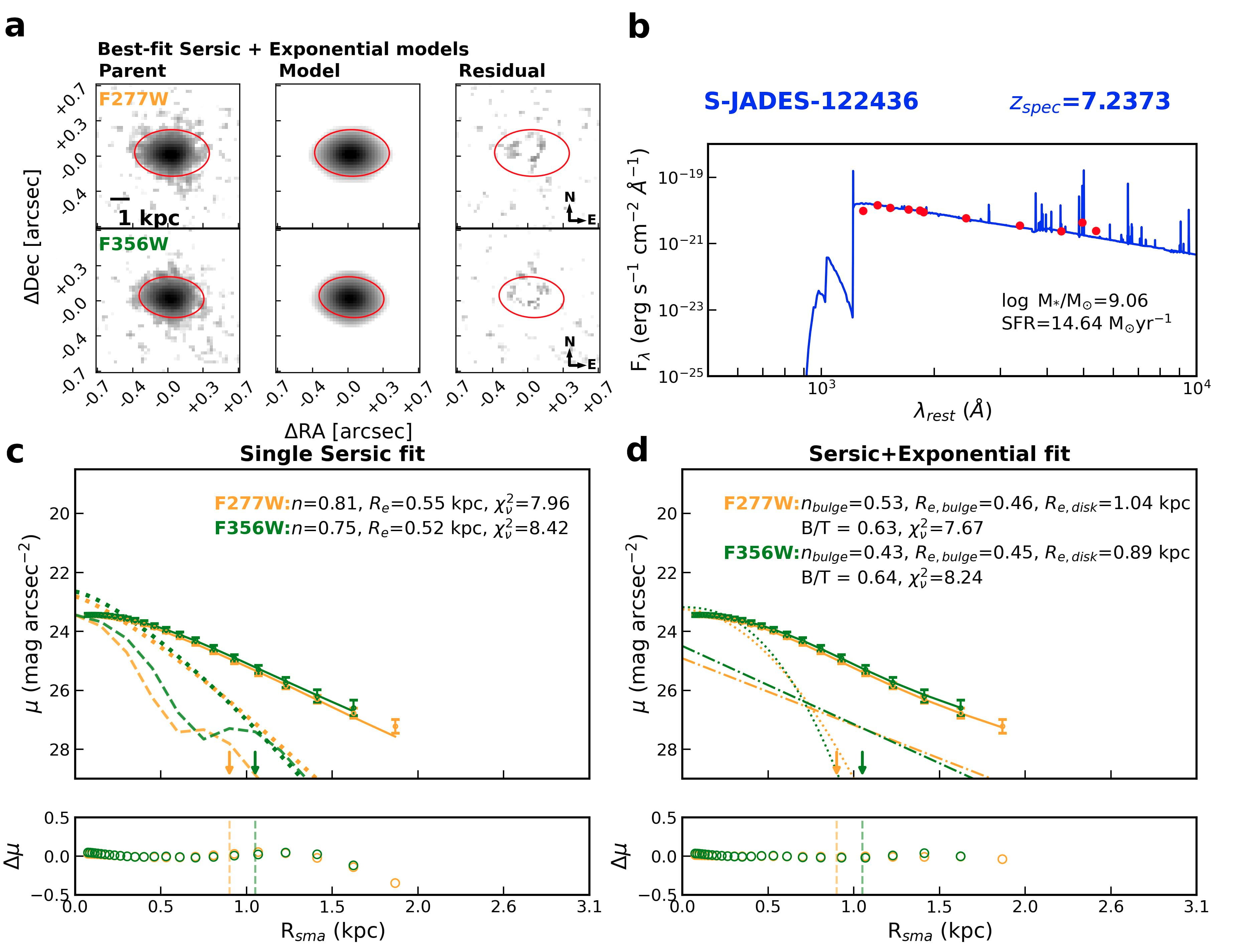}
\figsetgrpnote{See Figure \ref{fig:representative} for description.}
\figsetgrpend

\figsetgrpstart
\figsetgrpnum{1.17}
\figsetgrptitle{S-JADES-128502}
\figsetplot{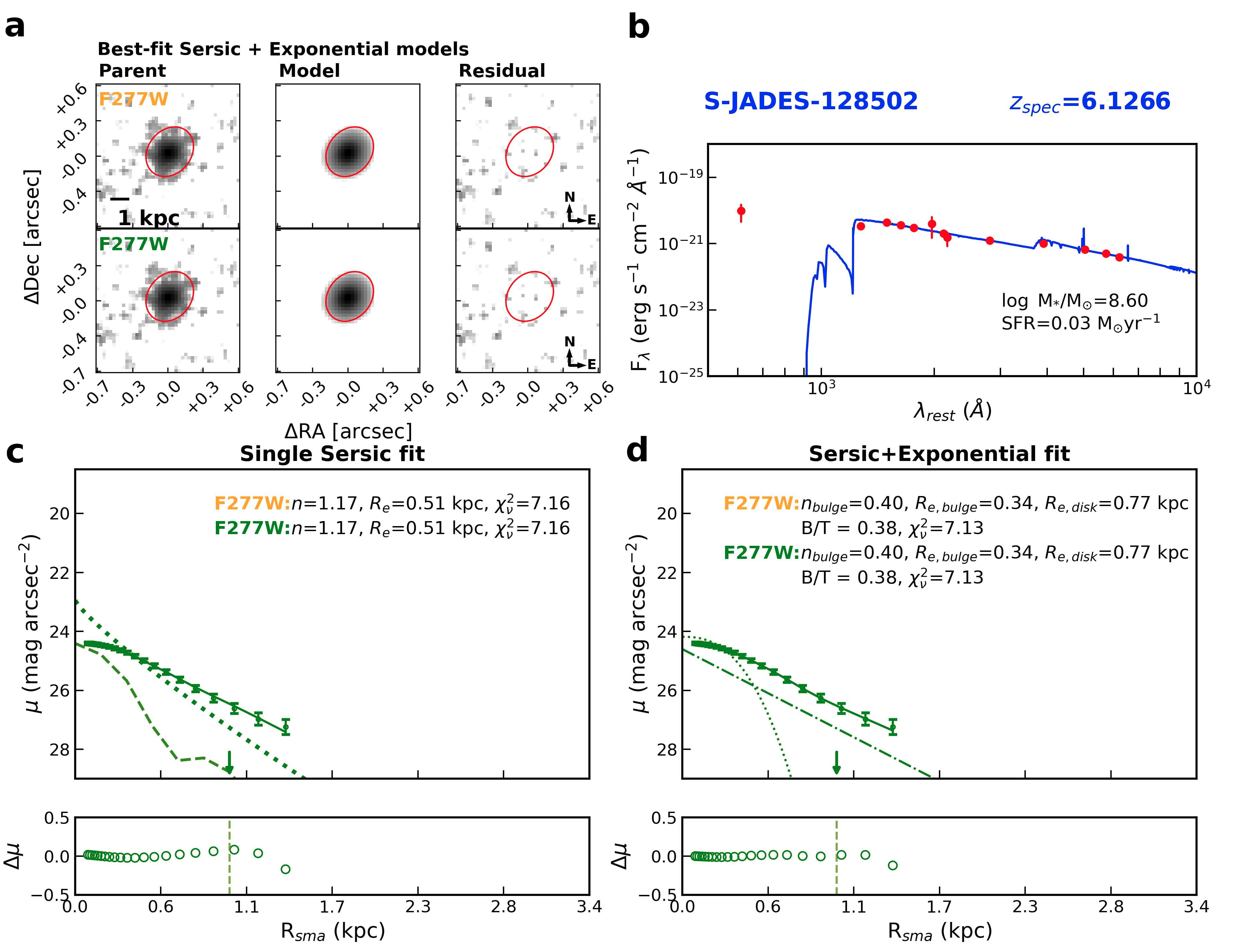}
\figsetgrpnote{See Figure \ref{fig:representative} for description.}
\figsetgrpend

\figsetgrpstart
\figsetgrpnum{1.18}
\figsetgrptitle{S-JADES-132845}
\figsetplot{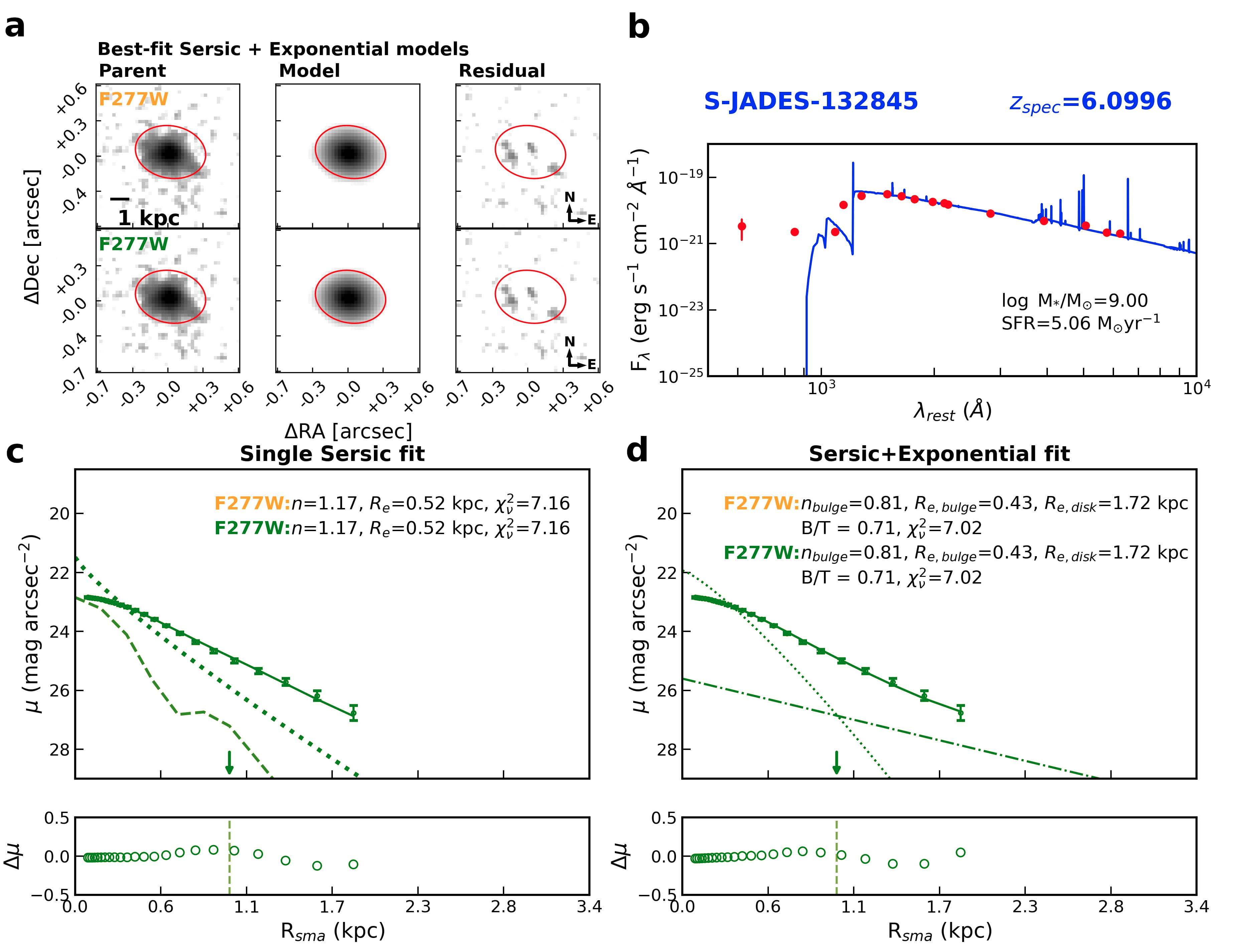}
\figsetgrpnote{See Figure \ref{fig:representative} for description.}
\figsetgrpend

\figsetgrpstart
\figsetgrpnum{1.19}
\figsetgrptitle{S-JADES-147779}
\figsetplot{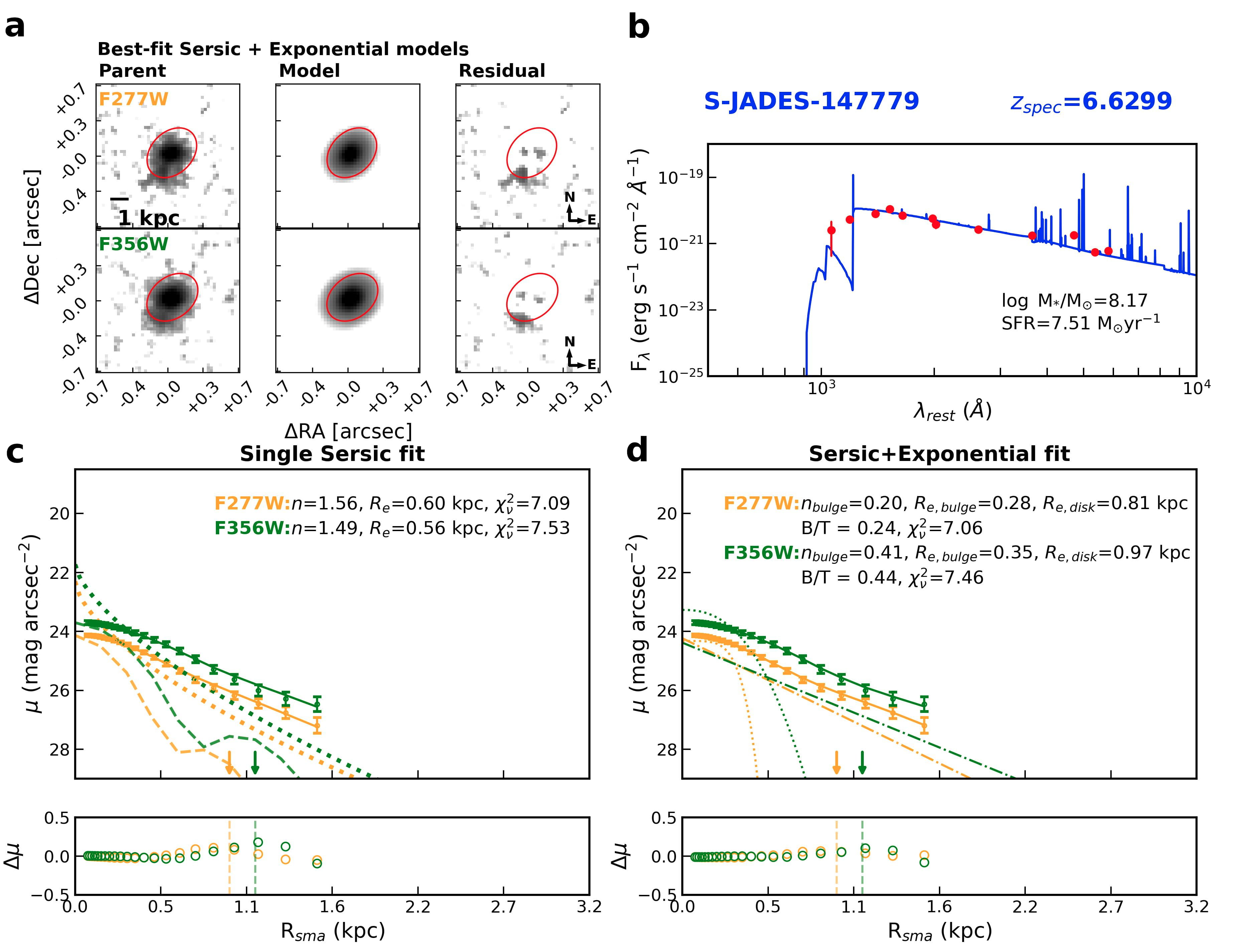}
\figsetgrpnote{See Figure \ref{fig:representative} for description.}
\figsetgrpend

\figsetgrpstart
\figsetgrpnum{1.20}
\figsetgrptitle{S-JADES-197888}
\figsetplot{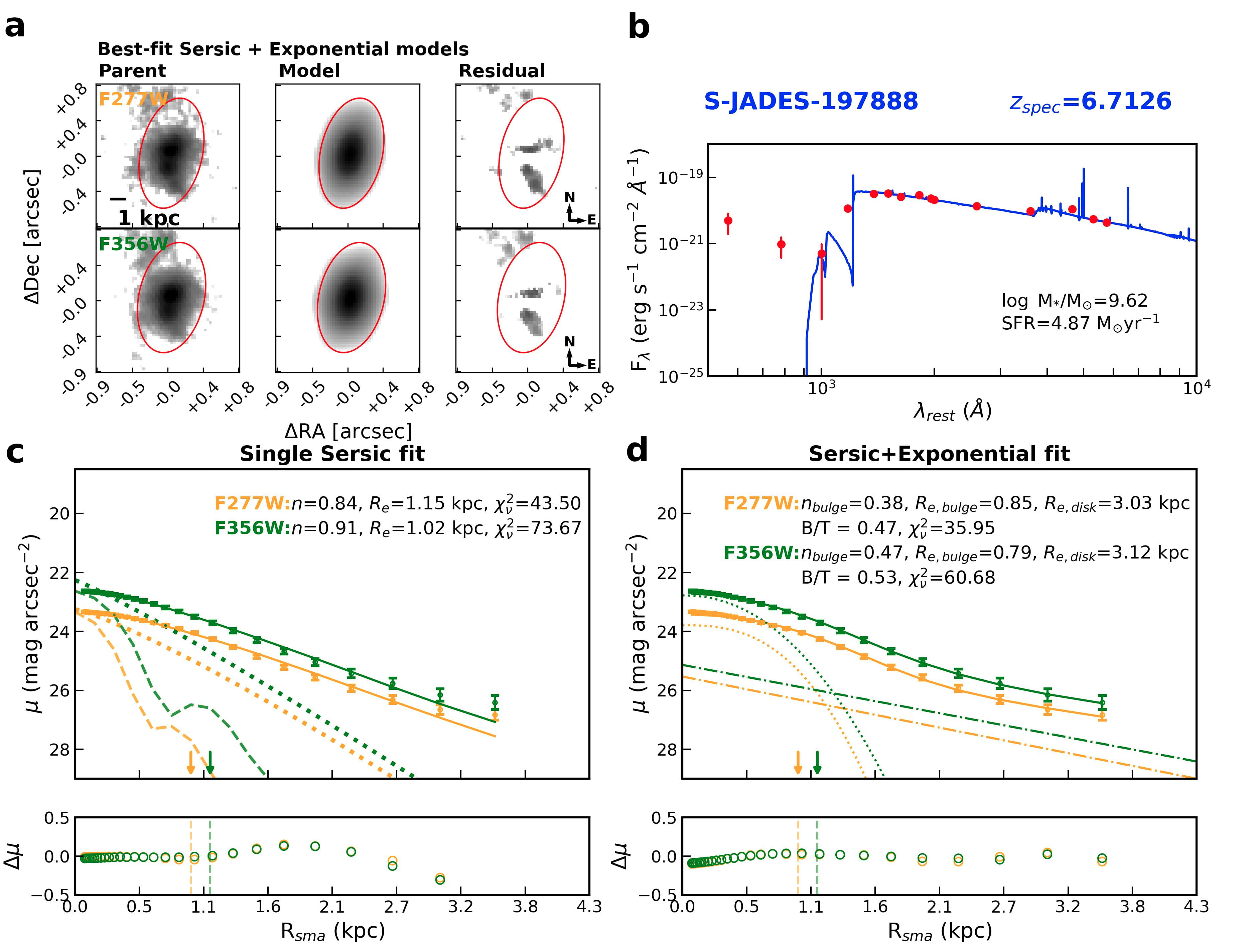}
\figsetgrpnote{See Figure \ref{fig:representative} for description.}
\figsetgrpend

\figsetend

\vspace{1cm}

\subsection{Double-component fit}

\begin{figure*}
\figurenum{2}
    \centering
    \includegraphics[width=0.8\textwidth]{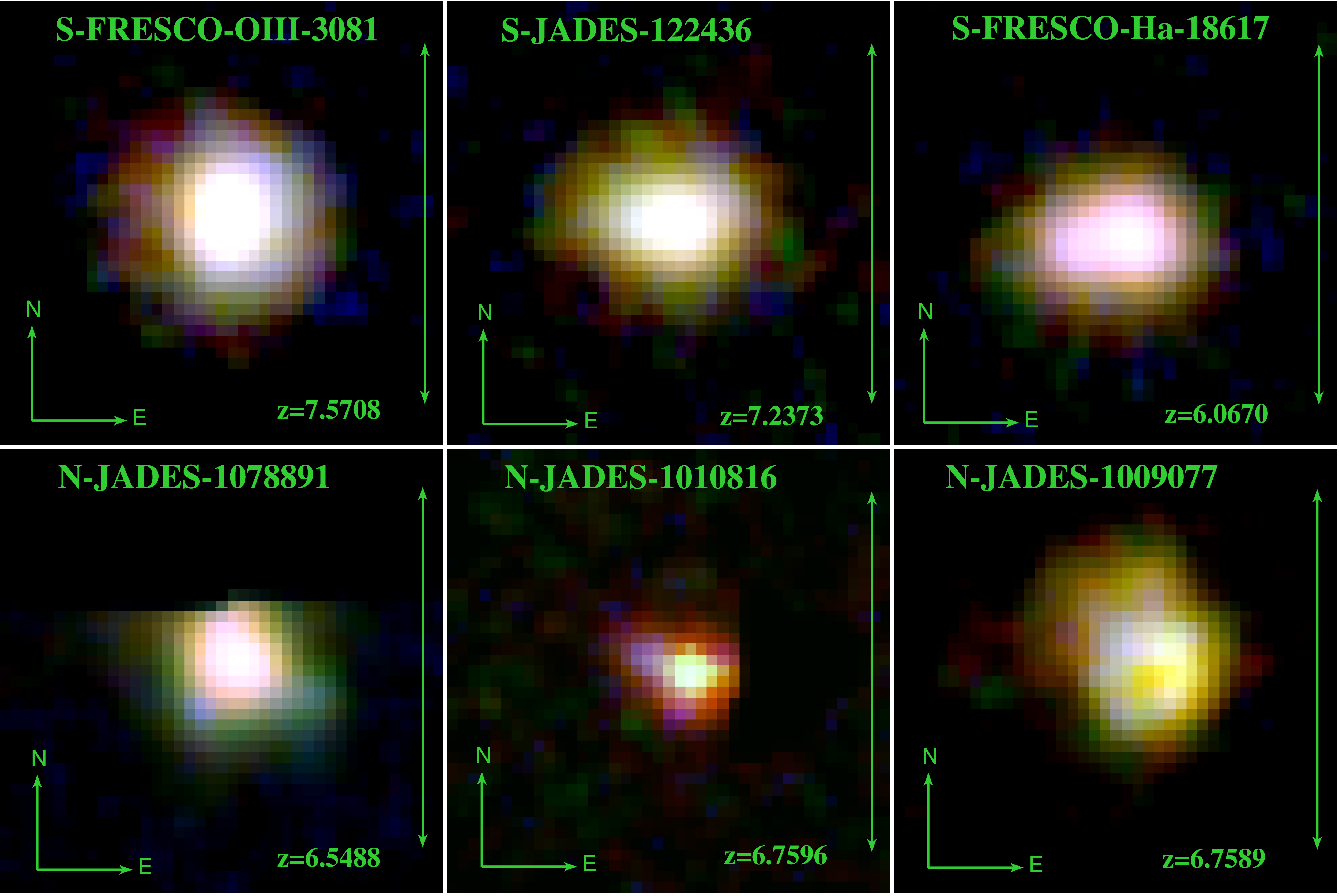}
    
    \caption{False-colour R (F444W), G (F277W or F356W), B (F115W) images of a few best cases of galaxies modeled with an inner Sersic and underlying disk component. The vertical lines in the images span an angular size of 1''. }
    \label{fig:color-images}
\end{figure*}

We perform a visual assessment of the residual-maps from the single component modeling along with the 1D residual profiles. Based on this, we discard a few of the galaxies with clear unsatisfactory fits and that did not have coverage of rest-frame \textit{B}-band emission. The rejected objects are compact with sizes comparable to the PSF, showing large deviations in their residual profiles and/or the best-fit parameters that converge to the imposed parameter boundaries, indicating unreliable fits. This quality control step leaves us with a final sample of 187 galaxies spanning $z_{spec} \sim 6 - 11$. This inspection is important for removing unreliable structural measurements within the modeling framework adopted here. We note that the final sample may not be statistically complete and may be biased against very compact galaxies for which the structural parameters cannot be robustly constrained. Our aim is to identify those objects for which a single Sersic model is insufficient to model the observed outer light distribution. Among these, we find 95 objects for which the outer extent in the rest-frame \textit{B}-band (extent beyond which $\mu_{err}$ drops below 0.3 mag arcsec$^{-2}$) lie beyond a radius enclosing $\sim$80\% of the PSF flux. We mark this as a reference radius because beyond this extent the observed galaxy profiles should become progressively less dominated by the PSF core. This also ensures that the galaxy has sufficient signal-to-noise ratio. Therefore, it is possible that if a galaxy extends beyond the 80\% PSF-flux radius and if a single Sersic model is insufficient to model the outer light distribution, there may be an underlying stellar disk. We note that the decision to include another component is non-trivial and it is quite challenging to distinguish multiple components at such high redshifts, as highlighted in previous studies \citep[e.g.,][]{meert-etal2015, lange-etal2016}. For robustness, we re-run \textsc{GALFIT} on the full sample by including an additional exponential disk component. For an exponential disk, equation \ref{eq:sersic_function} is modified accordingly for a value of $n$=1.

In our two-component modeling, we use the best-fit parameters from the single-Sersic fits as initial guesses and use the same parameter bounds except for effective radius of the inner Sersic component (R$_{e,~bulge}$). For brevity, we refer to the inner Sersic component as the bulge. Here, we explore the possibility of an outer stellar disk and not a compact disk embedded within the bulge component. As explored in \citet{lange-etal2016}, unphysical solutions may arise when the bulge component is larger than the disk component. With this motivation, we introduce appropriate constraints. Accordingly, in the second run, we restrict the upper bound of R$_{e,~bulge}$ to the value obtained during the single-component fit. We further configure our constraint file such that R$_{e,~bulge}$ remain smaller than or equal to the effective radius of the exponential disk component (R$_{e,~disk}$). During the fit, we fix the center, position angle (PA) and axis ratio ($q$) values to those obtained from the single-Sersic modeling. Thus, the exponential-disk component essentially contain two free parameters -- its magnitude and half-light size. We leave the Sersic index of the bulge component ($n_{bulge}$) free during the fit. We emphasize that keeping $n_{bulge}$ free is appropriate, given these central structures are in the process of ongoing assembly. This is important because different physical processes may be at play that give rise to a morphologically different bulge type. Hence, we do not assume a de Vaucouleurs bulge to be already in place during our two-component analysis, which is in contrast to recent morphological studies using \textit{JWST} observations \citep[e.g.,][]{genin-etal2025}. A representative example of two-component morphological decomposition is shown in Figure \ref{fig:representative}. 

The requirement for a complex model over a simple one can be evaluated by using the statistical Bayesian Information Criterion \citep[BIC,][]{schwarz1978}. This aids in deciding if a complex model should be preferred over a simple one \citep[e.g.,][]{Paulino-Afonso-etal2019} and given by:
\begin{equation}
    \text{BIC} = \chi^{2} + k \times ln(\text{N})    
\end{equation}
where $\chi^{2}$ is the goodness of fit value from \textsc{GALFIT}, $k$ is the number of free parameters of the model and N is the number of datapoints used during the modeling (number of pixels). We measure this quantity for the single-Sersic (BIC$_{S}$) and Sersic+exponential disk (BIC$_{SE}$) model and compute the difference to obtain an estimate $\Delta$BIC as follows:
\begin{equation}
   \label{eq:delta-bic}
   \Delta \text{BIC} = \text{BIC}_{SE} - \text{BIC}_{S}
\end{equation}
Previous works have shown that a $\Delta$BIC $<$ -10 \citep[e.g.,][]{Paulino-Afonso-etal2019, Tang-etal2025} can be used to favour a more complex model over a simple one. However, literature also demonstrates that BIC measurements alone are insufficient to robustly justify the need for an additional structural component, as that may lead to unphysical results \citep[e.g.,][]{mendez-abreu-etal2017}. Additional model components can partially absorb residual systematics or irregular outer structures and produce a `better' fit without any physical significance. This behaviour was evident during the inspection of the 1D residual profiles of our sample. While a majority of galaxies satisfy $\Delta \text{BIC}<-10$, visual inspection of the 1D model and residual profiles indicated that for many, either the disk was unnecessary or unphysical. Here, we apply two conditions to make a blind selection of candidates that may have two components:
\begin{itemize}
    \item $\Delta \text{BIC}<-10$.
    \item 1D profiles extend beyond the 80\% PSF-flux radius.
\end{itemize}
We find that 84 objects satisfy both these conditions.

To further refine the selection of genuine two-component objects from the above subset, we inspect the 1D model and residual profiles to assess whether the addition of a disk component yields a physically and visually meaningful improvement. We examine whether the residual profiles exhibit systematic deviations in the single component fits and whether a second component drives the residual profiles to a flatter configuration. In cases where the residual profiles improve, we make sure that the underlying disk components are not unrealistically flat, unphysically extended, or hitting the boundary values in the chosen parameter space. Additionally, when the residual profiles of the single- and double-component fits are comparable, we favour the simpler single-component model. This procedure results in a final sample of 19 two-component systems. We further look into the remaining sample where the galaxy did not extend beyond the 80\% PSF-flux radius but still satisfied the $\Delta \text{BIC}$ condition. We found 1 such object that exhibits both an improved residual profile and a physically possible disk component. As highlighted in previous studies of galaxy morphological analysis, traditional minimisation techniques and $\chi^{2}$ based approaches have well-known limitations, motivating a shift towards Bayesian MCMC or machine learning frameworks\citep[e.g.,][]{Robotham-etal2017, Tuccillo-etal2018}. We present false-colour images of a few of the best cases of two-component galaxies in Figure \ref{fig:color-images}.

%-----recovery table-----------
\begin{table*}
    \centering
    \caption{Table showing the true parameters (mean value) to generate the Sersic+exponential disk mocks and the recovered parameters (mean value) after \textsc{GALFIT} modeling, in appropriate columns. The Mock$_{frac}$ represents the number of mocks that have sizes larger than the 80\% PSF-flux size. The errors represent the 1-$\sigma$ scatter of the distributions. }
    \begin{tabular}{ccccccccc}
        \hline
        \hline
         True& Mock$_{frac}$ & Recovered& True& Recovered & True& Recovered& True&Recovered \\ 
        $n_{bulge}$ &  &$\overline{n}$$_{bulge}$ & $\overline{R}$$_{e,~bulge}$ (pix) & $\overline{R}$$_{e,~bulge}$ (pix) & $\overline{R}$$_{s}$ (pix)& $\overline{R}$$_{s}$ (pix) & $\overline{B/T}$ & $\overline{B/T}$ \\
        \hline

        \multirow{4}{*}{0.5} & 62/500 &0.7$\pm$0.5       &  2.0$\pm$0.5      & 2.9$\pm$0.8       &  4.0$\pm$0.5     &  4.6$\pm$0.8      &  0.3$\pm$0.2      &  0.4$\pm$0.1      \\ \cline{2-9}
                         & 497/500 &0.5$\pm$0.1      &  5.0$\pm$0.5      &   5.1$\pm$1.0     &    7.0$\pm$0.5    &   7.2$\pm$1.1     &   0.5$\pm$0.1     &   0.5$\pm$0.1     \\ \cline{2-9}
                         & 500/500  &0.5$\pm$0.0     &   8.0$\pm$1.0     &   8.0$\pm$1.0     &   10.0$\pm$1.0     &   10.2$\pm$1.1     &    0.5$\pm$0.1    &   0.5$\pm$0.1     \\ \cline{2-9}
                         & 500/500 &0.5$\pm$0.0      &   15.0$\pm$1.0     &  15.0$\pm$1.0      &  20.0$\pm$1.0      &   19.6$\pm$1.1     &  0.5$\pm$0.1      &   0.5$\pm$0.1     \\ \hline
        \multirow{4}{*}{1.0} & 104/500 &1.2$\pm$0.3      & 2.0$\pm$0.5       &   2.9$\pm$0.7     & 4.0$\pm$0.5       & 4.7$\pm$1.0       &  0.4$\pm$0.1      &  0.5$\pm$0.2      \\ \cline{2-9}
                         & 500/500 &1.0$\pm$0.1      &   5.0$\pm$0.5     &    5.0$\pm$1.0    &    7.0$\pm$0.5    &  7.3$\pm$1.4      &   0.5$\pm$0.1     &    0.5$\pm$0.2    \\ \cline{2-9}
                         & 500/500  &1.0$\pm$0.0     &   8.0$\pm$1.0     &  7.8$\pm$1.1      &   10.0$\pm$1.0     &   10.1$\pm$1.5     &    0.6$\pm$0.1    &   0.6$\pm$0.1     \\ \cline{2-9}
                         & 500/500 &1.0$\pm$0.0      &   15.0$\pm$1.0     &   14.7$\pm$1.0     &    20.0$\pm1.0$    &  19.2$\pm$1.5      &   0.6$\pm$0.1     &   0.5$\pm$0.1     \\ \hline
        \multirow{4}{*}{1.5} & 144/500 &1.6$\pm$0.5      &  2.0$\pm$0.5      &  2.9$\pm$0.8      & 4.0$\pm$0.5       &  4.5$\pm$2.0      &  0.5$\pm$0.1      &    0.5$\pm$0.2    \\ \cline{2-9}
                         & 500/500 &1.5$\pm$0.2      &   5.0$\pm$0.5     &   5.1$\pm$1.2     &  7.0$\pm$0.5      &   7.3$\pm$1.7     &   0.6$\pm$0.1     &  0.6$\pm$0.2      \\ \cline{2-9}
                         & 500/500 &1.5$\pm$0.1      &   8.0$\pm$1.0     &  7.9$\pm$1.3      &   10.0$\pm$1.0     &  10.1$\pm$1.5      &    0.6$\pm$0.1    &    0.6$\pm$0.1    \\ \cline{2-9}
                         & 500/500  &1.5$\pm$0.1     &   15.0$\pm$1.0     &  14.5$\pm$1.2      &   20.0$\pm$1.0     &   19.5$\pm$1.1     &  0.6$\pm$0.1      &   0.6$\pm$0.1     \\ \hline
        \multirow{4}{*}{2.5} & 190/500 &2.5$\pm$0.8      &  2.0$\pm$0.5      &    2.5$\pm$0.8    &  4.0$\pm$0.5      &  4.1$\pm$0.6      &   0.5$\pm$0.1     &   0.5$\pm$0.2       \\ \cline{2-9}
                         & 500/500  &2.5$\pm$0.3     &   5.0$\pm$ 0.5    &    5.0$\pm$1.2    &    7.0$\pm$0.5   &   7.0$\pm$1.0     &   0.6$\pm$0.1     &  0.6$\pm$0.1      \\ \cline{2-9}
                         & 500/500 &2.5$\pm$0.1      &    8.0$\pm$1.0    &   7.9$\pm$1.2     &   10.0$\pm$1.0     &   10.0$\pm$1.0     &  0.7$\pm$0.1      &   0.7$\pm$0.1     \\ \cline{2-9}
                         & 500/500  &2.5$\pm$0.0     &   15.0$\pm$1.0    &  14.3$\pm$1.0      &    20.0$\pm$1.0    &   19.8$\pm$0.8     &   0.7$\pm$0.1     &   0.6$\pm$0.1     \\ \hline
    \end{tabular}

    \label{tab:mock-table}
\end{table*}

\subsection{Recovery tests}
\label{sec:recovery-tests}

In the challenging regime of morphological decomposition at high redshift, it is essential to understand the limitations and accuracy of our modeling. Given the compact nature of the objects, the modeling is highly susceptible to parameter degeneracies and assumptions. To validate and check the accuracy of our morphological decomposition, we run a set of recovery tests as described below. For this, we generated mock galaxies with a) single Sersic and b) Sersic + exponential disk component with following structural parameters:

\begin{itemize}
    
    \item For one-component mocks: the mean half-light radii (R$_{e}$) are 2 pixels with a gaussian $\sigma$=0.5, and 5, 8 and 15 pixels with a gaussian $\sigma$=1.0 pixels

    \item For two-component mocks: the mean half-light radii of the sersic component (R$_{e, ~bulge}$) are 2 pixels with a gaussian $\sigma$=0.5, and 5, 8 and 15 pixels with a gaussian $\sigma$=1.0 pixels. The mean scale-length of the exponential disk component (R$_{s}$) are 4 pixels with a $\sigma$=0.5, and 7, 10 and 20 pixels with a $\sigma$=1.0 pixel. When expressed as an exponential function, the size parameter of the exponential disk (R$_{s}$) is the radius at which the central surface brightness decreases by a factor of $e$. The equivalent effective radius of the disk is given by R$_{e,~disk} =$  1.678 R$_{s}$, corresponding to a Sersic profile with $n=1$. Throughout this work, we use R$_s$ when discussing the sizes of the exponential disk component in the mock galaxies.
    
\end{itemize}

\noindent For the Sersic component, four values of the Sersic indices were considered, namely \textit{n$_{bulge}$} = 0.5, 1.0, 1.5 and 2.5. Unlike the other parameters, the Sersic index was sampled using a discrete set of representative values rather than from a continuous distribution. This choice was made to keep the number of mock galaxies computationally manageable while still spanning the range of Sersic indices relevant to our sample. The Sersic component intensity, $I_{e}$ = 0.05 and the exponential disk intensity, $I_{0}$ = 0.04 were fixed for all the mock galaxies. There are 500 objects each for every value of $n$ or \textbf{$n_{bulge}$}. We adopt the F356W-PSF as the representative PSF and add gaussian noise representative of the F356W observations to the mock images. We then fit the mocks using \textsc{GALFIT} following the same procedure as described above in section 4.2. The summary of the exercise is as follows:

\begin{itemize}
    \item Comparison of the input and recovered parameter distributions for the one-component mock galaxies indicates that parameter recovery is generally robust, except for mocks with R$_{e}$=2 pixels. For example, for mocks with true \textit{n}=0.5 and R$_{e}$=2 pixels, we recover a mean \textit{n}=0.6 ($\sigma=0.6$) and a mean R$_{e}$=2.2 pixels ($\sigma=0.4$ pixels). The dispersion in the recovered parameters, particularly in \textit{n}, decreases for larger true R$_{e}$ values.
    
    \item Similarly, for two-component mock galaxies with mean R$_{e,~bulge}$ = 2 pixels, the recovery of \textbf{\textit{n$_{bulge}$}} is relatively poor, whereas R$_{e,~bulge}$, R$_{s}$, and B/T are recovered more reliably. Across all input values of $n_{bulge}$, the sizes and B/T are generally well recovered. For example, for mocks with true \textit{n$_{bulge}$}=0.5, R$_{e,~bulge}=$ 2 pixels with $\sigma=$ 0.5 pixels, R$_{s}$ = 4.0 with $\sigma$ = 0.5 pixels and B/T = 0.3 with $\sigma=$ 0.1, we recover \textit{n$_{bulge}$}=1.0 ($\sigma=$0.8), R$_{e,~bulge}$=2.6 pixels ($\sigma=0.8$ pixels), R$_{s}=4.6$ pixels ($\sigma=1.3$ pixels) and B/T = 0.4 ($\sigma=0.1$). As in the case of the one-component systems, the accuracy of recovery improves as the sizes of the components increase. We do not inspect the dependence of recovered parameters on galaxy magnitude.

    \item We examined the 1D radial surface brightness profiles of the mocks and found that the structural parameters are recovered more accurately when the outer extent (radius beyond which $\mu_{err}<$0.3 mag arcsec$^{-2}$) is larger than the radius that encloses 80\% of the PSF flux. For one-component mocks satisfying this condition, the recovery is nearly perfect. For example when mean R$_{e}=2$ pixels, out of 500 mocks for each \textit{n} = 1, 1.5 and 2.5, only 2--3 objects extended beyond the 80\% PSF-flux radius and had R$_{e}>3$ pixels. The parameters for these objects were recovered correctly. The recovery remained similarly robust for larger objects. For two-component mocks with true \textit{n$_{bulge}$} = 0.5, 62 out of 500 satisfied the condition for size. This subset now had true mean R$_{e,~bulge}=2.5$ pixels ($\sigma=0.5$ pixels), R$_{s}$ = 4.3 pixels ($\sigma$ = 0.5 pixels) and B/T = 0.3 ($\sigma=$ 0.2). The recovered distributions are substantially closer to the true ones: \textit{n$_{bulge}$}=0.7 ($\sigma=$0.5), R$_{e,~bulge}$=2.9 pixels ($\sigma=0.8$ pixels), R$_{s}=4.6$ pixels ($\sigma=0.8$ pixels) and B/T = 0.4 ($\sigma=0.1$). The recovery improves further with larger sizes.

    \item We further assessed the reliability of the \textsc{GALFIT} modelling on an object-by-object basis. For this, we compared the recovered and input structural parameters for each mock galaxy (Figures 3.1 and 3.5 in the online figure set) with sizes larger than the radius enclosing 80\% of the PSF flux. The recovered structural parameters closely follow the one-to-one relation over the full range of simulated parameters. The agreement improves with increasing galaxy size, while the smallest systems, as expected, exhibit the largest scatter. Overall, these tests demonstrate that our fitting procedure reliably recovers the structural parameters of galaxies, especially the two-component systems that satisfy our adopted size criterion. However, as shown in Figure 3.5, there is a significant scatter in the recovered R$_{e,~bulge}$ when the bulge sizes are close to 2-3 pixels, highlighting the challenges of obtaining accurate size measurements for structures approaching the size of the PSF. In contrast, R$_{s}$ (or equivalently R$_{e,~disk}$) and the B/T ratio are recovered well across all sizes. We note that our recovery tests were designed to assess the accuracy with which the structural parameters of a simulated galaxy population are recovered, rather than the completeness or purity of identifying two-component systems. The latter requires a dedicated validation of the model-selection procedure and is beyond the scope of the present work. In our analysis, however, the identification of two-component galaxies does not rely on the \textsc{GALFIT} decomposition alone. Instead, it is based on the combined application of the adopted size criterion, $\Delta$BIC model selection, and careful inspection of the surface-brightness and residual profiles, which together provide a robust basis for rejecting spurious two-component solutions.

    \item Thus, the level of noise (and hence the observed size) plays a crucial role in the accuracy of the modeled parameters. As stated above, to make sure that our derived parameters are closest to the true structural parameters of the galaxy, we select only those objects for two-component decomposition whose outer extent is larger than the size that encloses 80\% of the PSF flux.
    
\end{itemize}

\begin{figure*}[htpb]
\figurenum{3}
\includegraphics[width=\textwidth]{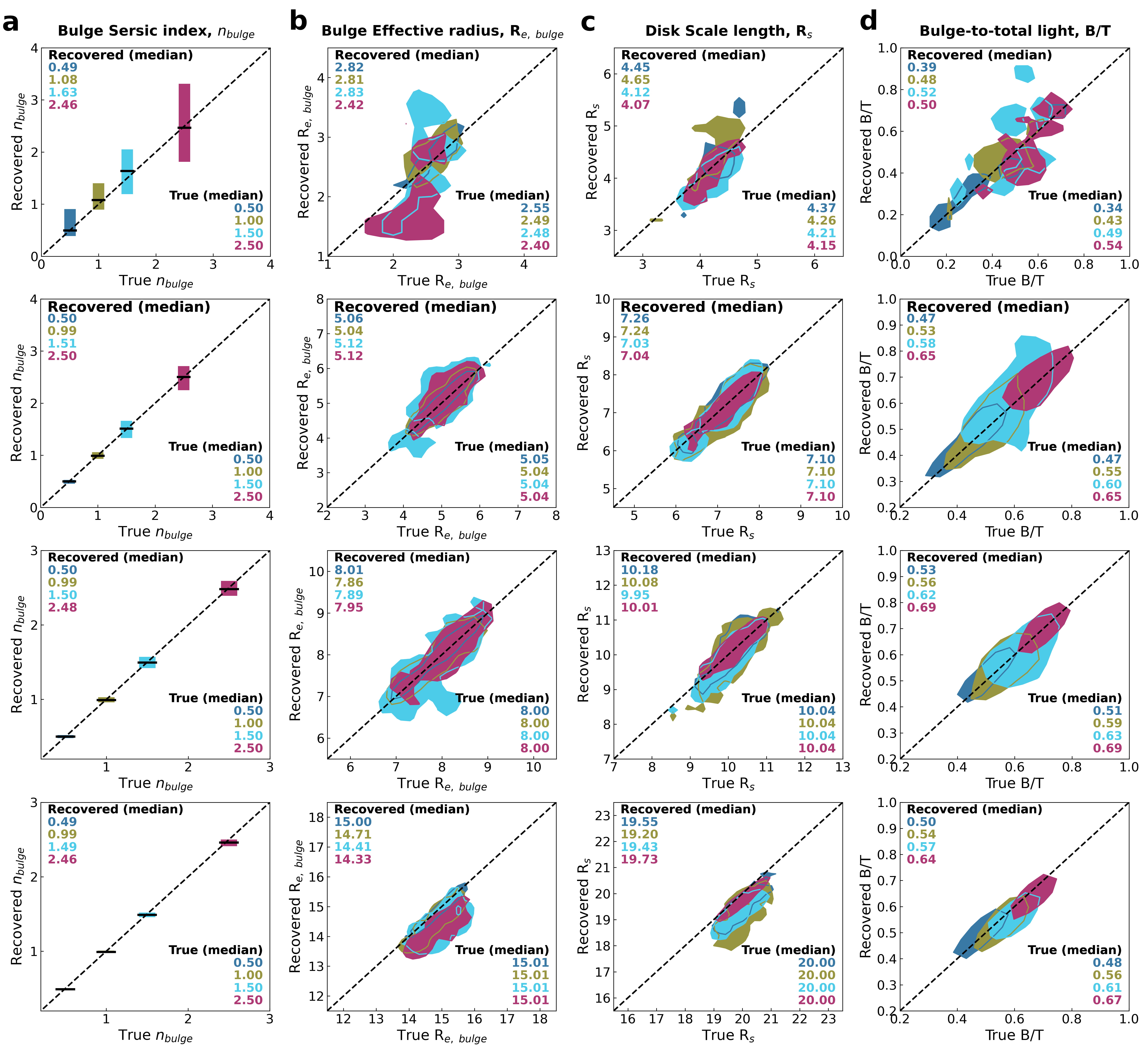}
\caption{\textbf{Recovery of structural parameters from two-component mock galaxies. This figure:} Comparison between the input and recovered parameters obtained from the \textsc{GALFIT} decomposition of simulated Sersic+exponential disk galaxies (Section~\ref{sec:recovery-tests}). Panels in columns a to d show the recovery of the bulge Sersic index ($n_{bulge}$), bulge effective radius (R$_{e,~bulge}$), disk scale length (R$_s$), and bulge-to-total light ratio (B/T), respectively. From top to bottom, the rows correspond to progressively larger input galaxy sizes, with mean R$_{e,~bulge}$ = 2, 5, 8, and 15 pixels and corresponding R$_s$ = 4, 7, 10, and 20 pixels, respectively. Panels in column \textbf{a} show the median recovered value of $n_{bulge}$, with the coloured bars spanning the 16$^{\rm th}$--84$^{\rm th}$ percentile range of the recovered distribution. For the remaining panels (columns b to d), the coloured contours represent the respective 1$\sigma$ levels corresponding to a 2D Gaussian distribution for four input bulge Sersic indices: $n_{bulge}=0.5$ (blue), 1.0 (olive), 1.5 (cyan), and 2.5 (purple). The dashed line denotes the one-to-one relation. In each panel, the numbers denote the median values of the true and recovered parameters, color-coded according to the input value of $n_{bulge}$. Note that, only mock galaxies with sizes larger than the radius enclosing 80\% of the PSF flux are shown. The complete figure set (11 images), additionally showing the recovery for one-component mocks and the recovered histograms for all simulated size combinations, is available in the online journal.}

\label{fig:recovery}
\end{figure*}

We present the results of this exercise for mock galaxies with two components, having sizes larger than that encloses 80\% of the PSF-flux, in Table \ref{tab:mock-table}. We present the recovery plots for both one- and two-component mocks in Figure set \ref{fig:recovery}. The summary statistics of the recovered parameters are provided in the legend of each panel.

%----------fig set for recovery histograms----
\figsetstart
\figsetnum{3}
\figsettitle{Recovery test histograms on mock galaxies with one- and two-components.}

%---------------------for one component------

\figsetgrpstart
\figsetgrpnum{3.1}
\figsetgrptitle{One-to-one recovery of structural parameters from simulated one-component galaxies.}
\figsetplot{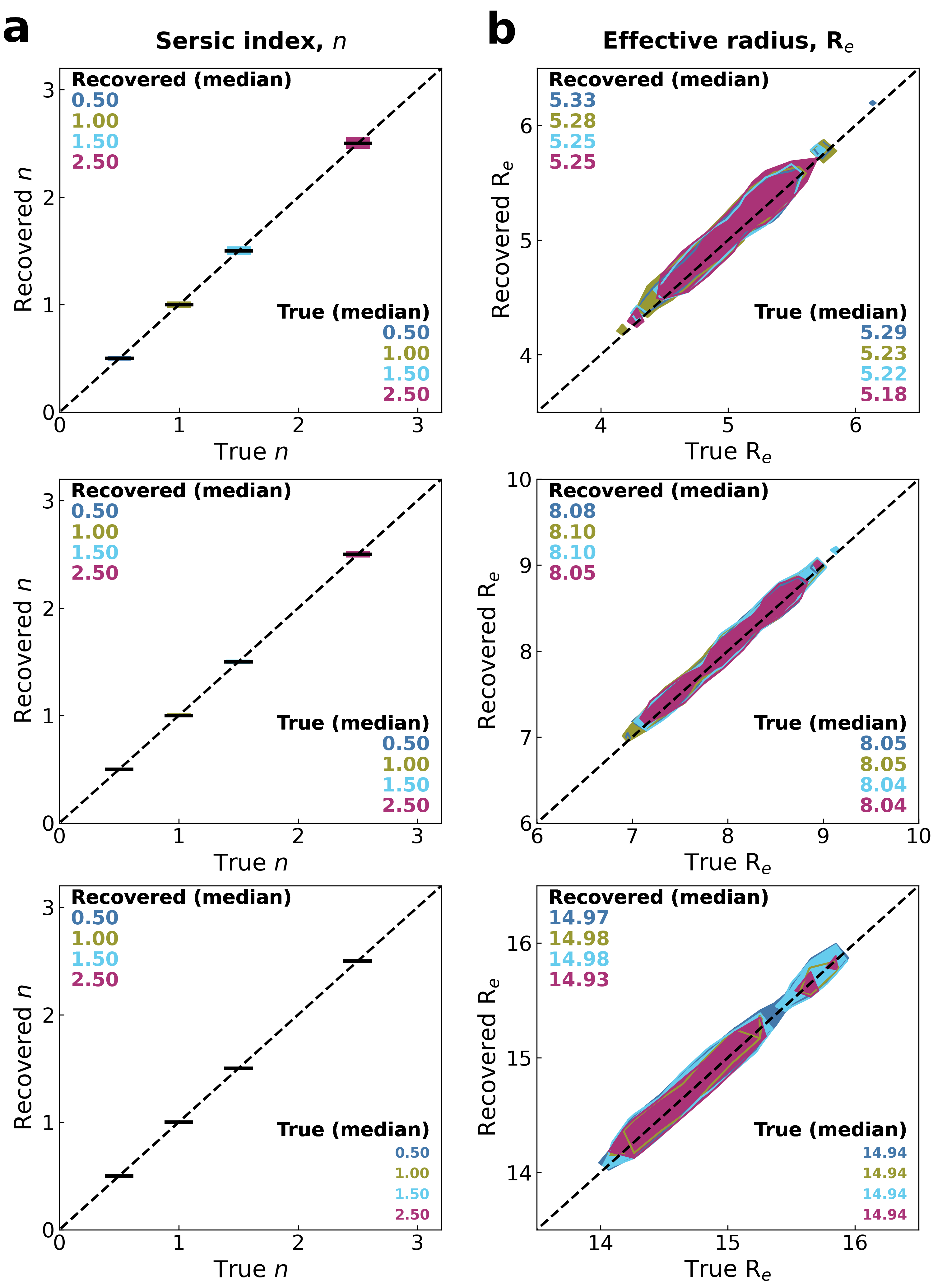}
\figsetgrpnote{\textbf{Comparison between the input and recovered parameters obtained from the \textsc{GALFIT} decomposition of simulated single-Sersic galaxies. Panels in columns a and b show the recovery of the Sersic index ($n$) and effective radius (R$_e$), respectively. From top to bottom, the rows correspond to progressively larger input galaxy sizes, with mean R$_e$ = 5, 8, and 15 pixels. Panels in column \textbf{a} show the median recovered $n$ value, with the coloured bars spanning the 16$^{\rm th}$--84$^{\rm th}$ percentile range of the recovered distribution. Panels in column \textbf{b} show the coloured contours that represent the respective 1$\sigma$ levels corresponding to a 2D Gaussian distribution for four input Sersic indices: $n=$ 0.5 (blue), 1.0 (olive), 1.5 (cyan), and 2.5 (purple). The dashed line indicates the one-to-one relation. In each panel, the numbers denote the median values of the true and recovered parameters, color-coded according to the value of $n$. Note that, only mock galaxies with sizes larger than the radius enclosing 80\% of the PSF flux are shown. Panel for mocks with R$_e$ = 2 pixels are not shown due to negligible objects being larger than the 80\% PSF flux radius.}}
\figsetgrpend

\figsetgrpstart
\figsetgrpnum{3.2}
\figsetgrptitle{Recovery histogram for one-component mocks when true R$_{e}$ = 2 pixels}
\figsetplot{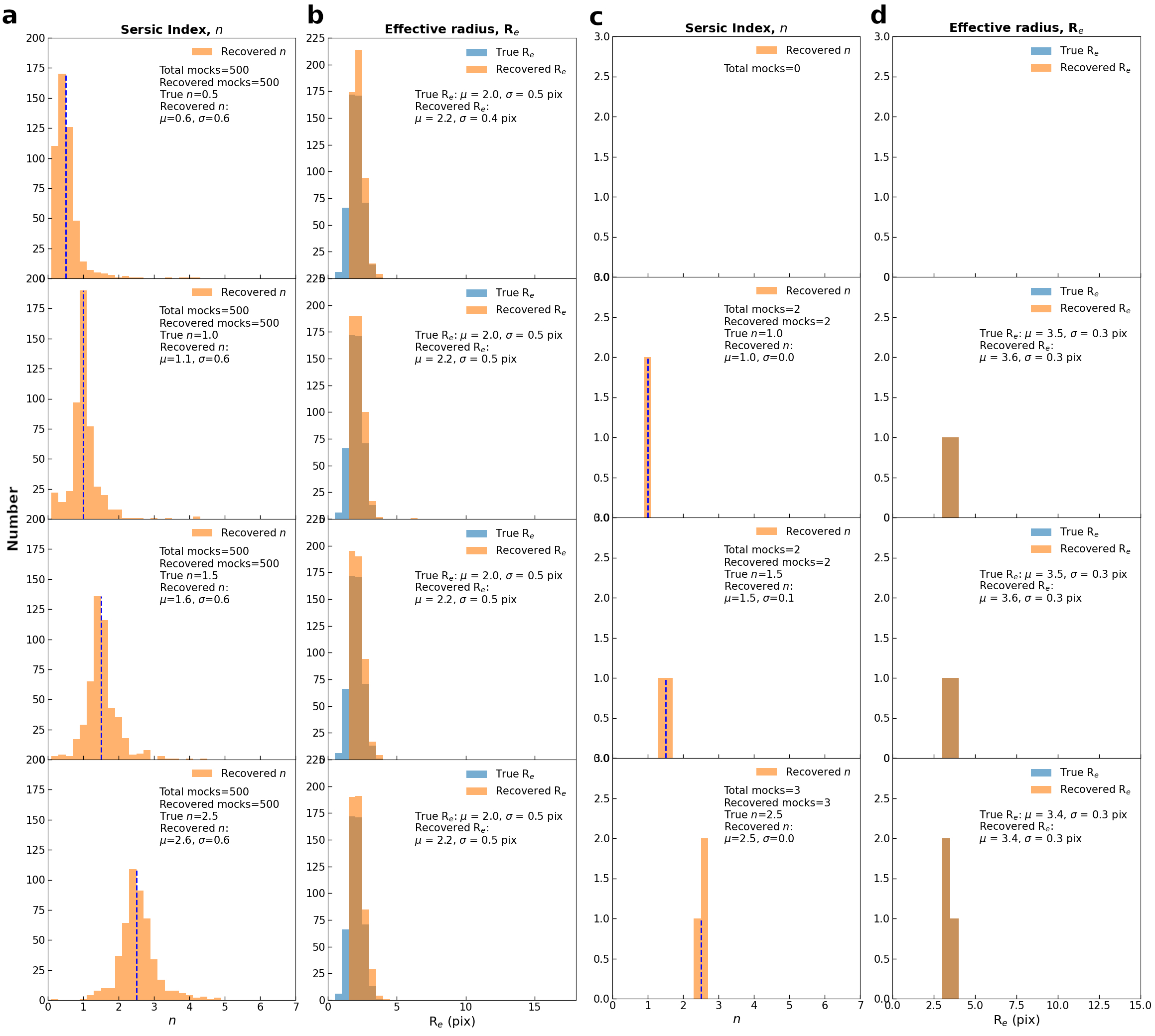}
\figsetgrpnote{\textbf{Comparison of input and recovered parameters from \textsc{GALFIT} modeling of one-component mock galaxies with true R$_{e}$ = 2 pixels. Each row corresponds to a specific input \textit{n} value of 0.5, 1, 1.5 and 2.5. Columns a and b represent all mocks without taking into consideration the extent of the surface brightness profiles. Columns c and d represent mocks whose sizes (based on threshold of $\mu_{err}<0.3$ mag arcsec$^{-2}$) extended beyond the size that encloses 80\% of the PSF flux. For each panel in column a, the vertical line marks the true $n$ value.}}
\figsetgrpend

\figsetgrpstart
\figsetgrpnum{3.3}
\figsetgrptitle{Recovery histogram for one-component mocks when true R$_{e}$ = 5 pixels}
\figsetplot{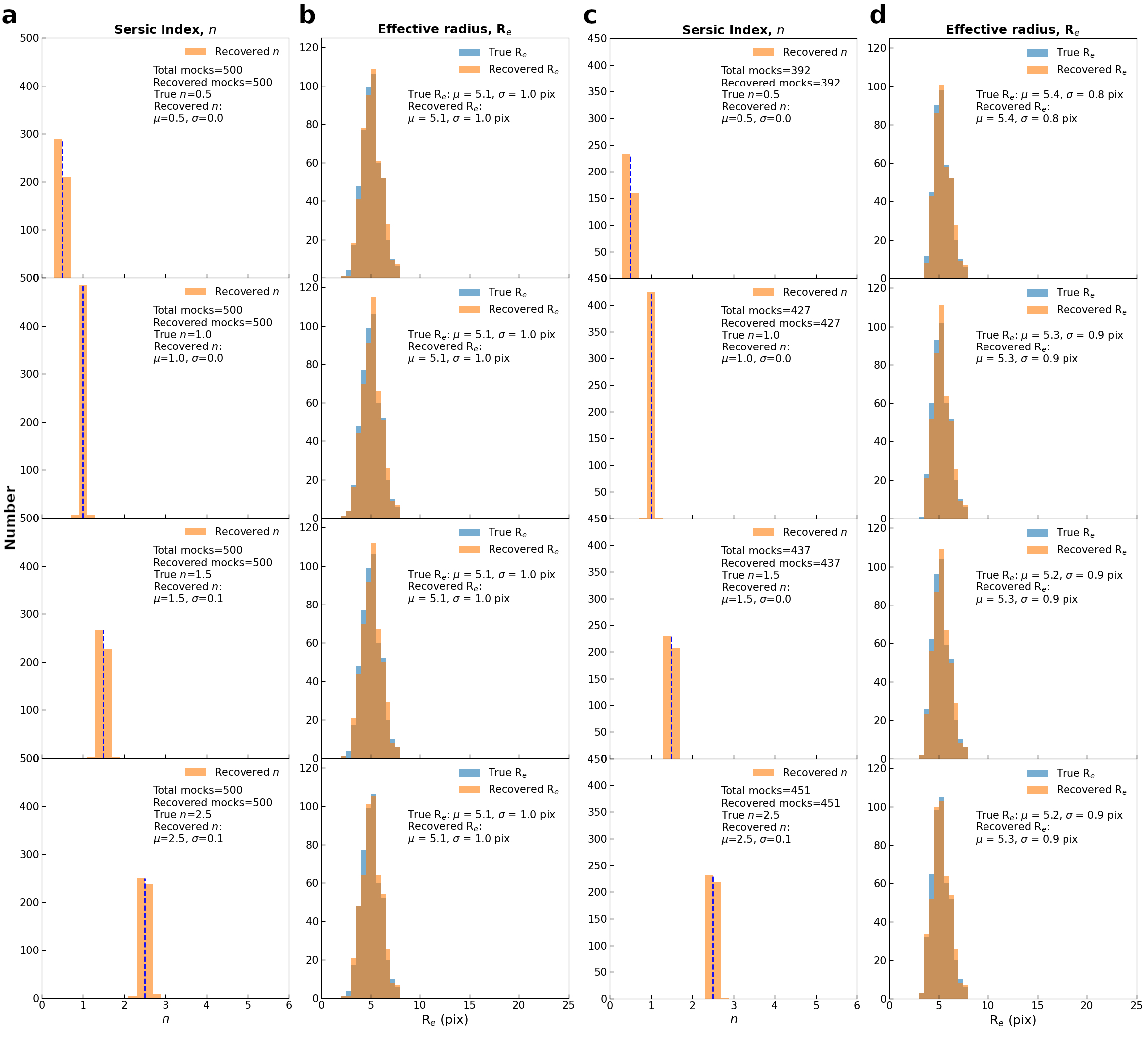}
\figsetgrpnote{\textbf{Same as Figure 3.2, but for mocks with true R$_{e}$ = 5 pixels.}}
\figsetgrpend

\figsetgrpstart
\figsetgrpnum{3.4}
\figsetgrptitle{Recovery histogram for one-component mocks when true R$_{e}$ = 8 and 15 pixels}
\figsetplot{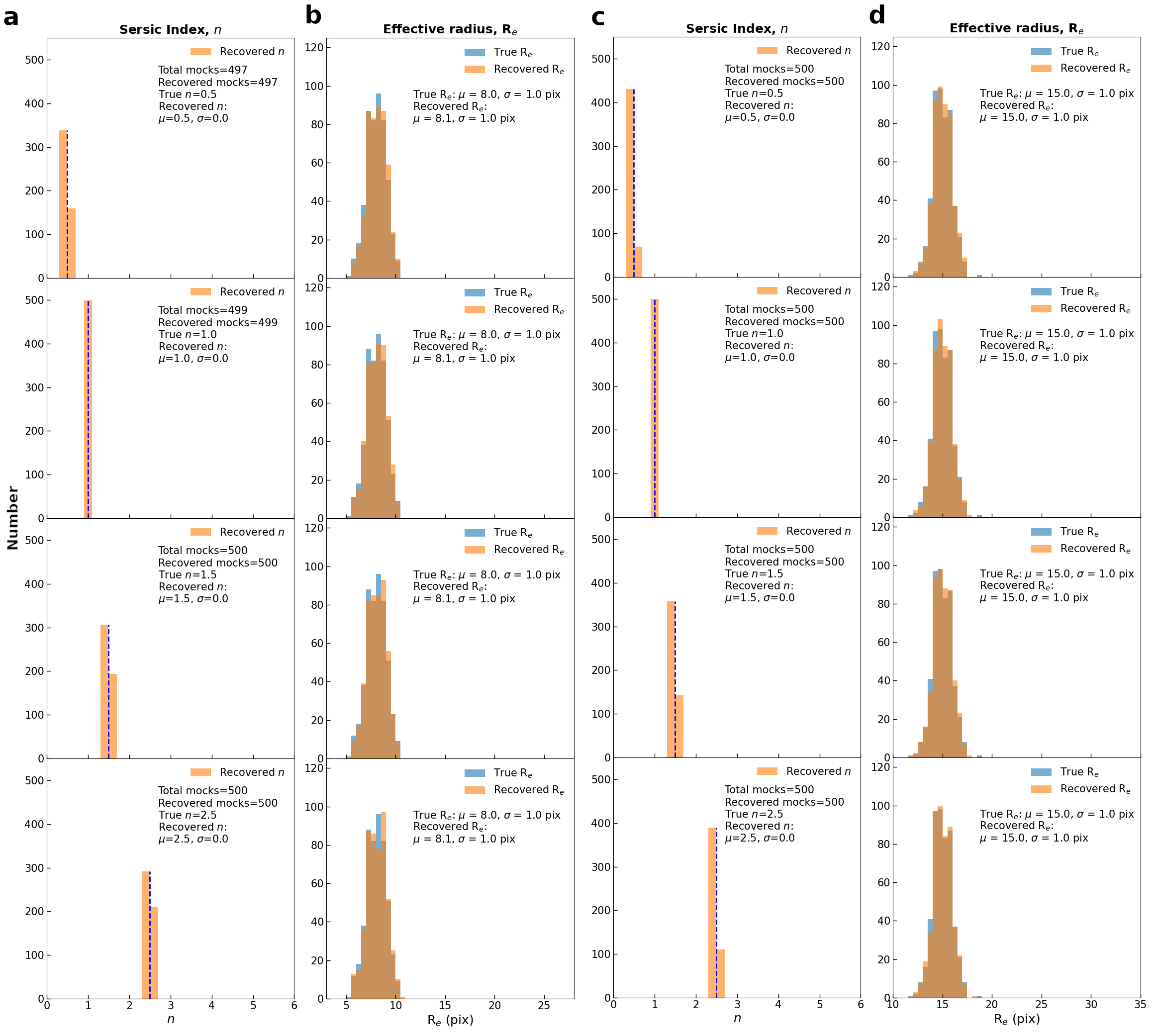}
\figsetgrpnote{\textbf{Comparison of input and recovered parameters from \textsc{GALFIT} modeling of one-component mock galaxies with true R$_{e}$ = 8 pixels (columns a and b) and 15 pixels (columns c and d). As earlier, each row corresponds to a specific input \textit{n} value of 0.5, 1, 1.5 and 2.5. Here, all mocks with sizes extending beyond the 80\% flux enclosing area of the PSF are shown.}}
\figsetgrpend

\figsetend

%----------------for two components------
\figsetgrpstart
\figsetgrpnum{3.5}
\figsetgrptitle{One-to-one recovery of structural parameters from simulated double-component galaxies.}
\figsetplot{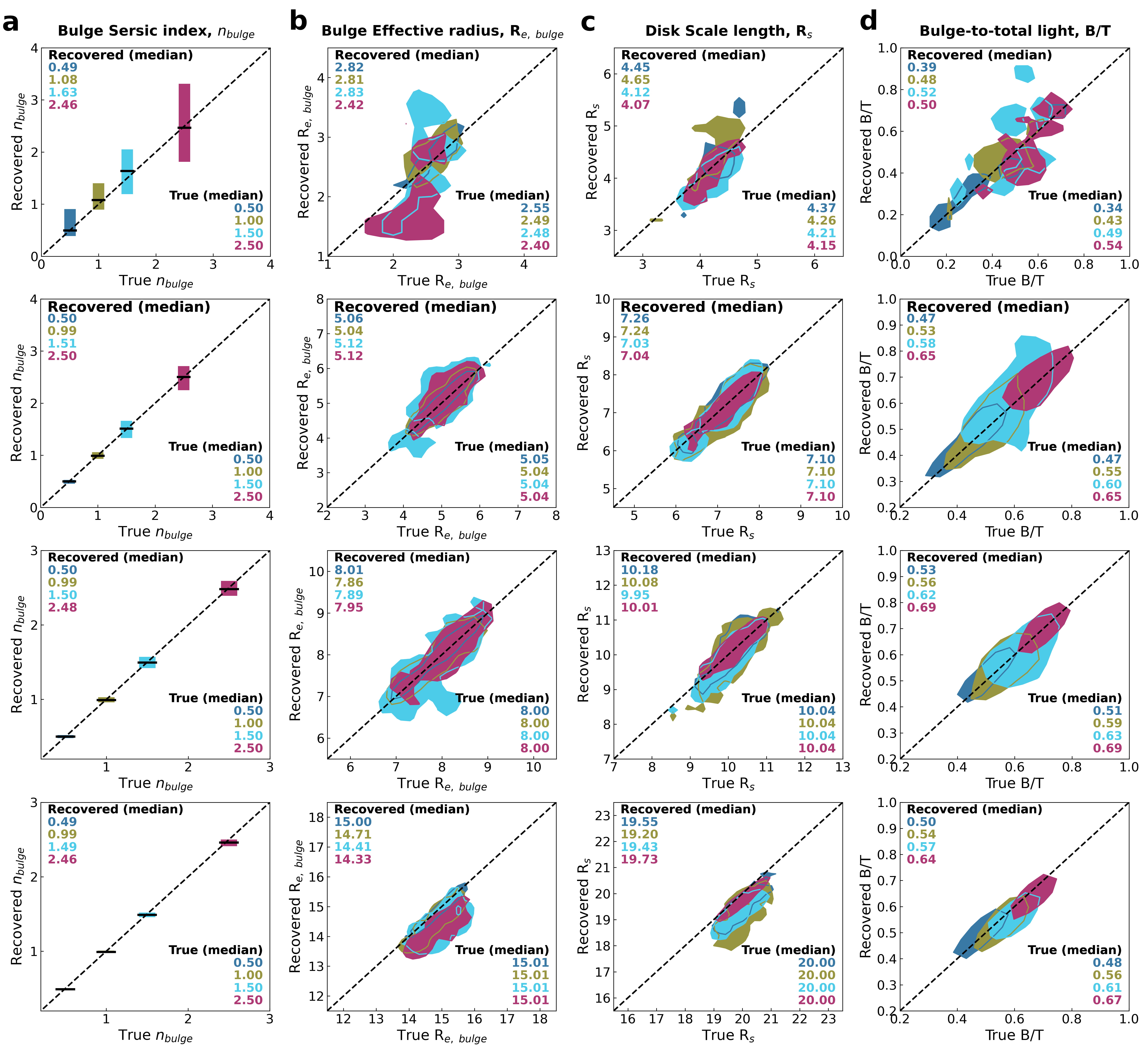}
\figsetgrpnote{\textbf{Comparison between the input and recovered parameters obtained from the \textsc{GALFIT} decomposition of simulated Sersic+exponential disk galaxies (Section~\ref{sec:recovery-tests})}. Panels in columns a to d show the recovery of the bulge Sersic index ($n_{bulge}$), bulge effective radius (R$_{e,~bulge}$), disk scale length (R$_s$), and bulge-to-total light ratio (B/T), respectively. From top to bottom, the rows correspond to progressively larger input galaxy sizes, with mean R$_{e,~bulge}$ = 2, 5, 8, and 15 pixels and corresponding R$_s$ = 4, 7, 10, and 20 pixels, respectively. Panels in column \textbf{a} show the median recovered value of $n_{bulge}$, with the coloured bars spanning the 16$^{\rm th}$--84$^{\rm th}$ percentile range of the recovered distribution. For the remaining panels (columns b to d), the coloured contours represent the respective 1$\sigma$ levels corresponding to a 2D Gaussian distribution for four input bulge Sersic indices: $n_{bulge}=0.5$ (blue), 1.0 (olive), 1.5 (cyan), and 2.5 (purple). The dashed line denotes the one-to-one relation. In each panel, the numbers denote the median values of the true and recovered parameters, color-coded according to the input value of $n_{bulge}$. Note that, only mock galaxies with sizes larger than the radius enclosing 80\% of the PSF flux are shown.}
\figsetgrpend

%--------------
\figsetgrpstart
\figsetgrpnum{3.6}
\figsetgrptitle{True R$_{e,~bulge}$ = 2 pixels, R$_{s}$ = 4 pixels}
\figsetplot{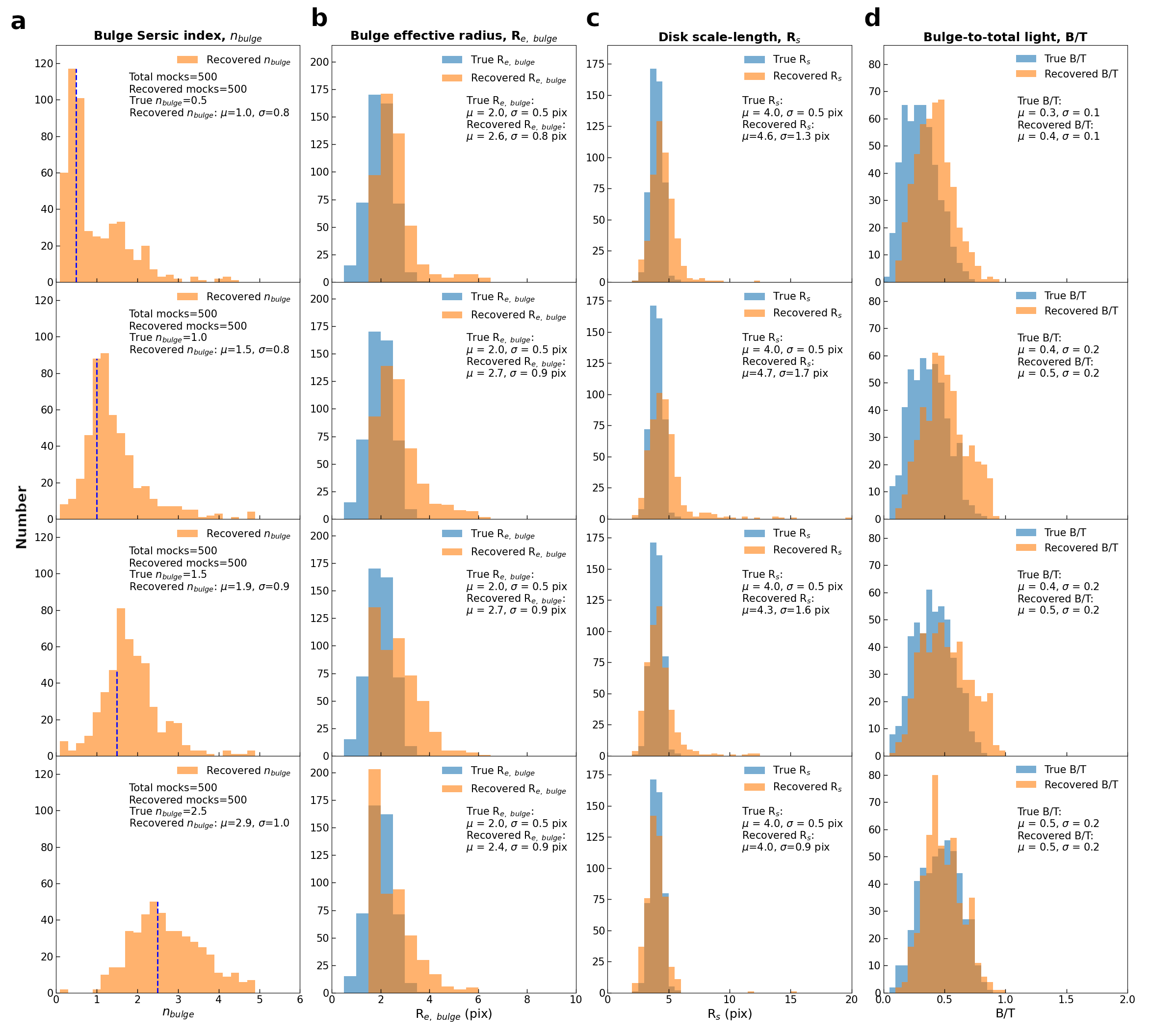}
\figsetgrpnote{Comparison of input and recovered parameters from \textsc{GALFIT} modeling of all \textbf{two-component} mock galaxies with true R$_{e,~bulge}$ = 2 pixels, true R$_{s}$ = 4 pixels. As earlier, each row corresponds to a specific input $n_{bulge}$ value of 0.5, 1, 1.5 and 2.5.}
\figsetgrpend

\figsetgrpstart
\figsetgrpnum{3.7}
\figsetgrptitle{True R$_{e,~bulge}$ = 2 pixels, R$_{s}$ = 4 pixels and mock sizes larger than 80\% of PSF-flux enclosing size}
\figsetplot{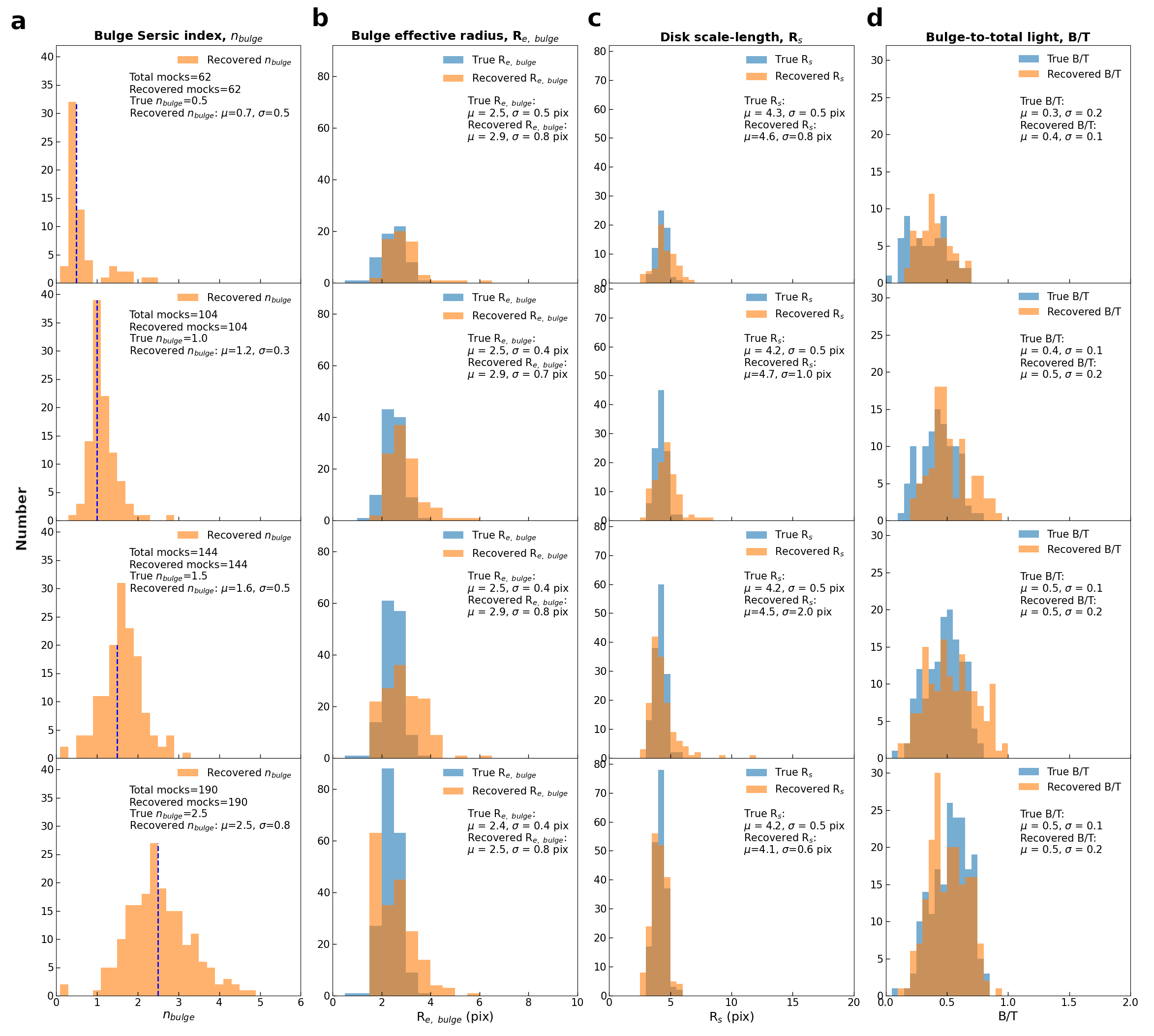}
\figsetgrpnote{Same as Figure 3.6 but using mocks which have sizes larger than the size that encloses 80\% of the PSF flux.}
\figsetgrpend

\figsetgrpstart
\figsetgrpnum{3.8}
\figsetgrptitle{True R$_{e,~bulge}$ = 5 pixels, R$_{s}$ = 7 pixels}
\figsetplot{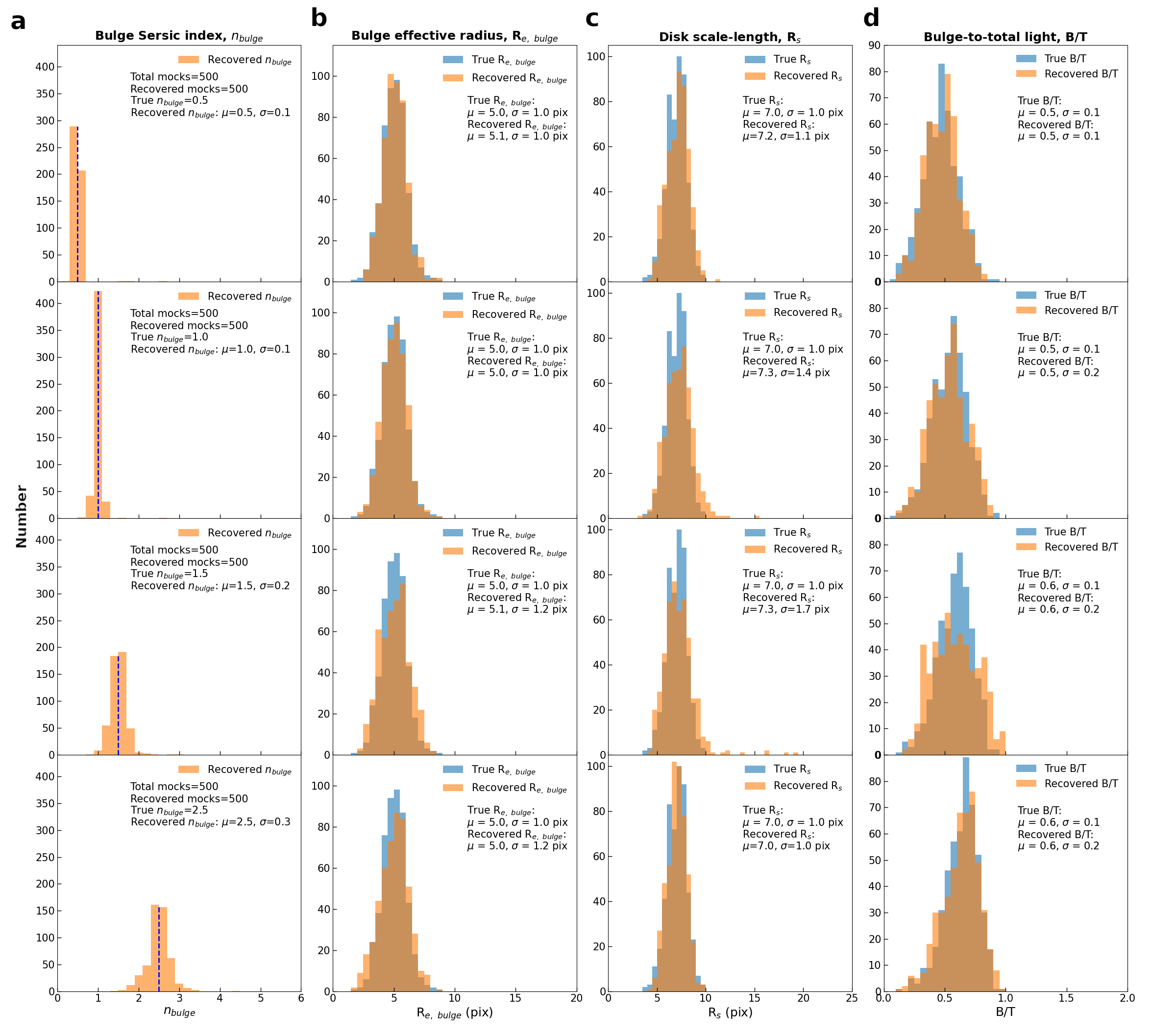}
\figsetgrpnote{Same as Figure 3.6 but true R$_{e,~bulge}$ = 5 pixels and true R$_{s}$ = 7 pixels.}
\figsetgrpend

\figsetgrpstart
\figsetgrpnum{3.9}
\figsetgrptitle{True R$_{e,~bulge}$ = 5 pixels, R$_{s}$ = 7 pixels and mock sizes larger than 80\% of PSF-flux enclosing size}
\figsetplot{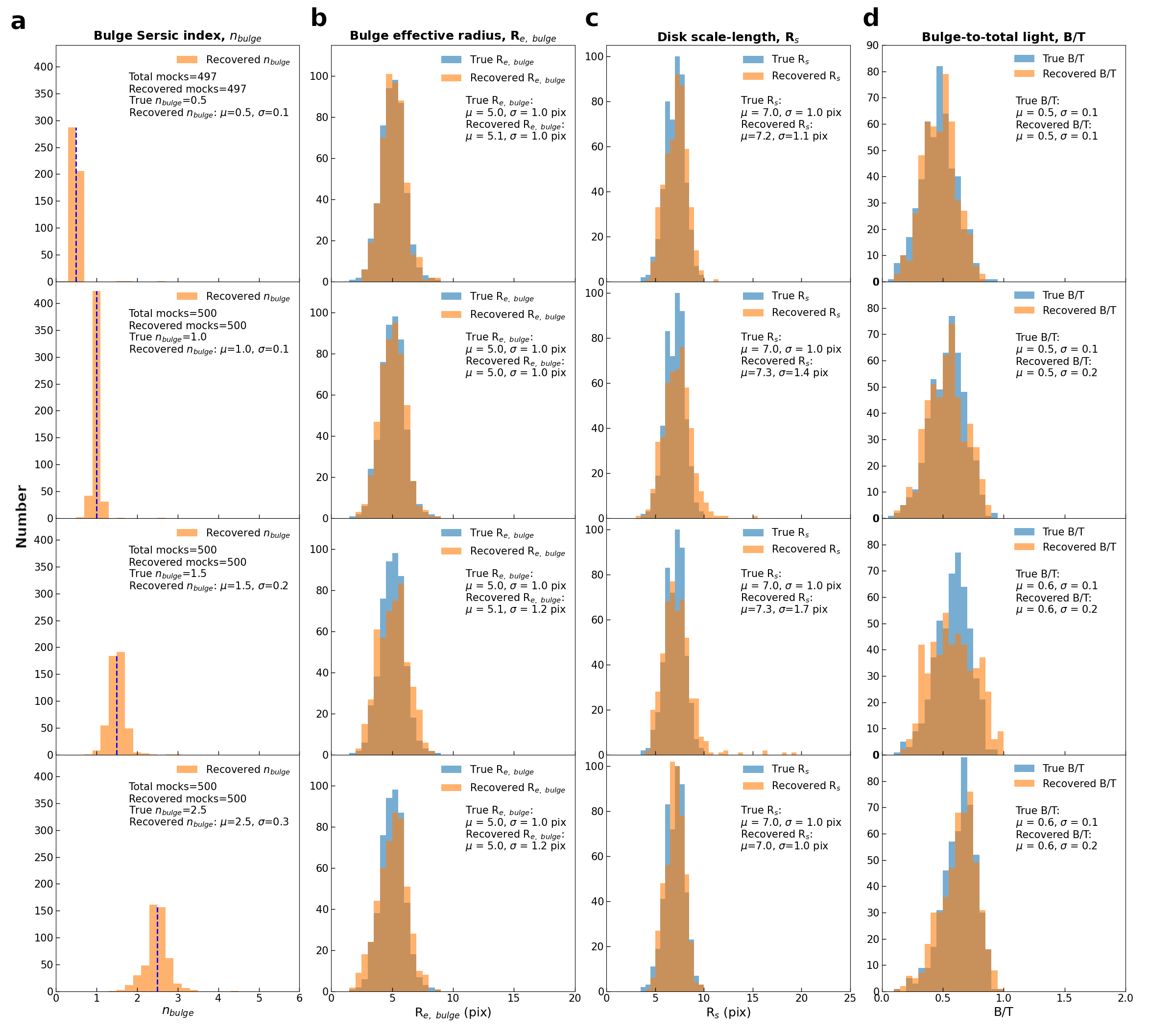}
\figsetgrpnote{Same as Figure 3.8 but using mocks which have sizes larger than the size that encloses 80\% of the PSF flux.}
\figsetgrpend

\figsetgrpstart
\figsetgrpnum{3.10}
\figsetgrptitle{True R$_{e,~bulge}$ = 8 pixels, R$_{s}$ = 10 pixels. All mocks are sufficiently larger than the PSF}
\figsetplot{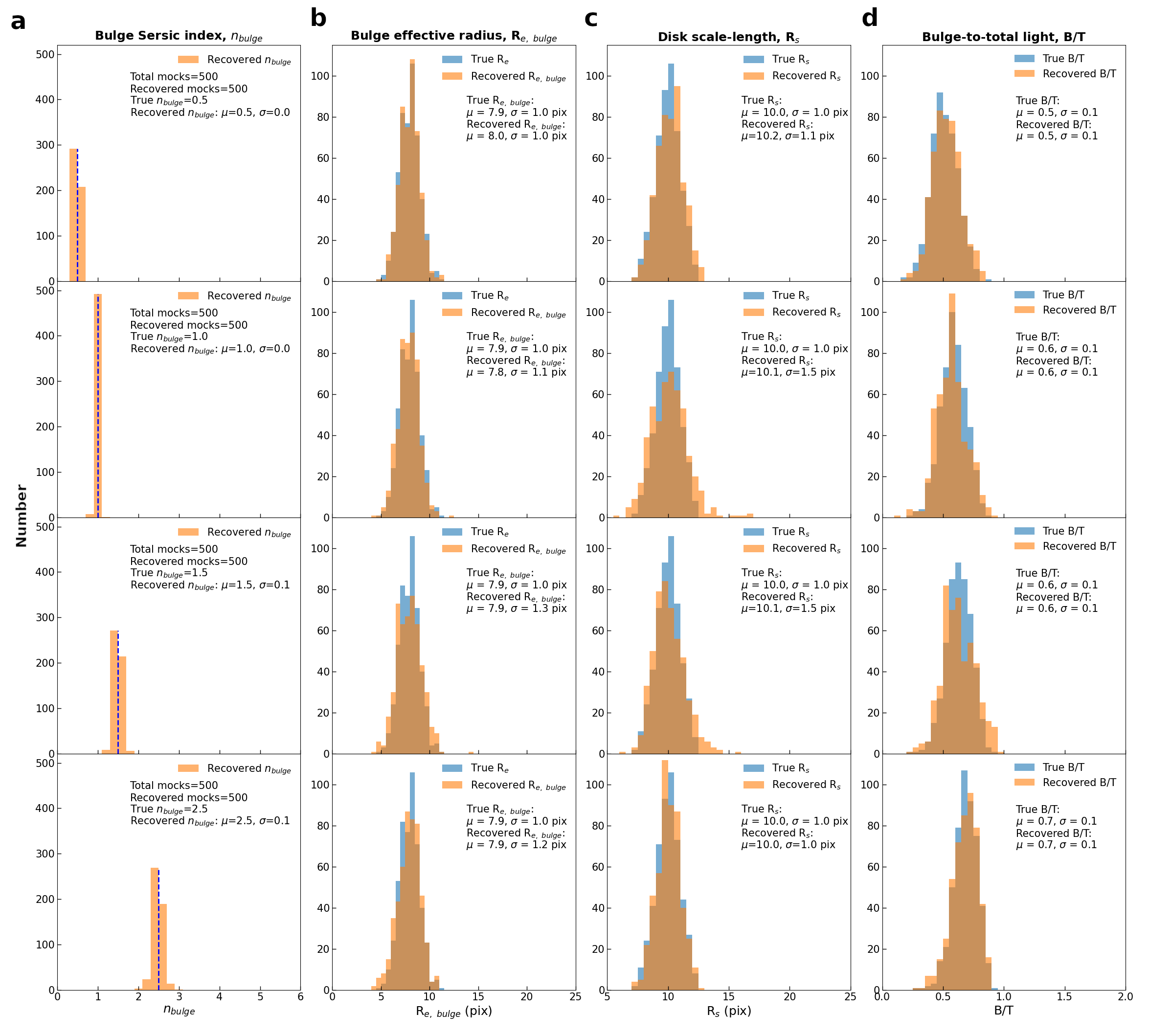}
\figsetgrpnote{Same as Figure 3.6 but true R$_{e,~bulge}$ = 8 pixels and true R$_{s}$ = 10 pixels. Here, all galaxies have sizes sufficiently larger than the PSF.}
\figsetgrpend

\figsetgrpstart
\figsetgrpnum{3.11}
\figsetgrptitle{True R$_{e}$ = 15 pixels, R$_{s}$ = 20 pixels. All mocks are sufficiently larger than the PSF}
\figsetplot{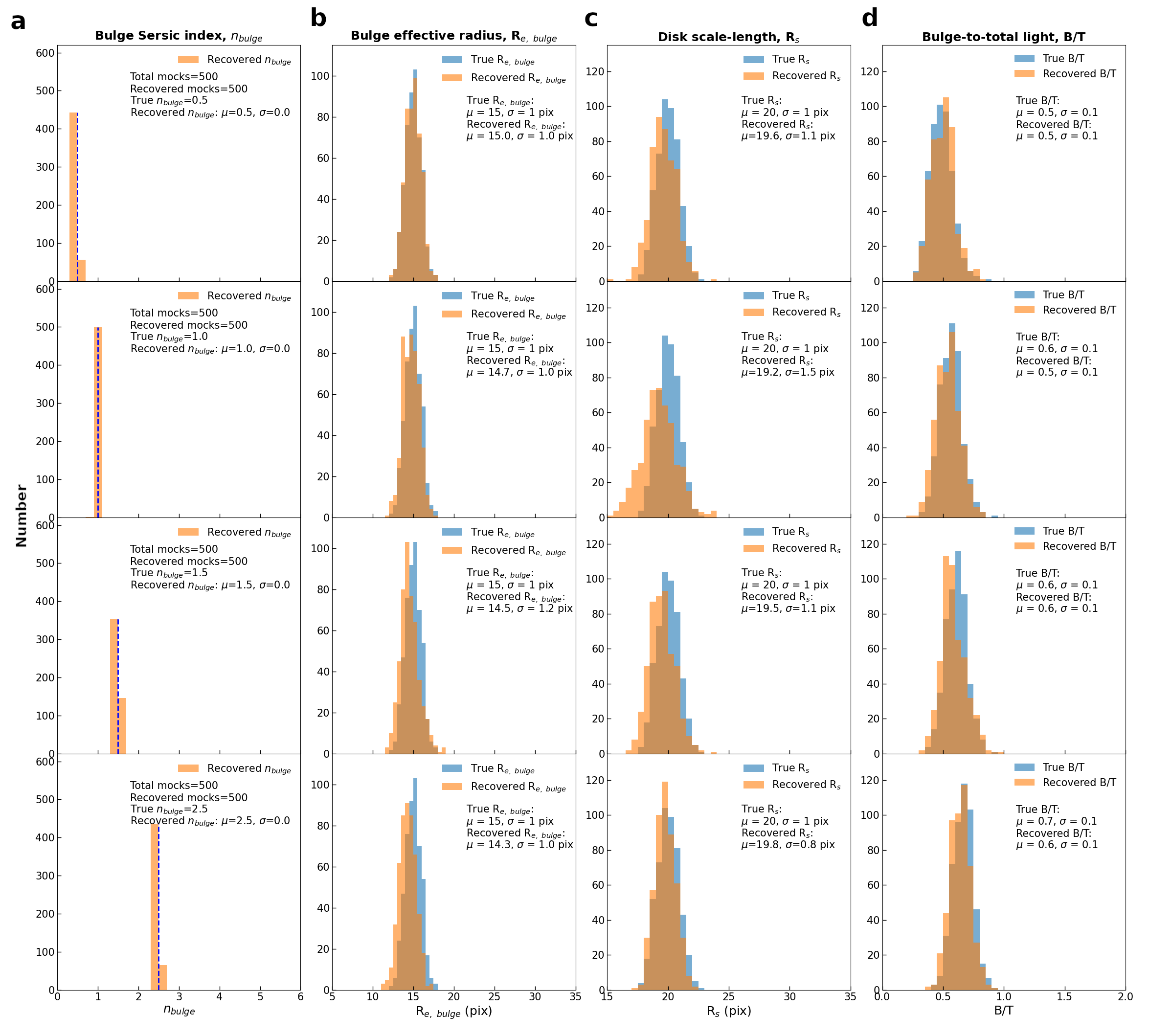}
\figsetgrpnote{Same as Figure 3.6 but true R$_{e,~bulge}$ = 15 pixels and true R$_{s}$ = 20 pixels. Here, all galaxies have sizes sufficiently larger than the PSF.}
\figsetgrpend

\figsetend

%------------------------------------------------

\section{Results}

\begin{figure*}[htp]
\figurenum{4}
\centering
\includegraphics[width=\textwidth]{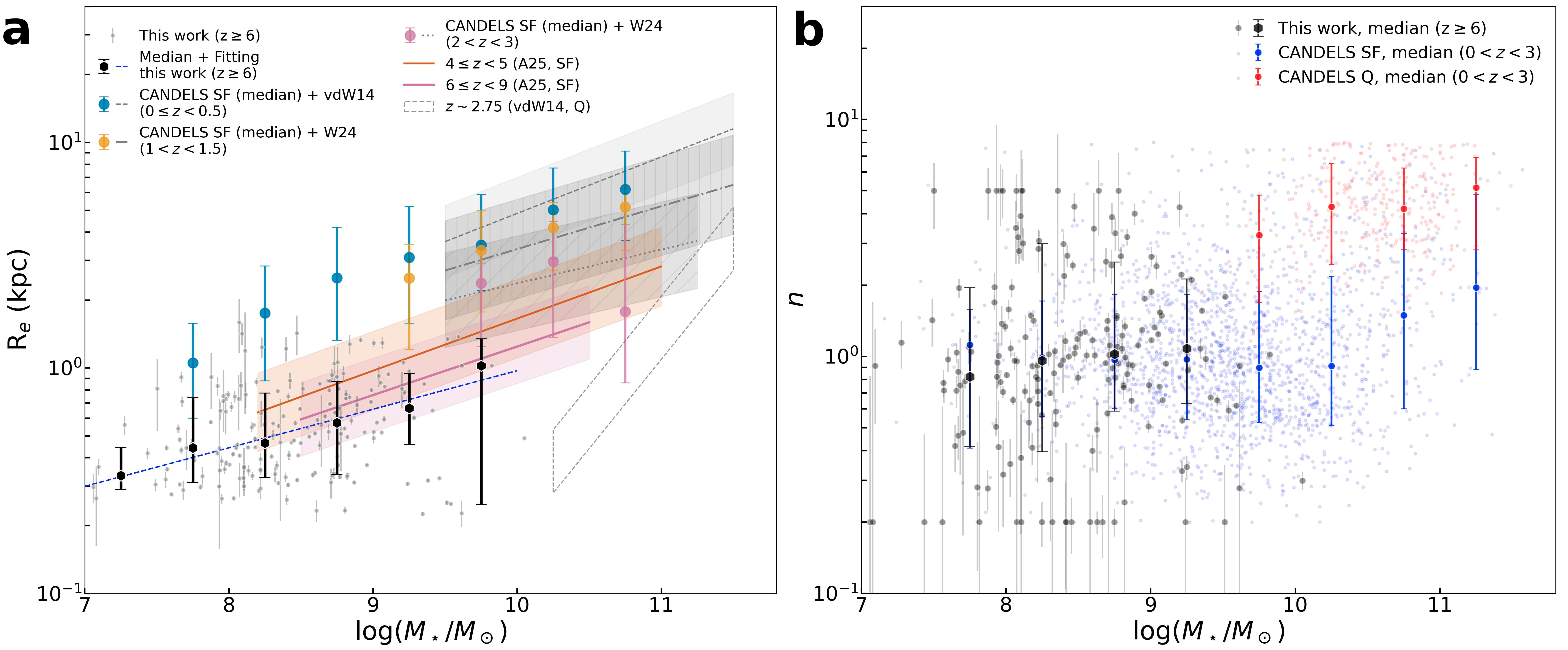}
\caption{\textbf{Comparing galaxy structure at $z \ge 6$ with previous studies spanning from high-$z$ to the local Universe. a:} The rest-frame optical size-mass ($M_\star - \rm{R}_e$) relation. Grey points denote individual galaxies in our sample, with sizes derived from single Sersic modeling, while black hexagons represent stellar mass-binned medians. For visual clarity, we show medians for star-forming (SF) galaxies in CANDELS at $0 \le z < 0.5$, $1 \le z < 1.5$, and $2 \le z < 3$. The dashed blue curve represents the best-fit relation through the median values of R$_{e}$ at $z\geq6$. We overlay the established optical size-mass relations for SF galaxies at $z \sim 0.5$ \citep[vdW14;][]{van-der-wel-etal2014}, $1 < z < 1.5$, $2 < z < 3$ \citep[W24;][]{ward-etal2024}, $4 \le z < 5$, and $6 \le z < 9$ \citep[A25;][]{allen-etal2025}, as well as quiescent galaxies at $z \sim 2.75$ \citep[vdW14,Q;][]{van-der-wel-etal2014}. The associated $1\sigma$ scatter of the relations is indicated by shaded and dashed-box regions. \textbf{b:} A comparison of rest-frame optical Sersic index ($n$) distributions. Most galaxies in the $z\ge6$ sample (grey points and black medians) exhibit a distribution consistent with low-$n$ values, similar to CANDELS SF galaxies at $z < 3$ (blue individual points and medians). The red symbols represent individual and median values for CANDELS quiescent (Q) galaxies. In both panels, error bars on individual points represent $1\sigma$ uncertainties, while error bars on median values denote the 16$^{\rm th}$ and 84$^{\rm th}$ percentiles of the distribution within each bin. We used a minimum threshold of 10 data points (5 for $z\geq6$ galaxies, owing to the smaller sample size) per mass bin to calculate medians.}  
\label{fig:size-shape-mass}
\end{figure*}

As presented in Figure \ref{fig:size-shape-mass}, a clear size-mass relation (R$_{e}\propto$ M$_{\star}^{\alpha}$) is observed in $z\geq6$ galaxies which are compact (median R$_{e}$ = 0.51 kpc) and exhibit lower Sersic indices (median $n$ = 0.96) at a given stellar mass (median $\log$M$_{\star}$/M$_{\odot}$ = 8.42). This provides an indication that the galaxies are steadily growing in mass via star-formation. We fit a linear relation of the form $\log\mathrm{R}_e=$ $\log$ A + $\alpha \log m_{*}$ , where $m_{*}$ = M$_{\star}$/($1\times10^{9})$ \citep{allen-etal2025}, to the binned median values which yields a best-fit slope of the size-mass relation ($\alpha$) = 0.1709 $\pm$ 0.0255.

We compare the structural properties of the galaxies at Cosmic Dawn, obtained from our single Sersic modeling, to their low-\textit{z} counterparts in the CANDELS fields \citep{dimauro-etal2018} as shown in Figure \ref{fig:size-shape-mass}. Throughout this work, all CANDELS galaxies are taken from the catalog provided by \citet{dimauro-etal2018}. We cross-matched this catalog with the 3D-HST catalog \citep{brammer-etal2012, momcheva-etal2016} to obtain spectroscopic redshifts, star-formation rates (SFRs), specific star-formation rates (sSFRs), and stellar masses. Once we have the redshifts, we directly use the structural measurements, based on single Sersic modeling, from the catalog. The measurements correspond to the filter that samples the rest-frame \textit{B}-band emission. We separate the CANDELS sample into star-forming (SF) and quiescent galaxies using the rest-frame \textit{UVJ} colors; following the prescription of \cite{whitaker-etal2011}. We apply an additional $\log$ sSFR threshold of -11 to avoid contamination from galaxies with very low SFR but that lie in the SF cloud as per \citet{whitaker-etal2011}. We note that while \citet{dimauro-etal2018} provide their own stellar mass estimates, we use the masses from the 3D-HST catalog to maintain internal consistency with the other physical properties used in this work. The low-$z$ ($0 <z< 3$) counterparts have substantially larger rest-frame sizes than the high-$z$ sample. The median values for star-forming (quiescent) galaxies are R$_{e}$ = 2.47 kpc (1.67 kpc), $n$ = 0.72 (2.93) and $\log$M$_{\star}$/M$_{\odot}$ = 9.61 (10.57).
\citet{van-der-wel-etal2014} showed that late-type galaxies grow as R$_{e} \propto (1+z)^{-0.75}$ and early-type galaxies grow faster as R$_{e} \propto (1+z)^{-1.48}$, at fixed stellar mass (rest-frame wavelength 5000 $\AA$), over $0<z<3$. At the same time, a consistent R$_{e}-$M$_{\star}^{\alpha}$ relation where $\alpha=0.22$ and 0.75 for late-type and early-type galaxies respectively, has been observed in the same redshift range. \citet{dimauro-etal2019} found an evolution of the slope to shallower values towards higher $z$, up to $\sim2.75$, possibly due to a sample selection effect \citep{nedkova-etal2021}. 

Subsequent studies of SF galaxies at higher redshifts, based on \textit{JWST} observations, also observe a similar redshift evolution of R$_{e}$, characterised by a slope of $\sim-0.7 ~\text{to}~ -0.8$ up to $z\sim9$ \citep[e.g.,][]{ward-etal2024, ormerod-etal2024, allen-etal2025} and consistent with an extrapolation of the trend in \citet{van-der-wel-etal2014}. Surprisingly, the size-mass relation for star-forming galaxies is roughly consistent out to $z\sim14$ or so \citep[e.g.,][]{morishita-etal2024, ward-etal2024, allen-etal2025}. If we extrapolate the R$_{e}-z$ relation of \citet{van-der-wel-etal2014} to $z\sim7$ (median for our sample), not considering the stellar mass effect, it predicts that SF (and quiescent) galaxies at $z\sim7$ should grow by a factor of $\sim4$ (and $\sim15$) up to $z\sim0.25$. The median R$_{e}$ for the CANDELS sample at $0<z<0.5$ is $\sim$2.8 kpc (and $\sim$2.6 kpc) for SF (and quiescent) galaxies. Our findings imply that galaxies at $z\ge6$ have grown by a factor of $\sim5$ or more up to $z\sim0.25$, which is similar to what is expected at these redshifts. However, there are indications of steeper slopes, from \textit{JWST} observations and simulations, of the R$_{e}-z$ and R$_{e}-$M$_{\star}$ relations at $z\gtrsim3$ \citep{sun-et-al2024, costantin-etal2023, miller-etal2025}. This may point to a transition in the galaxy morphology at later epochs, where massive compact systems begin to build extended structures.

\begin{figure*}
\figurenum{5}
\includegraphics[width=\linewidth]{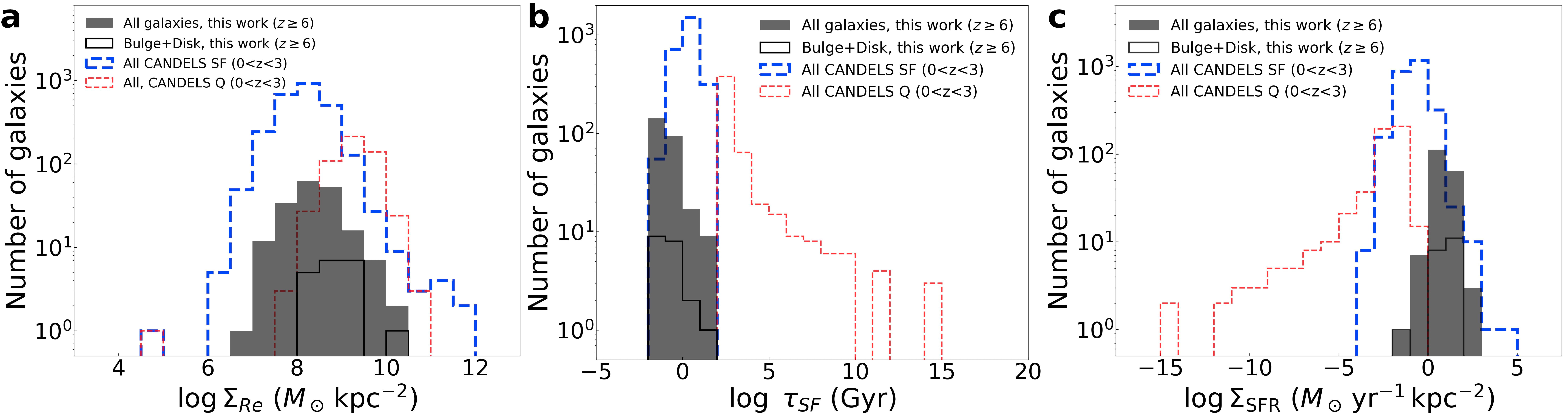} 
\caption{\textbf{Comparison of global galaxy properties $z \geq 6$ galaxy sample with lower-redshift CANDELS galaxies from \citet{dimauro-etal2018}.} The grey shaded histograms show the distributions for all galaxies in our sample, while the solid black histograms represent the subset of galaxies identified with two-components. The blue dashed and red dotted histograms show the distributions for CANDELS SF and quiescent (Q) galaxies at $0<z<3$, respectively. \textbf{a:} Distribution of stellar mass surface density, $\Sigma_{Re}$ calculated within circularised R$_{e}$. \textbf{b:} Distribution of star-formation timescales, $\tau_{SF}$. \textbf{c:} Distribution of star-formation rate surface density, $\Sigma_{ SFR}$.}
\label{fig:histograms}   
\end{figure*}

We now bring our attention to those galaxies that host a bulge and an underlying disk component. Previous studies of bulge properties and their evolution in the low-$z$ Universe \citep[e.g.,][]{sachdeva_saha2017, Sachdevaetal2019} highlight $z\sim$ 2 as an important period of bulge assembly within disks, or of disk growth around pre-existing spheroids \citep{Sachdevaetal2019}. Previous studies observe that the fraction of galaxies that host a stellar bulge and a disk increases from 46\% at 2 $<z<$ 4 to 70\% at 1.5 $<z<$ 2 \citep{Sachdevaetal2019}. As we extend this census to $z\geq$ 6, our results show that the mechanisms driving such morphological transformation may already be in place at very early times. We found that a subset of 20 galaxies (approximately 10\% of the parent sample) show clear evidence for galaxies with a bulge and a disk. We present the distributions of their stellar mass density ($\Sigma_{Re}$), galaxy star-formation timescale ($\tau_{SF}$) and star-formation rate surface density ($\Sigma_{SFR}$) in Figure \ref{fig:histograms}. To estimate $\Sigma_{Re}$ and $\Sigma_{SFR}$, we used the circularised R$_e$ (effective radius based on single-Sersic fitting) for all and adopted the F160W sizes for the CANDELS galaxies, as these provide a better representation of the underlying stellar population at lower-$z$. We find that the $z\geq6$ galaxies have comparable $\Sigma_{Re}$ to low-$z$ quiescent galaxies, $\tau_{SF}$ to low-$z$ SF galaxies and occupy extreme ends of the $\Sigma_{SFR}$ distribution when compared to their low-$z$ counterparts. These $z\geq$ 6 galaxies exhibit median $\Sigma_{Re}$ = 7.82$\times$10$^{8}$ M$_{\odot}$kpc$^{-2}$, rapid star formation as reflected in their low $\tau_{SF}$ (median $\tau_{SF}$ = 0.11 Gyr) and high $\Sigma_{SFR}$ (median $\Sigma_{SFR}$ = 1.21$\times$10$^{1}$ M$_{\odot}$yr$^{-1}$kpc$^{-2}$). We note in Figure \ref{fig:histograms} that the two-component systems have similar global properties to the rest of the galaxies in our sample. The high stellar-mass densities and star-formation rate surface densities and short star-formation timescales, observed across the full $z\ge 6$ sample are consistent with a rapid stellar-mass assembly. Additionally, the presence of galaxies with a bulge and a disk within this sample is consistent with the inside-out scenario of galaxy growth. Moreover, multiple studies have shown that an increasing central stellar mass density can act as a driver of quenching  \citep[e.g.,][]{fang-etal2013, lang-etal2014}. The high central densities that we measure in our $z\ge$ 6 two-component galaxies are similar to some of the densest stellar systems known in the nearby Universe \citep{hopkins-etal2010}. 

\subsection{Bulge size vs galaxy stellar mass}

\begin{figure}[htp]
\figurenum{6}
\centering
\includegraphics[width=\linewidth]{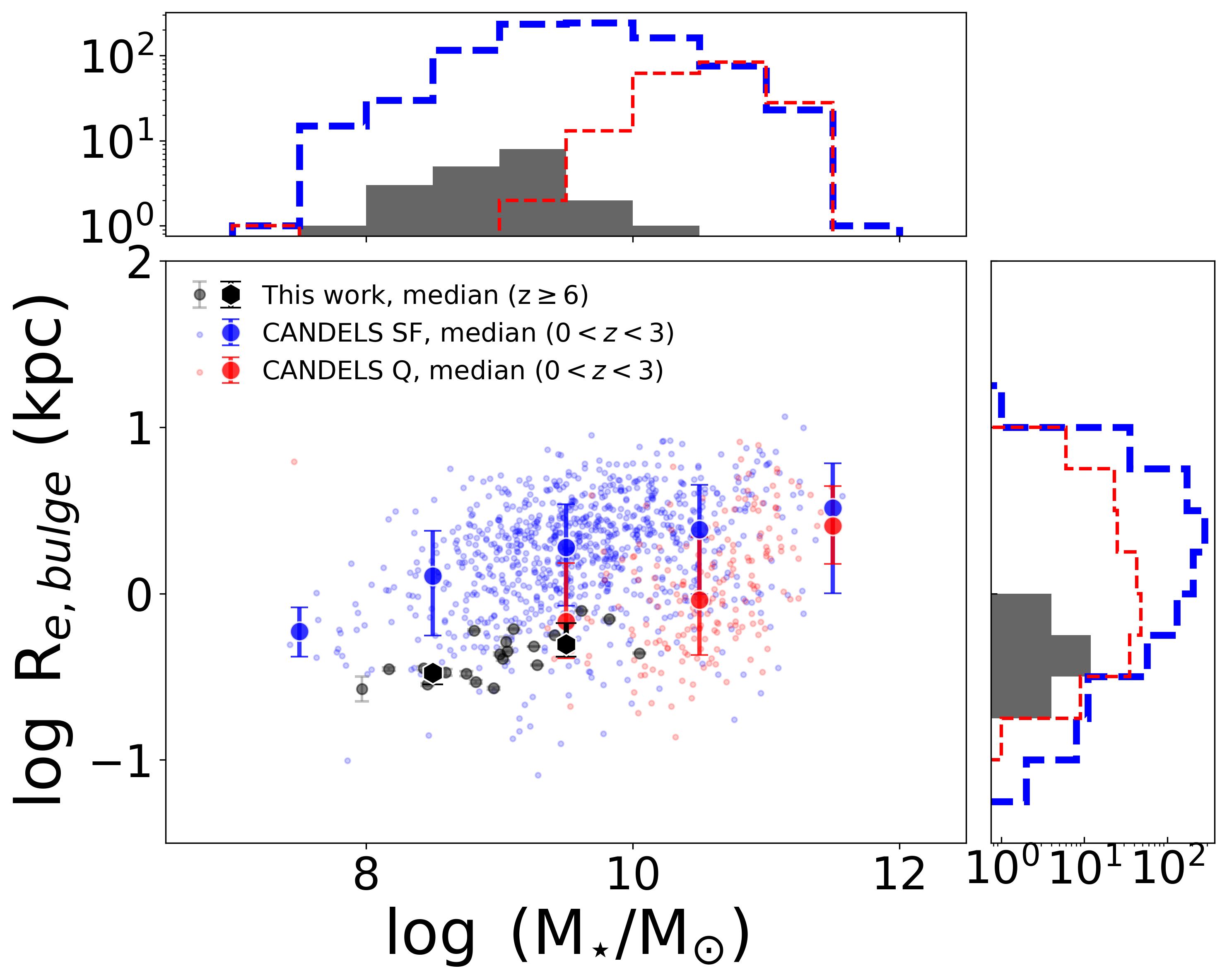}
\caption{\textbf{The rest-frame optical R$_{e,~bulge}$-to-total stellar-mass trends for the two-component galaxies.} The grey (and black) hexagons show the R$_{e,~bulge}$ (and median values), of galaxies identified with two-component structures in this work. The blue and red points represent the sizes of bulges in CANDELS galaxies and their corresponding median values for SF and quiescent (Q) galaxies at $0<z<3$, respectively. The upper and right panels show the distributions of stellar mass and R$_{e,~ bulge}$, respectively. The grey histograms correspond to the galaxies from this work, while the blue dashed and red dotted histograms represent the CANDELS SF and quiescent populations. The error bars on median values denote the 16th and 84th percentiles of the distribution within each stellar-mass bin.}  
\label{fig:bulge_size_galaxy_mass}
\end{figure}

 The basic measured quantities of the two-component galaxies are presented in Table \ref{tab:bulge-disk-table}. We find that the median optical effective radius of bulges (R$_{e,~bulge}$) at $z\geq6$ in the two-component systems is 0.42 kpc. We investigate whether the extreme star-formation, as indicated by the above results, contributes significantly to bulge growth at Cosmic Dawn. For that, we examine R$_{e,~bulge}$ in relation to the total galaxy stellar-mass as shown in Figure \ref{fig:bulge_size_galaxy_mass}. It shows that the bulges are more compact for a given galaxy stellar-mass and their sizes increase with increasing stellar-mass. We obtain the Spearman correlation coefficients to be 0.73, 0.33 and 0.45 for our $z\geq6$ sample, CANDELS SF and CANDELS quiescent galaxies, respectively. The positive size-mass correlation for the $z\ge6$ sample is consistent with ongoing star-formation contributing to the rapid bulge growth alongside the host galaxy stellar mass assembly. Despite a high positive correlation coefficient for the $z\ge6$ sample, we caution against a strong claim for a bulge-size and host-galaxy mass relation given the limited sample size. Our measurements are broadly consistent with recent observations at high-$z$. A stellar bulge as compact as $<0.1$ kpc was discovered at $z\sim7.4$ \citep{baker-etal2025}. Based on \textit{JWST} observations, \citet{genin-etal2025} found that R$_{e,~bulge}$ in massive (M$_{\star}>10^{10}$ M$_{\odot}$) star-forming galaxies is $\sim0.6$ kpc and for quiescent galaxies is $\sim0.4$ kpc at $z\sim5$. The study also shows a mildly decreasing trend in the bulge size towards higher redshift, up to $z\sim6$ for both SF and quiescent galaxies. We note in Figure \ref{fig:bulge_size_galaxy_mass} that the CANDELS star-forming and quiescent medians are offset from one another by roughly 0.5 dex across most of the stellar mass range, with bulges in star-forming galaxies being systematically larger. However, this offset narrows toward the highest masses. In fact, at later epochs ($1<z<3$), \citet{bruce-etal2014} found that stellar bulges in massive galaxies (M$_{\star}>10^{11}$ M$_{\odot}$) have similar sizes in both star-forming and quiescent galaxies, consistent with the convergence we see at the high-mass end of the CANDELS sample.

\begin{figure*}
\figurenum{7}
\includegraphics[width=\linewidth]{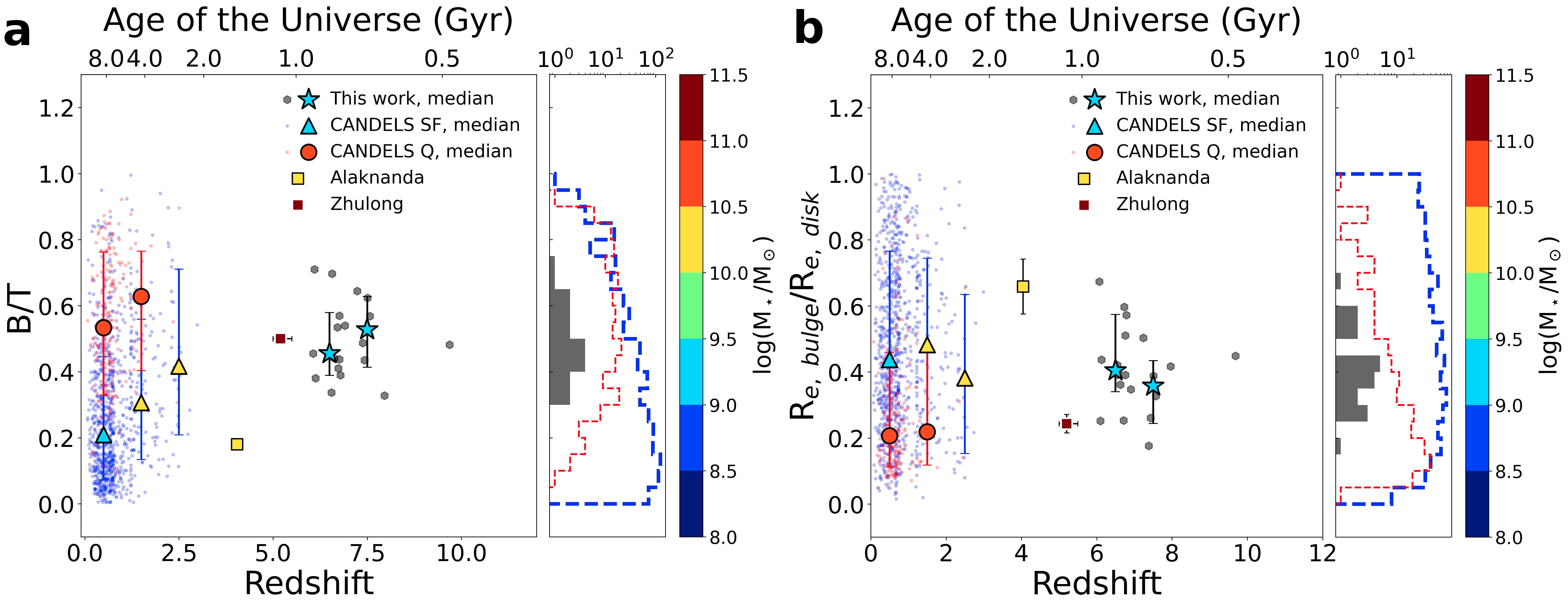}
\caption{\textbf{The evolution of bulge-to-total light ratio (B/T, panel a) and bulge-to-disk size ratio (R$_{e,~bulge}$/R$_{e,~disk}$, panel b) with redshift.} The grey hexagons are the galaxies in this study. The CANDELS SF and quiescent (Q) galaxies are represented by blue and red datapoints respectively. The redshift-binned median values are shown as stars (this work), triangles (CANDELS SF), and circles (CANDELS Q), and are color-coded according to the median stellar mass of each redshift bin. The error bars on the median values denote the 16$^{\rm th}$ and 84$^{\rm th}$ percentiles of the distribution within each redshift bin. We plot the recently discovered grand-design spirals -- Alaknanda \citep{jain_wadadekar2024} and Zhulong \citep{xiao-etal2025}, color-coded by their stellar masses, for comparison. Histograms of B/T and R$_{e,~bulge}$/R$_{e,~disk}$ are shown on the right of their respective scatter plots.}   
\label{fig:bulge_growth_with_time}   
\end{figure*}

\subsection{Bulge growth and bulge-to-total light (B/T)}

We present the rest-frame optical bulge-to-total light ratios (B/T) and bulge-to-disk size ratio (R$_{e,~bulge}$/R$_{e,~disk}$) for the two-component galaxies in our sample in Figure \ref{fig:bulge_growth_with_time}, and compare them with those of low-$z$ systems. The median values are color-coded according to the median stellar mass of each redshift bin to illustrate the possible influence of stellar mass on the observed structural evolution. We use the PSF-corrected \textsc{GALFIT} model magnitudes of the Sersic and exponential disk components to estimate the B/T values for our sample. For the CANDELS galaxies, we take care to select only those objects from the catalog \citep{dimauro-etal2018} which have R$_{e,~bulge}~\leq$ R$_{e,~disk}$, while excluding unclassified objects and stars. Additionally, we use the B/T measurements obtained in the HST-F160W filter, which is a better representative of a galaxy's bulge mass. We note that this is a comparison between inhomogeneous measurements. However, this choice ensures that the emission is traced by the older stars in a galaxy at a given redshift and should be representative of the underlying mass. In our $z\geq$ 6 sample, the rest-frame optical B-band emission trace the longer lived stars as compared to the bluer U-band light. 

Figure \ref{fig:bulge_growth_with_time} shows that galaxies at $z\ge$ 6 exhibit high B/T values (median B/T = 0.47). It also shows that such high B/T is observed at lower stellar masses than in the low-$z$ CANDELS population. It is quite possible that the core of the galaxy formed first, or it is forming rapidly as compared to its stellar disk, which also points towards the inside-out growth scenario at these redshifts \citep[e.g.,][]{baker-etal2025}. The high B/T attained early on and at even lower stellar masses further suggests that bulge assembly at these epochs proceeds efficiently and the disk assembly is yet to catch up. As we present in Table \ref{tab:bulge-disk-table}, the restframe \textit{U-} to \textit{B-} band size ratios for bulges and disk components in almost half of the two-component galaxy sample is greater than 1; indicating an inside-out growth scenario. 

Notably, recent detection of grand-design spirals at $z_{phot}$ = 4.0 with M$_{\star}$/M$_{\odot}$ = 10$^{10.17}$, B/T = 0.18, R$_{e,~bulge}$ = 2.2 kpc \citep{jain_wadadekar2024} and at $z_{phot}$ = 5.2 with M$_{\star}$/M$_{\odot}$ = 10$^{11.03}$, B/T = 0.44, R$_{e,~bulge}$ = 0.9 kpc \citep{xiao-etal2025} evidently indicates that evolved structures can indeed exist at such epochs. The high B/T values observed for our $z\ge$ 6 galaxies and their shorter $\tau_{SF}$ (Figure \ref{fig:histograms}) provide further evidence that a rapid build-up of their bulges is happening in the early Universe. Their elevated stellar mass densities also imply that the drivers of quenching were already in place within the first Gyr. Previous observations have identified quenched galaxies at $z>$ 4 with compact and dense bulges that lead to their quenching in less than a Gyr \citep{barro-etal2013, dekel_burkert2014}. The increasing number of such quenched galaxies at high-$z$, as revealed by newer observations \citep{tanaka-etal2019,nanayakkara-etal2025, antwi-danso-etal2025} also suggest an early onset of quenching, which is crucial to explain the emergence of quiescent galaxies at these epochs. Together, the high-B/T systems we identify at $z \geq 6$ indicate that significant bulge growth is already underway within the first billion years of cosmic history, and that these systems may represent plausible progenitor candidates of the bulge-dominated, and subsequently quiescent, galaxies observed at later cosmic times.

\begin{table*}[htp]
    \centering
    \caption{The sample of galaxies modeled with a bulge and a disk component.  $^{*}$ N- (S-) indicates that object is present in the GOODS North (South) field. The identifier JADES (FRESCO-H$\alpha$, FRESCO-OIII) indicates that the object ID is the NIRCAM ID taken from the JADES catalog (IDs from the FRESCO catalogs of H$\alpha$ and OIII emitters). Columns 5 and 6 present the wavelength-dependent size ratios for the disk and bulge components, defined as R$_{e,~disk,1}$/R$_{e,~disk,2}$ and R$_{e,~bulge,1}$/R$_{e,~bulge,2}$, respectively. Here, the subscripts 1 and 2 refer to the shorter- and longer-wavelength filters. Column 8 refers to the difference in BIC value as described in equation \ref{eq:delta-bic}. Other columns represent properties with their usual meanings.}
    \begin{tabular}{cccccccc}
    \hline
    \hline
       $^{*}$ID  & z$_{spec}$ & $\log$ M$_{\star}$/M$_{\odot}$ & SFR (M$_{\odot}$yr$^{-1}$) & R$_{e,~disk, 1}$/R$_{e,~disk, 2}$ \footnote{Objects which have an arbitrarily large disk size in any of the filters is marked as **. We do not quote a value if both the rest-frame wavelength ranges fall within a single filter.\\$^{\dagger}$ As we inspect the galaxy in the bluest filters, a clumpy morphology is evident.\\ $\#$The profile of this galaxy do not extend beyond the 80\% PSF-flux enclosing radius, but a two-component model results in a better fit in the rest-frame \textit{B}-band and satisfies the $\Delta$BIC condition.} & R$_{e,~bulge, 1}$/R$_{e,~bulge, 2}$ & B/T &$\Delta$BIC\\
       \hline
    N-FRESCO-Ha-1324 &6.5690&8.82&9.61&1.78&    1.07   &0.70&-83.1\\
    N-FRESCO-OIII-2203 &7.5055&8.75&28.43&0.48&  0.70   &0.62&-73.4\\
    N-JADES-1009077 &6.7589&9.82&4.16&1.08&   1.17     &0.44&-1126.1\\
    N-JADES-1010816 &6.7596&10.05&7.36&**&    **      &0.57&-240.7\\
    N-JADES-1022943 &6.7328&8.81&7.54&1.37&   1.07     &0.41&-155.9\\
    N-JADES-1078455 &6.5478&9.05&6.97&1.23&   1.14     &0.34&-213.3\\
    N-JADES-1078891 &6.5488&9.26&15.44&1.28&    1.02    &0.46&-12910.2\\
    S-FRESCO-Ha-18617 &6.0670&9.41&14.99&-&     -    &0.46&-865.5\\
    S-FRESCO-OIII-3081 &7.5708&9.03&20.27&1.38&  1.05   &0.57&-1569.3\\
    S-FRESCO-OIII-23077 &7.3803&9.28&5.40&1.12&  1.05   &0.49&-2722.8\\
    S-FRESCO-OIII-26147 &6.9084&8.96&11.83&0.83&  0.89  &0.54&-1229.2\\
    S-FRESCO-OIII-28631 &7.9606&9.11&11.78&-&   -    &0.33&-4643.5\\
    S-JADES-116352 &6.7901&8.43&1.71&0.85&    0.83      &0.39&-238.2\\
    S-JADES-117602 &7.4302&8.46&14.61&**&     **      &0.43&-291.9\\
    S-JADES-122436 &7.2373&9.06&14.64&1.17&    1.02     &0.64&-1247.2\\
    S-JADES-128502 &6.1266&8.60&0.03&-&      -       &0.38&-193.5\\
    S-JADES-132845 &6.0996&9.00&5.06&-&       -      &0.71&-941.4\\
    S-JADES-147779 &6.6299&8.17&7.51&0.83&     0.8     &0.44&-478.7\\
    $^{\dagger}$S-JADES-197888 &6.7126&9.61&4.87&0.97&  1.07   &0.53&-63382.5\\
    $\#$S-JADES-107324 &9.6860&7.97&4.73&**&     **       &0.48&-68.8\\
    \hline
    \end{tabular}
    \label{tab:bulge-disk-table}
\end{table*}

Figure \ref{fig:bulge_growth_with_time} reveals two apparent sequences in the B/T values as a function of $z$, corresponding to SF and quiescent galaxies. It shows that quiescent galaxies have higher B/T ratios (median B/T = 0.54, MAD = 0.14) than SF galaxies (median B/T = 0.23, MAD = 0.13) for the combined CANDELS sample at $0<z\lesssim3$). We observe that the CANDELS populations, particularly the SF population, exhibit lower B/T values at lower redshift, where the median stellar masses are also lower. However, such generalisations may be affected by the large uncertainties arising from incompleteness and therefore a larger sample over the full redshift range is required to disentangle the relative roles of stellar mass and cosmic evolution. Interestingly, the relative size of a bulge to its disk in SF galaxies is larger (median R$_{e,~bulge}$/R$_{e,~disk}$ = 0.44, MAD = 0.19) than that observed for quiescent ones (median R$_{e,~bulge}$/R$_{e,~disk}$ = 0.21, MAD = 0.09). The median values of the rest-frame R$_{e,~bulge}$/R$_{e,~disk}$ appear to be non-evolving for SF galaxies with redshift. We extend the sample size beyond $z=$ 6 and find that the median R$_{e,~bulge}$/R$_{e,~disk}$ = 0.40, similar to low-$z$ SF galaxies. These results indicate that the onset of these two sequences inevitably began with the formation of bulges within the first Gyr. Such rapid bulge formation is possible via a ``blue-nugget'' phase - which happens within a few 100 Myr \citep{dekel_burkert2014,zolotov-etal2015}. This is likely to be accompanied by inward migration of massive clumps, rapid gas inflows or major mergers within timescales $<$1 Gyr \citep{noguchi1999, dekel-etal2009, hopkins-etal2009, ceverino-etal2010, tachella-etal2016a}. Subsequently, the halo gas or accreting gas can settle into stable rotating disks. As the stellar bulge quenches and the disk continues to form stars, the B/T ratio decreases along with the decrease in R$_{e,~bulge}$/R$_{e,~disk}$. Alternatively, intense star formation, or AGN activity in young bulges can exhaust, expel, or ionize the pristine halo gas. This thereby suppresses subsequent disk growth or quenches the system completely, resulting in high B/T ratios observed in quiescent galaxies or potentially evolve into elliptical galaxies. This dual evolutionary pathway is consistent with theoretical expectations \citep{dekel_burkert2014, zolotov-etal2015, tachella-etal2016a}, and is supported by studies of ``blue nuggets'' and their quenched descendants at $z \gtrsim 2$ \citep{barro-etal2013, van-dokkum-etal2015}. It is interesting to note that the B/T and R$_{e,~bulge}$/R$_{e,~disk}$ measurements for quiescent galaxies, based on the CANDELS sample, remain broadly similar up to $z\sim3$ or so within errorbars, i.e., we find no evidence for strong evolution in these quantities. However, we caution that this indication is limited by the sample size, and also the true evolutionary picture may only emerge once other relevant factors are properly accounted for. For example, whether the bulge is of a \textit{classical} or \textit{pseudo} type could play an important role, since the two point to distinct formation and evolutionary pathways. Our above results may also indicate a possible co-evolution of the bulge and disk components in these systems, which will be explored in a future work. We do not rule out this possibility for star-forming galaxies but further data is required to look into this.

\begin{figure*}
\figurenum{8}
\centering
   \includegraphics[width=0.9\textwidth]{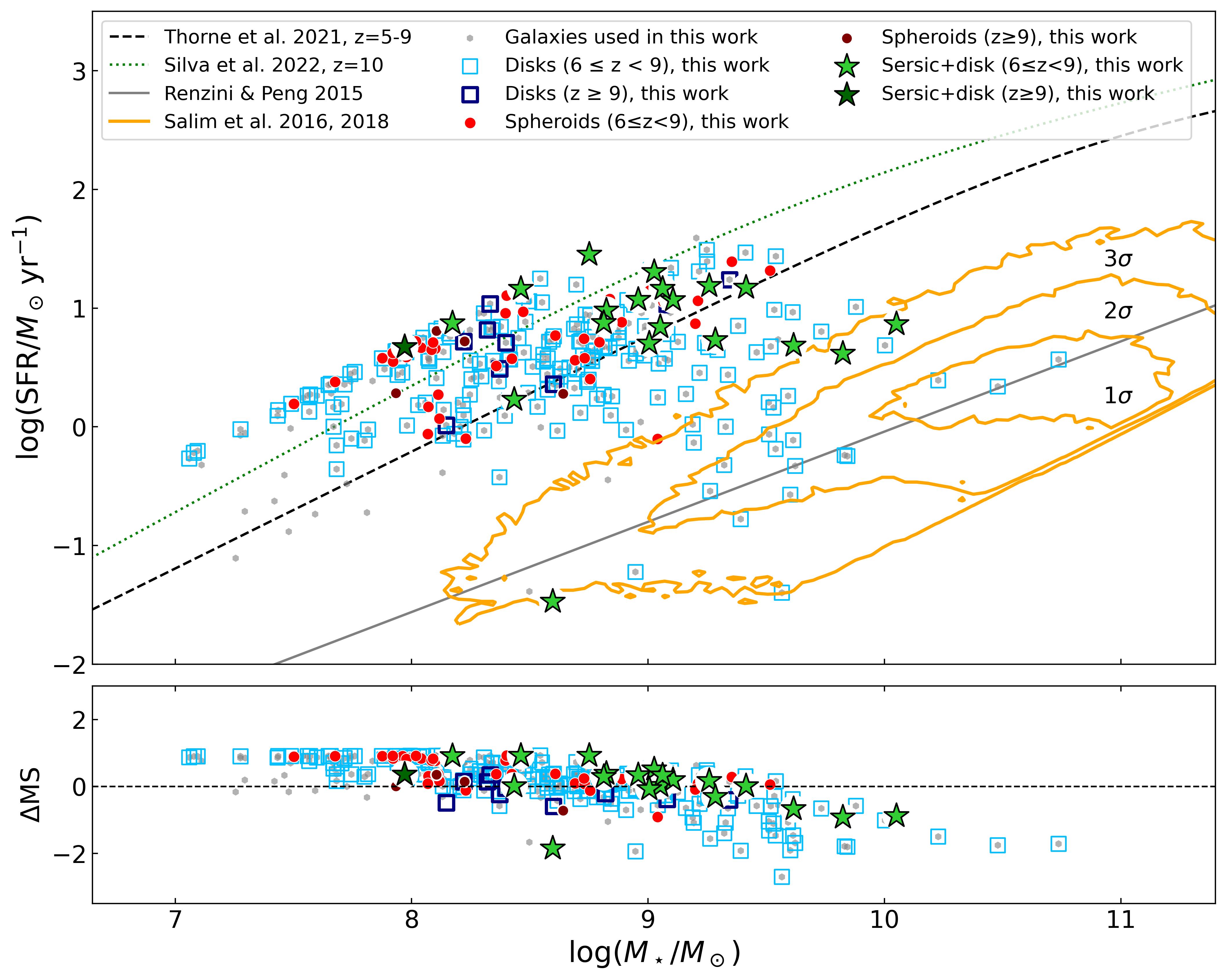}
\caption{\textbf{The galaxy main-sequence and $\Delta$MS for the sample and comparison with local star-forming galaxies.} The morphological information is based on rest-frame \textit{B}-band emission for our sample and SDSS-\textit{g} band for local galaxies. The yellow contours represent the distribution of the local SDSS star-forming galaxies from \citet{salim-etal2016, salim-etal2018}, showing the 1$\sigma$, 2$\sigma$, and 3$\sigma$ regions. Different main-sequence relations obtained from observations and predicted from simulations are shown by different curves. Black dashed curve represents the observed main sequence relation for high-$z$ galaxies from \citet{thorne-etal2021}. The green dotted curve represents the predicted relation for high-$z$ galaxies from the FLARES-\textit{JWST} simulations \citep{dsilva-etal2023}. The solid gray curve represents the observed relation in the local Universe \citep{renzini_peng2015}.}   
\label{fig:MS-plot}   
\end{figure*}

\subsection{Galaxy main sequence at $z>$ 6}

We examine the SF main sequence (MS) for our galaxy sample at $z\ge$ 6 and compare it with the relations in the local and high-$z$ Universe. The best-fit stellar masses and SFRs for our sample, together with those of local galaxies \citep{salim-etal2016, salim-etal2018}, are shown in Figure \ref{fig:MS-plot}. The local SF galaxies are selected using a threshold $\log$ sSFR $=$ -11 \citep{salim-etal2018}. As predicted by simulations, the MS relation is already in place during Cosmic Dawn, driven by highly efficient and self-regulated star formation \citep[e.g.,][]{ceverino-etal2018, lovell-etal2021, dsilva-etal2023, Ceverino-etal2024}. Such high efficiency is expected to result from high gas fractions, short depletion timescales, merger-driven starbursts and/or clumpy gas accretion \citep{ceverino-etal2018, dsilva-etal2023}. However, some models also suggest that the assembly of the first massive galaxies may deviate from this expected scenario \citep[e.g.,][]{qin-etal2023}. We find that the majority of galaxies in our sample agree well with the MS predictions at high-$z$ \citep{dsilva-etal2023}, exhibiting high sSFRs as compared to present-day systems.

We also identify the morphologies of our sample in Figure \ref{fig:MS-plot}. A major fraction of the galaxies are morphologically disk-like ($n<$1.5), as probed in the rest frame optical \textit{B}-band emission. This finding is consistent with previous observations at high-$z$ that report a high fraction of disk-like galaxies \citep[e.g.,][]{ferreira-etal2023, kartaltepe-etal2023}. However, such low-$n$ systems need not necessarily be rotating disks as observed in previous studies \citep[e.g.,][]{Krajnovic-etal2013}. For brevity, we call them disk-like galaxies. As we can see in Figure \ref{fig:MS-plot}, a smaller fraction of galaxies, all with log M/M$_{\odot} \lesssim$9.5, have higher Sersic indices ($n>$1.5), which we categorize as spheroid-like. As these galaxies evolve, the disk-like galaxies may build their stellar bulges or conversely, the spheroid-like galaxies may build their outer disks at later times \citep[e.g.,][]{sachdeva_saha2018}. However, at this stage there is no clear dichotomy in morphology with stellar mass. While the spheroid-like galaxies are confined to the lower-mass regime, the disk-like galaxies span the full stellar-mass range probed by our sample, and both morphological types are present at low stellar masses. This is expected of galaxies at these epochs as they are still in the process of assembly.
 We note that all the galaxies in the sample that contain a bulge and an exponential disk have stellar mass log M/M$_{\odot} \gtrsim$ 8.

We further find that, a galaxy with a bulge and disk component have lower SFRs which are comparable to local galaxies, suggestive of an early onset of morphological quenching due to bulge formation. While this is a possibility, we also find a few disk-like galaxies located in the same region below the high-$z$ MS. This is better noticeable in the $\Delta$MS plot. This likely reflects the bursty nature of star-formation or self-regulation in these early systems \citep{tachella-etal2016a,ceverino-etal2018}.

We note that a small number of galaxies in the sample appear to lie along a line on the main sequence, a well-known artefact arising from parameter gridding in SED-fitting tools such as CIGALE. Our grid, consisting of $\sim$6 million models per object, was chosen to minimise this effect while maintaining computational feasibility and efficiency. We compare the best-fit and Bayesian-marginalised stellar masses and SFRs, and find robust agreement between the two estimates, as presented in the Appendix.

\section{Summary and Conclusions}

In this work, we study a sample of spectroscopically confirmed galaxies that existed when the Universe was less than a Gyr old. We perform broadband SED modeling of the galaxies to derive and gain insight into their star formation properties. Importantly, we perform morphological analysis of the sample using \textsc{GALFIT} and identify a subset hosting growing stellar bulges. The key findings of our work are summarised as follows:
\begin{itemize}

    \item The $z\ge$ 6 galaxies are compact, with a majority of them exhibiting disk-like morphologies and median half-light radius of $\sim$ 0.51 kpc and Sersic index of $\sim$0.96.

    \item The first galaxies (at $z\ge$ 6) are undergoing intense star-formation, reaching $\Sigma_{SFR}\sim$ 12.1 M$_{\odot}$yr$^{-1}$kpc$^{-2}$. Such elevated star-formation activity naturally implies rapid stellar mass assembly, for instance, galaxies acquiring $\sim10^{9}$ M$_{\odot}$ within timescales of $\sim$100 Myr.

    \item Our results reveal the rapid inner assembly and growth of bulges in $\sim$ 10\% of $z\ge$ 6 galaxies. These galaxies have rapidly grown their cores and exhibit stellar mass densities as high as $\sim$ 10$^{10}$ M$_{\odot}$kpc$^{-2}$. Their relatively high bulge-to-total light ratios (B/T$ \sim 0.47$) suggest that their stellar disks are still in the process of assembly and that they will evolve into their lower-$z$ counterparts at later stages.

    \item A fraction of $z\geq6$ galaxies, predominantly those with $n<1.5$, exhibit comparatively reduced SFRs, possibly reflecting the bursty star-formation histories expected at these early epochs. Interestingly, this subset also includes a galaxy modeled with a Sersic and exponential disk component. While some of these systems exhibit high B/T ratios and elevated stellar mass densities -- properties that may be consistent with the possibility of morphological quenching, our current sample does not allow us to distinguish between the onset of such a process or bursty star formation activity. A larger sample of galaxies with similar structural properties will be required to robustly assess the role of morphological quenching at these redshifts.

\end{itemize}

The presence of dense stellar bulges at these early times can provide key insights into the existence of quiescent galaxies at high redshift. These compact structures may also create favorable conditions for the early growth of supermassive black holes (SMBHs) \citep[e.g.,][]{dubois-etal2012}. We do not investigate the environment of these galaxies, but dense environments could also play a role in the accelerated assembly of bulges through frequent interactions and mergers, as suggested by simulations \citep{lapiner-etal2023}. The rapid growth of bulges that we observe at $z > 6$ implies that the dominant processes must be highly dissipative and efficient, consistent with wet mergers and compaction-driven channels.
As these systems evolve, they may separate into star-forming or quiescent populations of galaxies depending on whether their disks are able to stabilize and continue to grow via sustained gas accretion, or whether feedback mechanisms quench further growth. In either scenario, our findings of bulge and disk components in galaxies at $z > 6$ provide crucial evidence for rapid central mass assembly and the onset of inside-out quenching during Cosmic Dawn.

\section*{Acknowledgments}
This work is based on observations taken by the 3D-HST Treasury Program (GO 12177 and 12328) with the NASA/ESA HST, which is operated by the Association of Universities for Research in Astronomy, Inc., under NASA contract NAS5-26555. This work is based on observations made with the NASA/ESA/CSA James Webb Space Telescope. The data were obtained from the Mikulski Archive for Space Telescopes at the Space Telescope Science Institute, which is operated by the Association of Universities for Research in Astronomy, Inc., under NASA contract NAS 5-03127 for \textit{JWST}. These observations are associated with programmes 1180, 1181, 1210, 1286, 1287, 1895, 1963, and 3215.

% jades doi: doi:10.17909/8tdj-8n28
% fresco doi: doi:10.17909/gdyc-7g80

\bibliography{AAS70074R2_new.ms}
\bibliographystyle{aasjournal}
\appendix

\section{SED modeling of the sample}

\begin{figure*}[htpb]
\figurenum{A1}
    \centering
    \includegraphics[width=\linewidth]{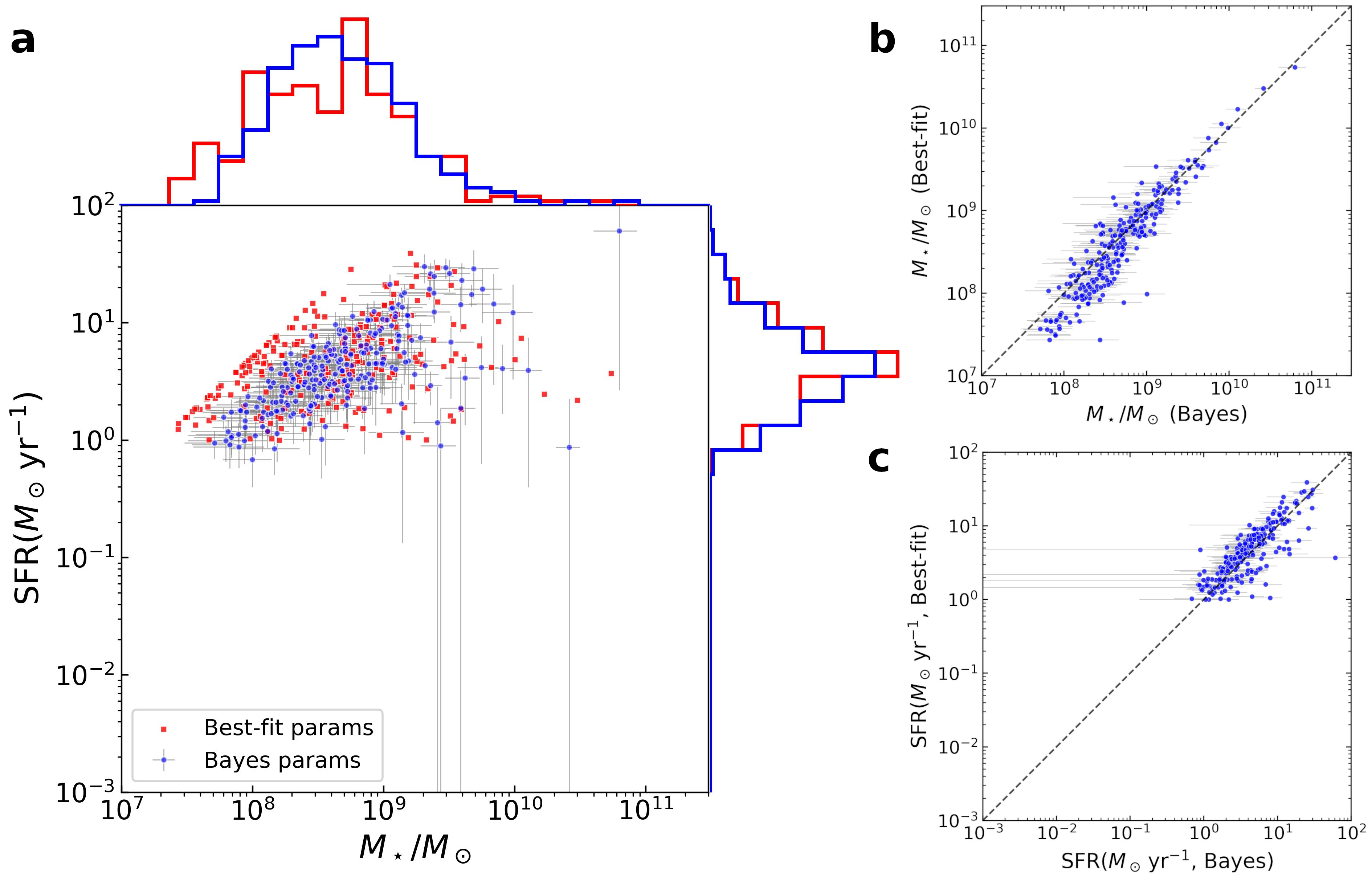}
    \caption{\textbf{Comparison of best-fit and bayesian galaxy stellar masses and star formation rates derived from SED modeling using CIGALE. a:} galaxy main-sequence plot of our sample. Red squares denote the best-fit parameters, while blue circles with gray error bars represent the bayesian estimates and their associated 1$\sigma$ uncertainties. The corresponding distributions of stellar mass and SFR for both methods are shown by the histograms, with same colour coding, along the top and right axes, respectively. \textbf{b, c:} Comparison of the best-fit and bayesian stellar mass and SFR estimates. The dashed lines indicate the one-to-one relation.}
    \label{fig:MS-plot-app}
\end{figure*}

During the modeling, we observe that coarser grids produce a striping pattern in the SFR–stellar mass plane. This can be mitigated by adopting a finer grid space with physically and observationally motivated parameter ranges appropriate for these redshifts. However, this increases the model grid space manifold.  Our final grid was chosen to balance optimal grid resolution with computational efficiency. As a sanity check, we compare the best-fit SFRs and stellar masses with their corresponding Bayesian marginalized estimates and find good agreement between the two, confirming that the grid resolution does not significantly bias our results. This comparison is presented in Figure \ref{fig:MS-plot-app}, which illustrates the consistency between the two sets of estimates while also highlighting the differences arising from grid discretization in the best-fit solutions. Further refinement of the grid could yield even more accurate results; however, precise constraining of all physical parameters is beyond the scope of the current work.

CIGALE also permits the inclusion of emission line fluxes in the SED modeling. A subset of our sample appears in the FRESCO H$_{\alpha}$ and OIII-emitter catalogs, which provide total emission line fluxes measured with the NIRCam/grism instrument. However, the JADES spectroscopic catalog, based on NIRSpec observations, provides fluxes whose slit-loss corrections are optimized for point sources. Combining these two inhomogeneous flux measurements would introduce systematic uncertainties that are difficult to characterize uniformly across the sample. We therefore adopt a homogeneous broadband photometric approach throughout this work.

\clearpage
\section{Morphological Decomposition of two-component galaxies}

\begin{figure*}[htpb]
\figurenum{A1.1}
\centering
\includegraphics[width=0.7\textwidth]{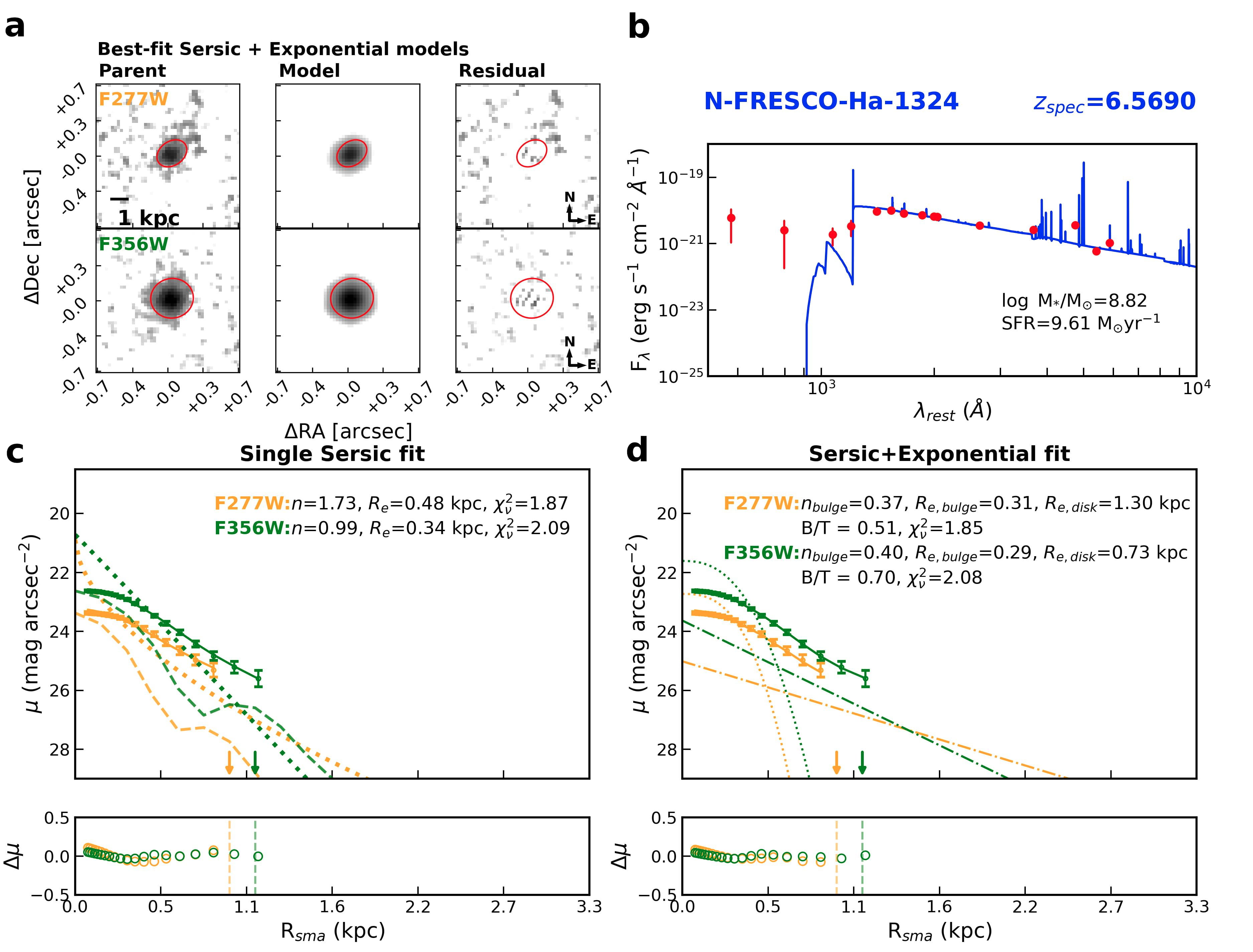}
\caption{Morphological decomposition of N-FRESCO-Ha-1324. See Figure~\ref{fig:representative} for a description of the panels.}
\label{fig:app_N-FRESCO-Ha-1324}
\end{figure*}

\begin{figure*}
\figurenum{A1.2}
\centering
\includegraphics[width=0.7\textwidth]{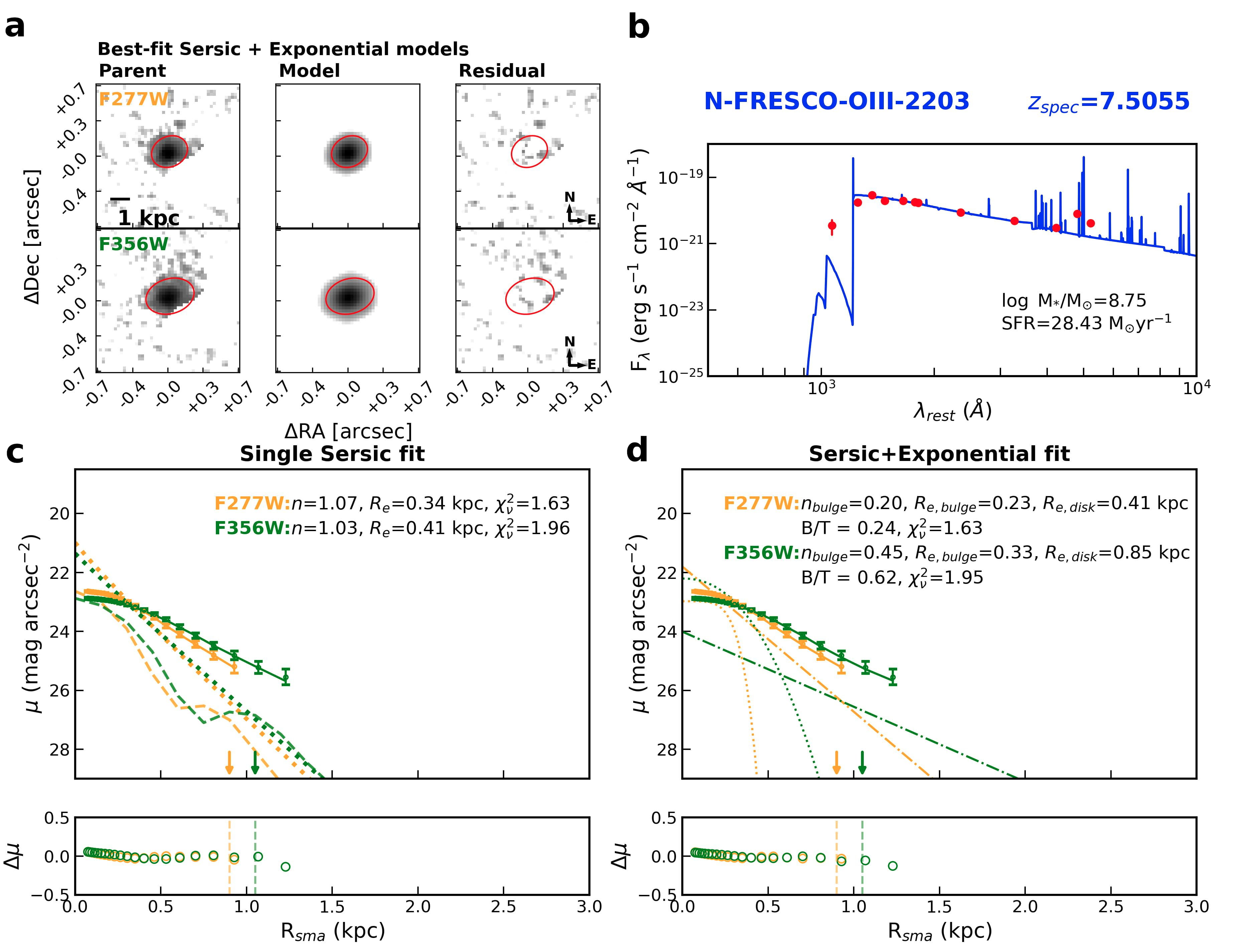}
\caption{Morphological decomposition of N-FRESCO-OIII-2203. See Figure~\ref{fig:representative} for a description of the panels.}
\label{fig:app_N-FRESCO-OIII-2203}
\end{figure*}

\begin{figure*}
\figurenum{A1.3}
\centering
\includegraphics[width=0.7\textwidth]{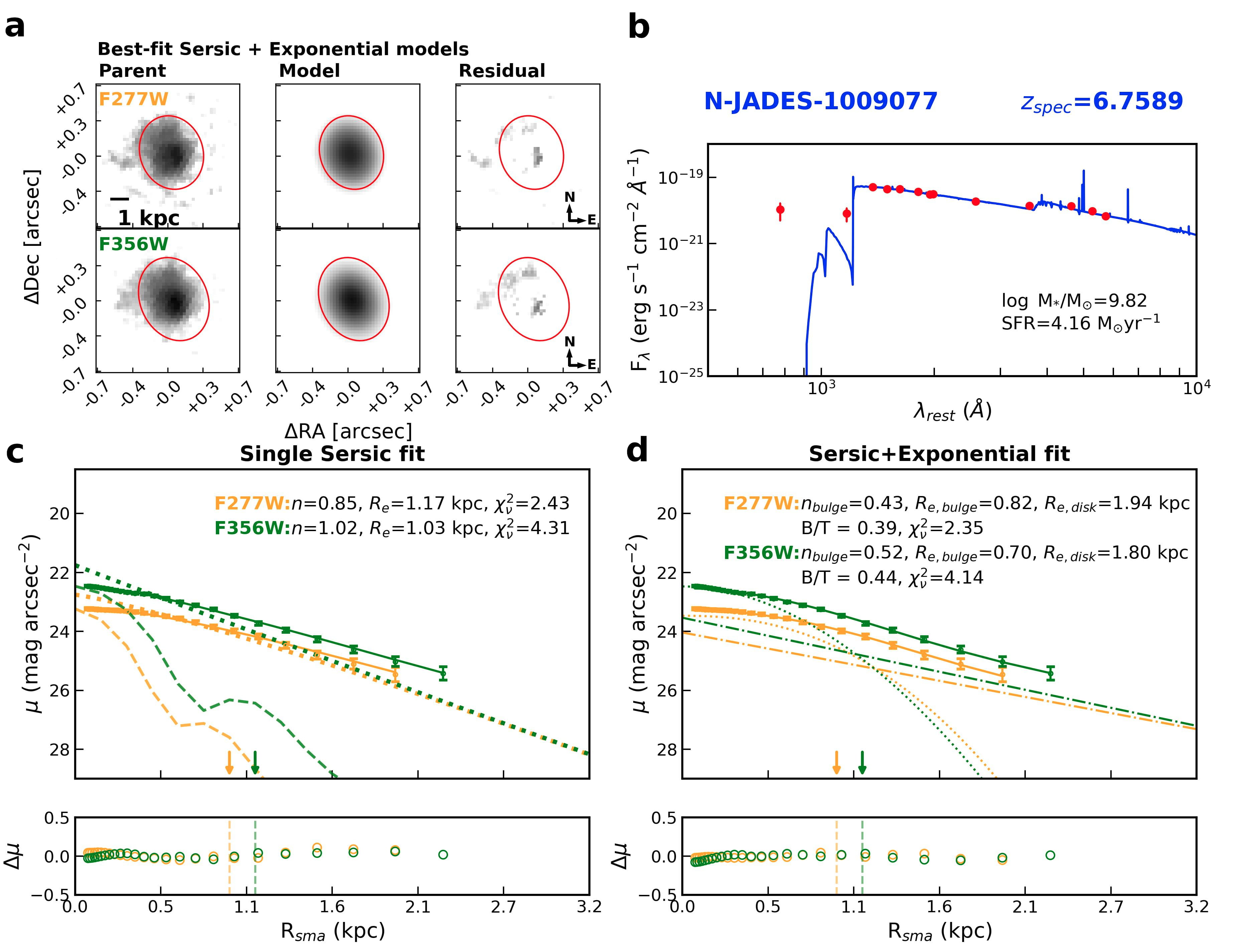}
\caption{Morphological decomposition of N-JADES-1009077. See Figure~\ref{fig:representative} for a description of the panels.}
\label{fig:app_N-JADES-1009077}
\end{figure*}

\begin{figure*}
\figurenum{A1.4}
\centering
\includegraphics[width=0.7\textwidth]{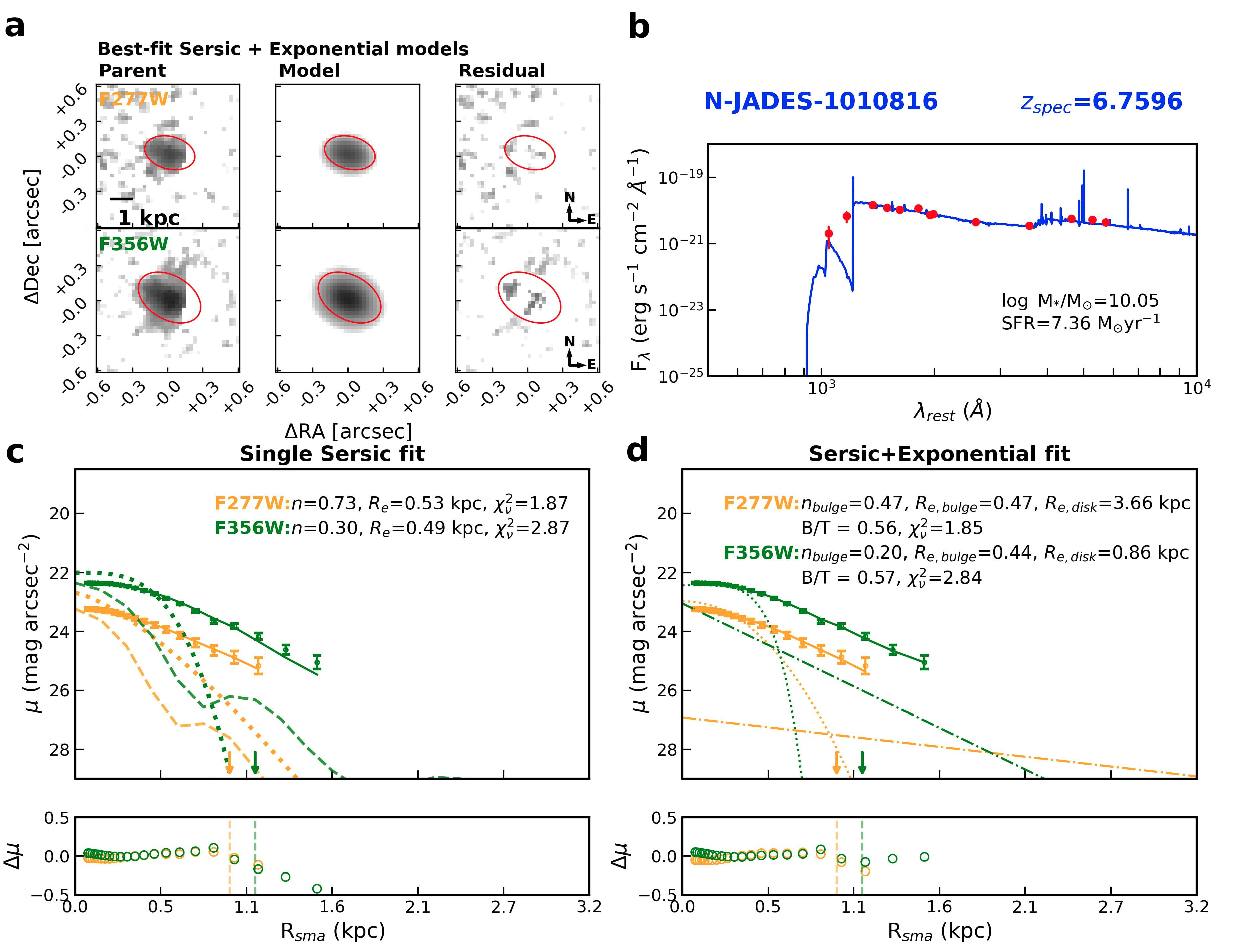}
\caption{Morphological decomposition of N-JADES-1010816. See Figure~\ref{fig:representative} for a description of the panels.}
\label{fig:app_N-JADES-1010816}
\end{figure*}

\begin{figure*}
\figurenum{A1.5}
\centering
\includegraphics[width=0.7\textwidth]{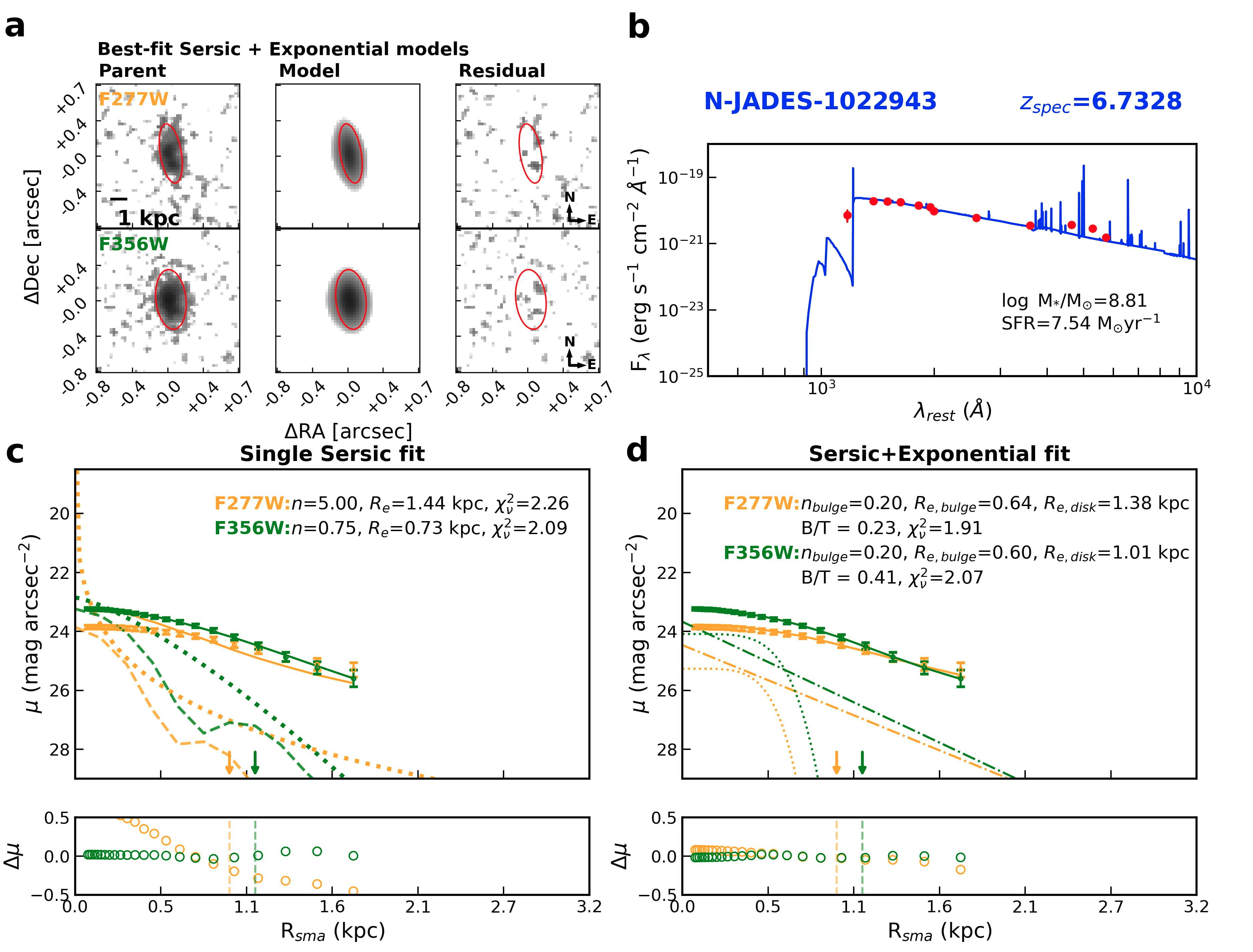}
\caption{Morphological decomposition of N-JADES-1022943. See Figure~\ref{fig:representative} for a description of the panels.}
\label{fig:app_N-JADES-1022943}
\end{figure*}

\begin{figure*}
\figurenum{A1.6}
\centering
\includegraphics[width=0.7\textwidth]{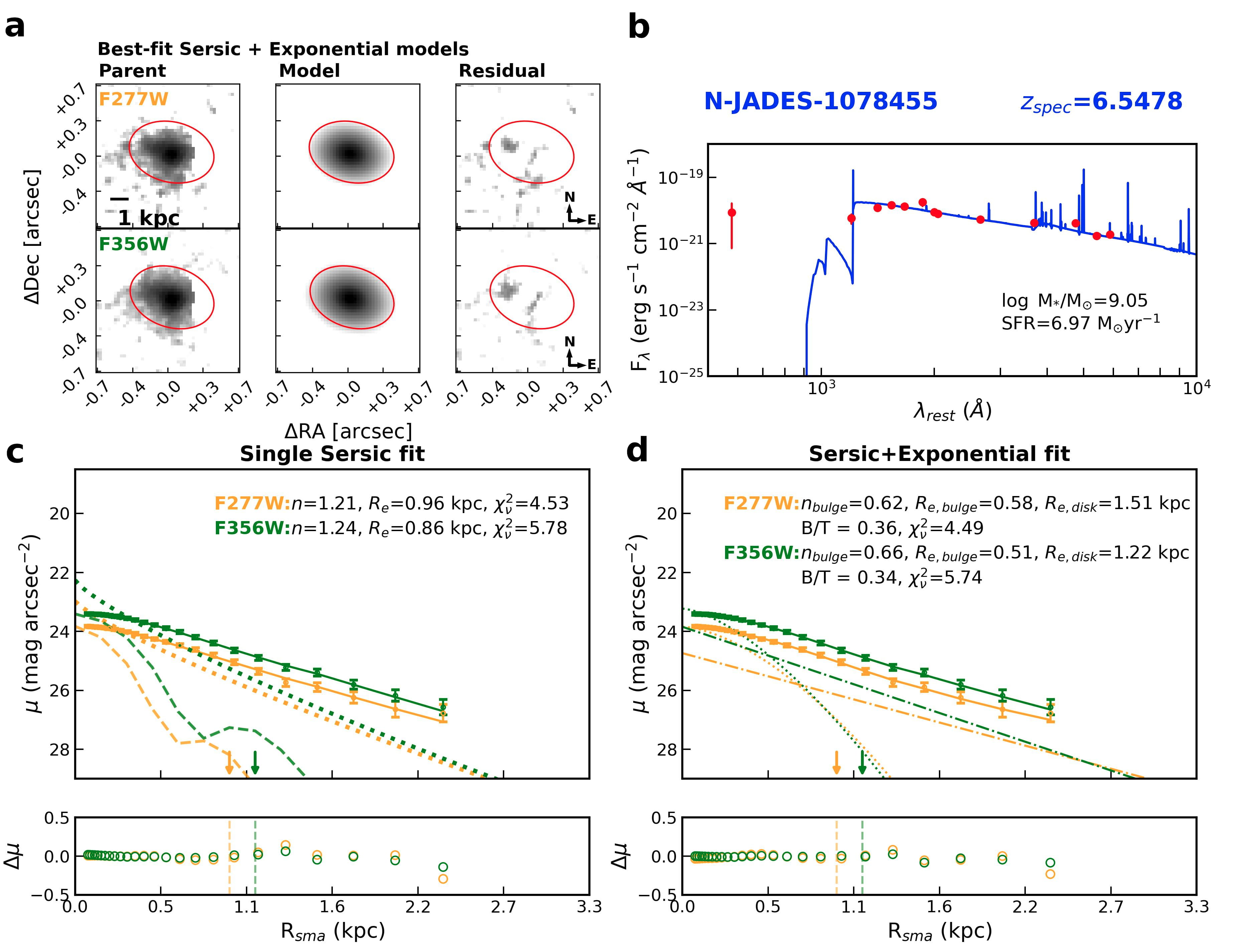}
\caption{Morphological decomposition of N-JADES-1078455. See Figure~\ref{fig:representative} for a description of the panels.}
\label{fig:app_N-JADES-1078455}
\end{figure*}

\begin{figure*}
\figurenum{A1.7}
\centering
\includegraphics[width=0.7\textwidth]{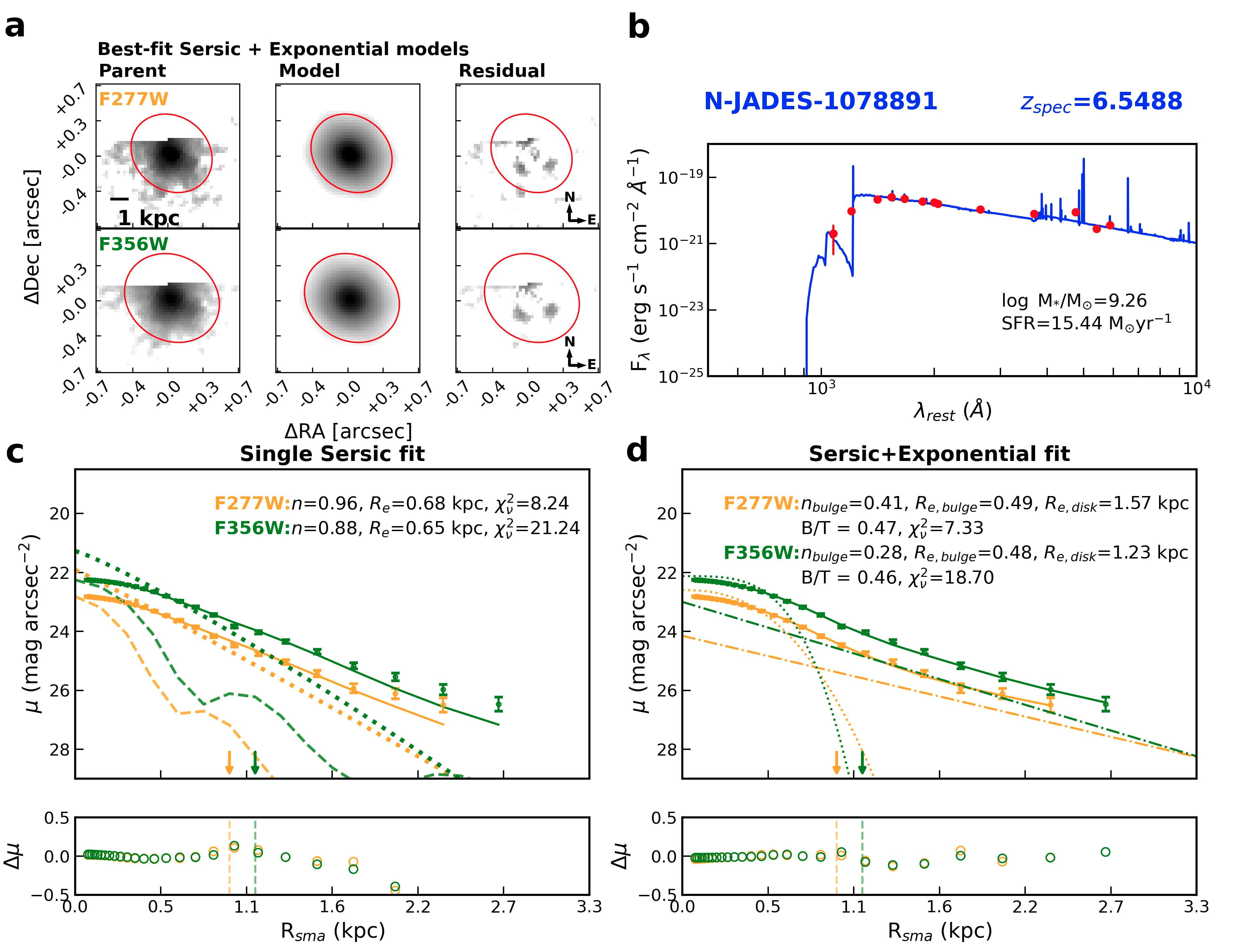}
\caption{Morphological decomposition of N-JADES-1078891. See Figure~\ref{fig:representative} for a description of the panels.}
\label{fig:app_N-JADES-1078891}
\end{figure*}

\begin{figure*}
\figurenum{A1.8}
\centering
\includegraphics[width=0.7\textwidth]{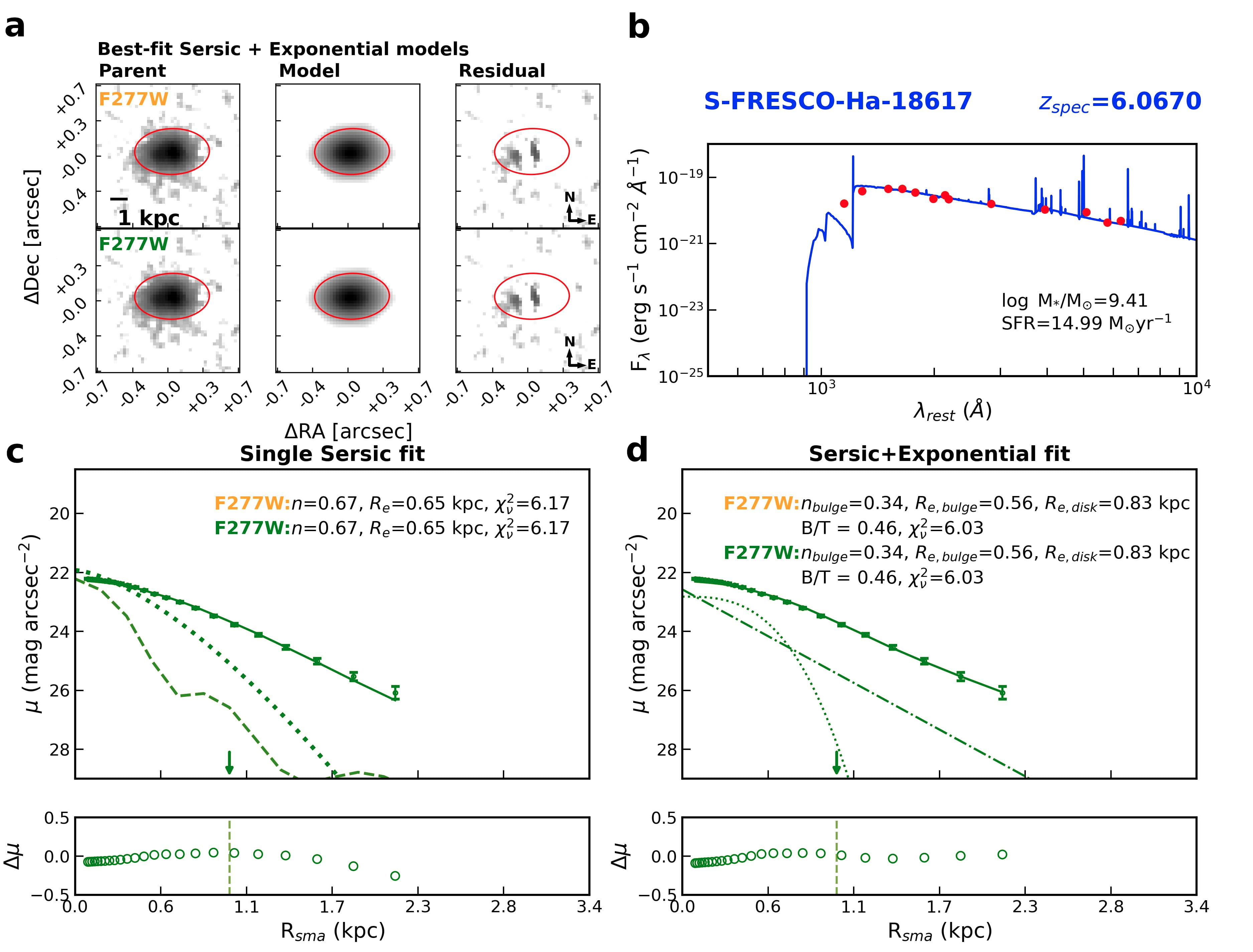}
\caption{Morphological decomposition of S-FRESCO-Ha-18617. See Figure~\ref{fig:representative} for a description of the panels.}
\label{fig:app_S-FRESCO-Ha-18617}
\end{figure*}

\begin{figure*}
\figurenum{A1.9}
\centering
\includegraphics[width=0.7\textwidth]{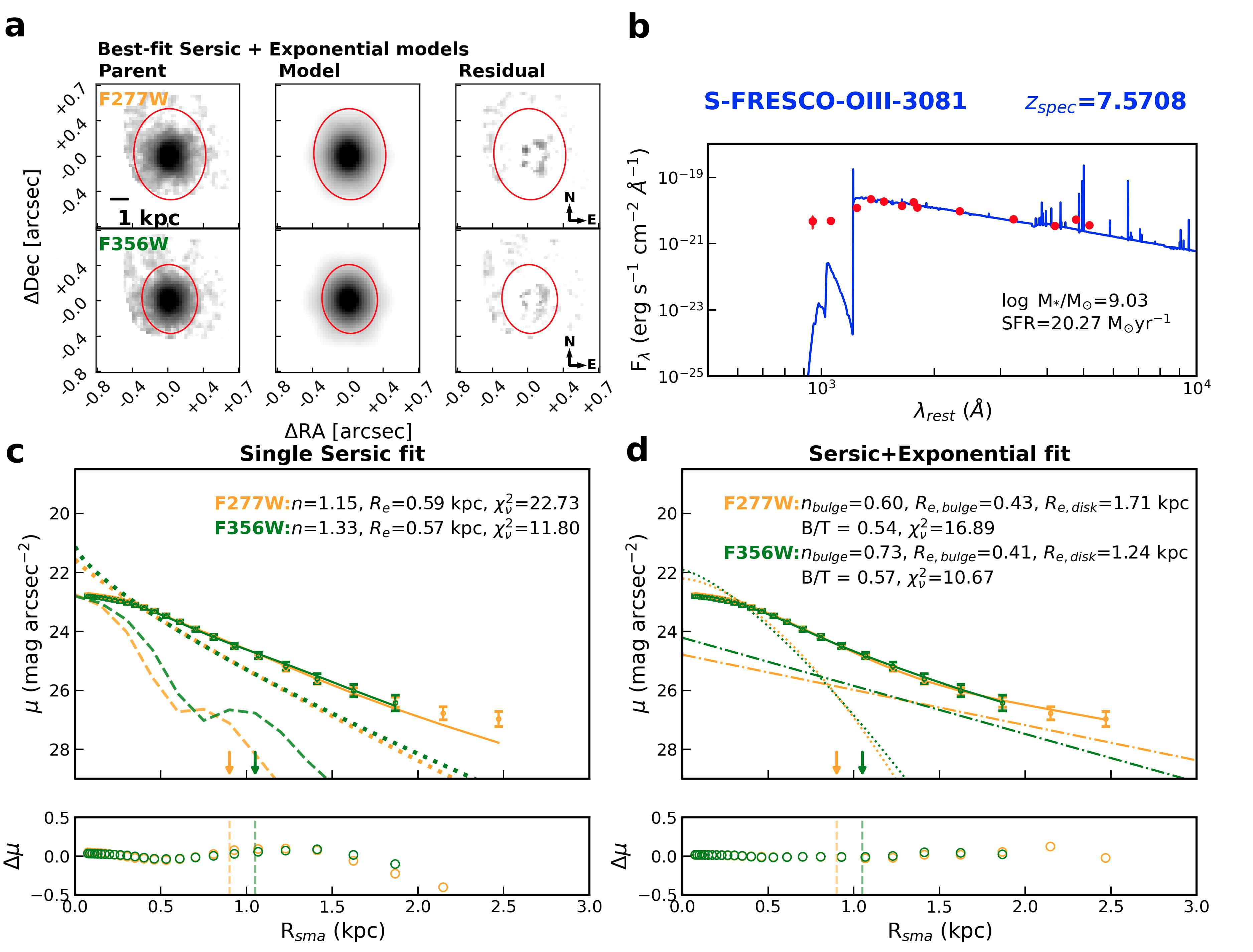}
\caption{Morphological decomposition of S-FRESCO-OIII-3081. See Figure~\ref{fig:representative} for a description of the panels.}
\label{fig:app_S-FRESCO-OIII-3081}
\end{figure*}

\begin{figure*}
\figurenum{A1.10}
\centering
\includegraphics[width=0.7\textwidth]{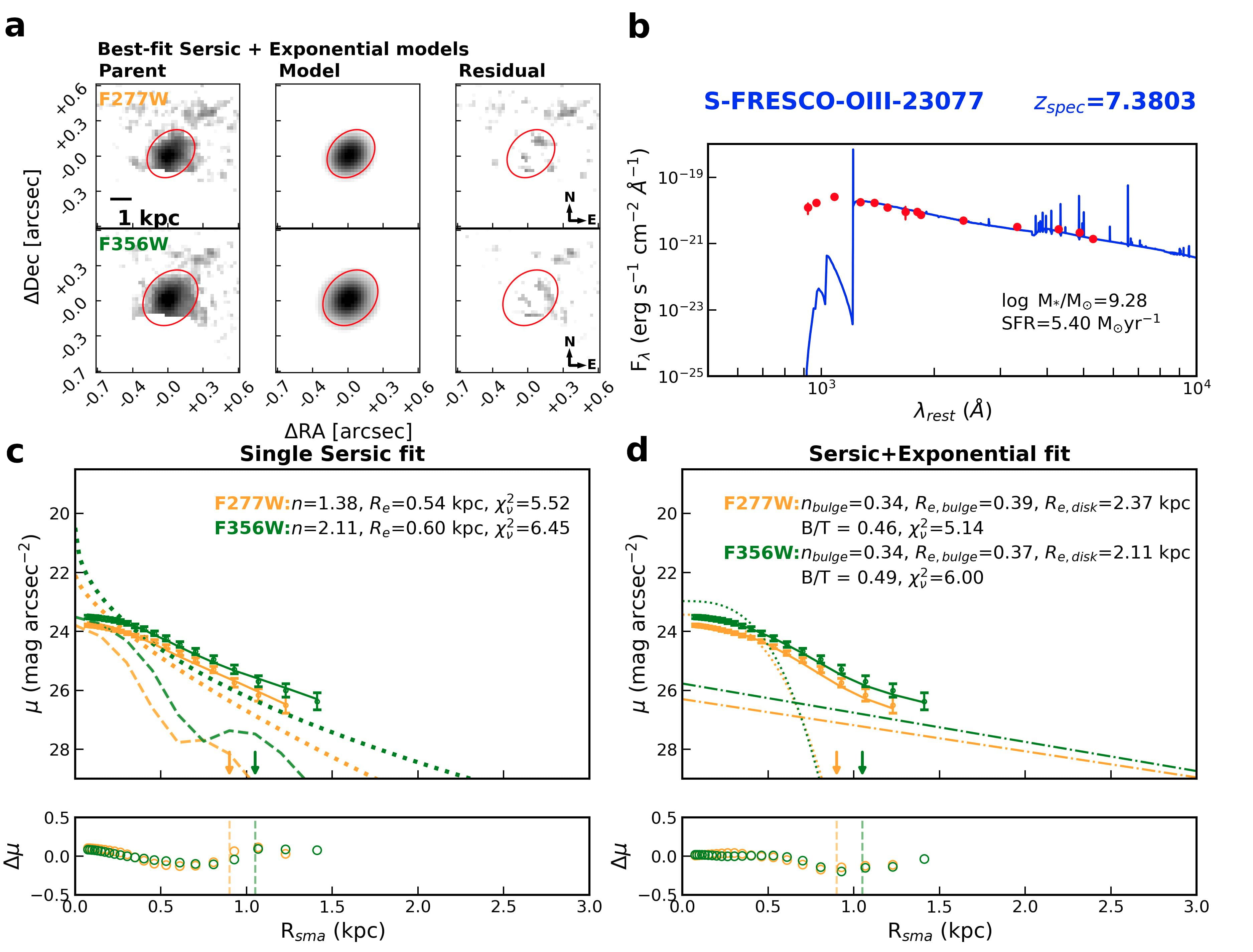}
\caption{Morphological decomposition of S-FRESCO-OIII-23077. See Figure~\ref{fig:representative} for a description of the panels.}
\label{fig:app_S-FRESCO-OIII-23077}
\end{figure*}

\begin{figure*}
\figurenum{A1.11}
\centering
\includegraphics[width=0.7\textwidth]{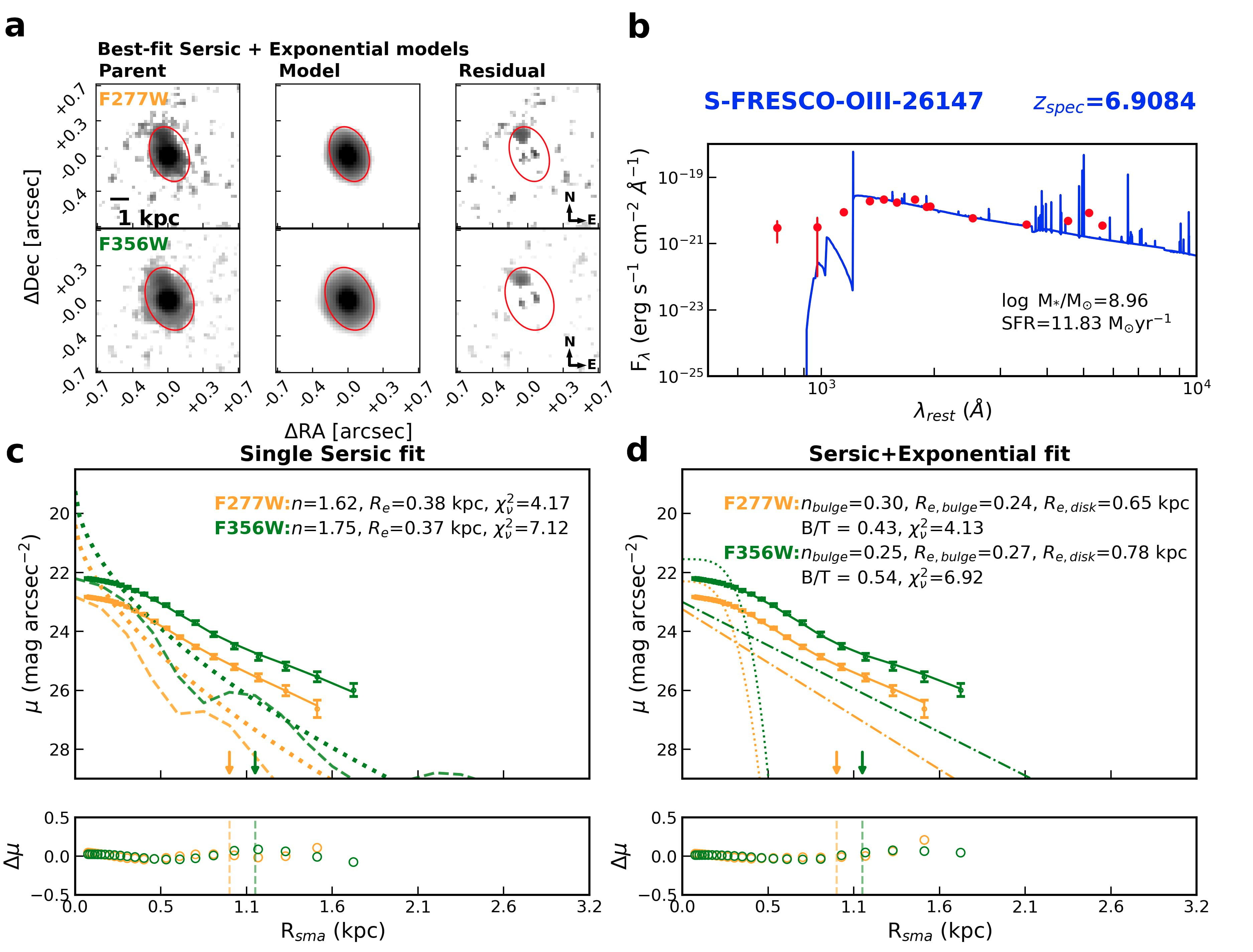}
\caption{Morphological decomposition of S-FRESCO-OIII-26147. See Figure~\ref{fig:representative} for a description of the panels.}
\label{fig:app_S-FRESCO-OIII-26147}
\end{figure*}

\begin{figure*}
\figurenum{A1.12}
\centering
\includegraphics[width=0.7\textwidth]{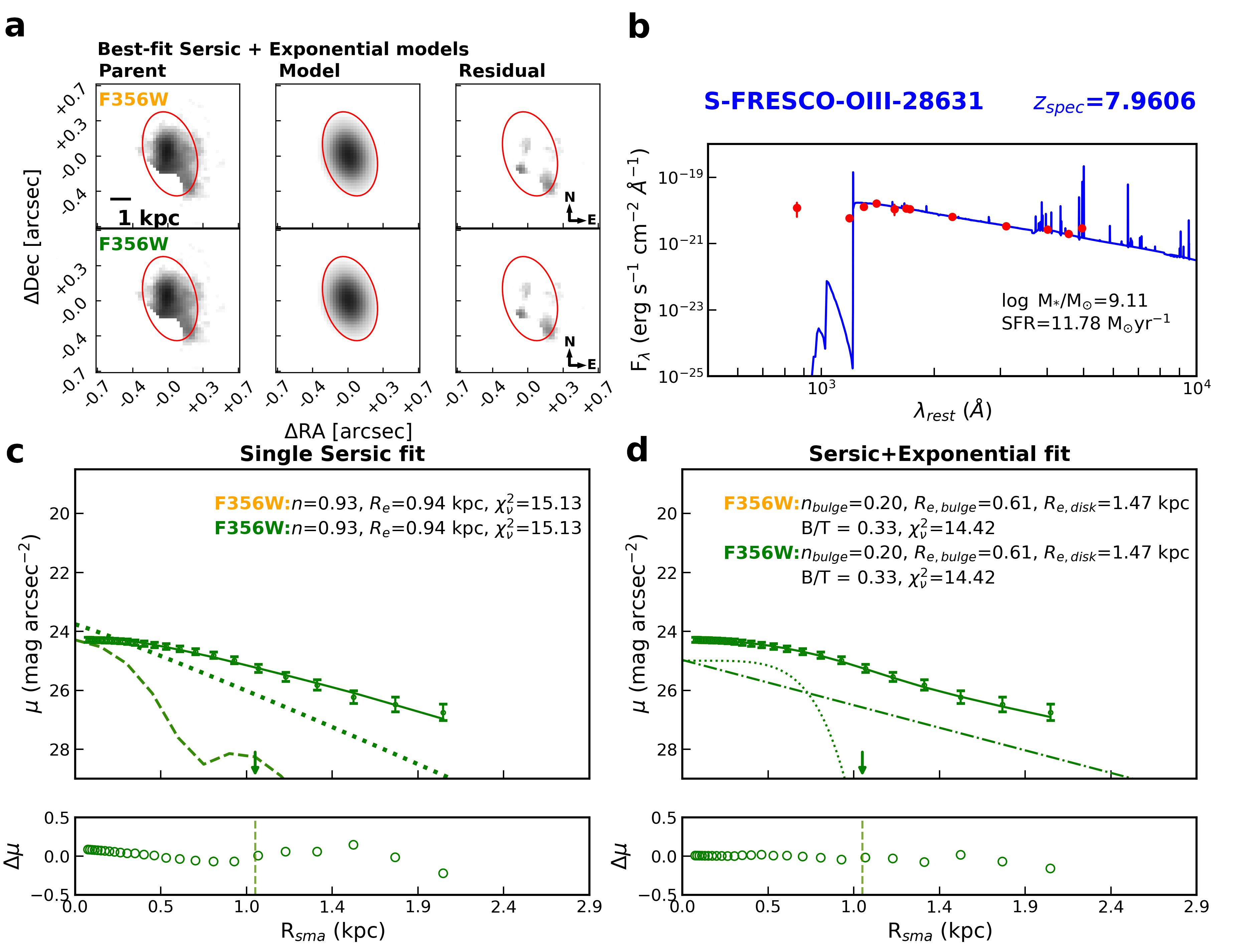}
\caption{Morphological decomposition of S-FRESCO-OIII-28631. See Figure~\ref{fig:representative} for a description of the panels.}
\label{fig:app_S-FRESCO-OIII-28631}
\end{figure*}

\begin{figure*}
\figurenum{A1.13}
\centering
\includegraphics[width=0.7\textwidth]{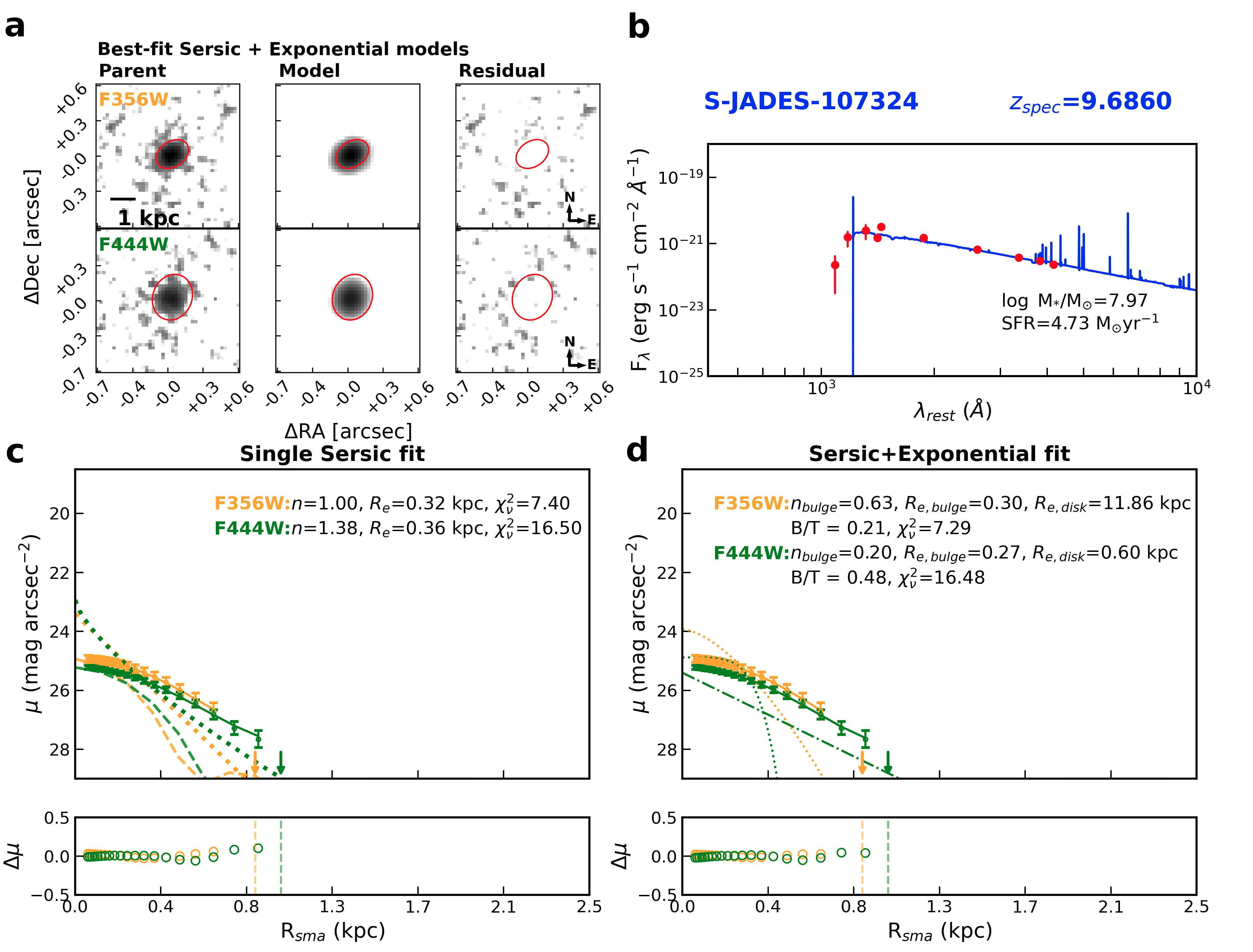}
\caption{Morphological decomposition of S-JADES-107324. See Figure~\ref{fig:representative} for a description of the panels.}
\label{fig:app_S-JADES-107324}
\end{figure*}

\begin{figure*}
\figurenum{A1.14}
\centering
\includegraphics[width=0.7\textwidth]{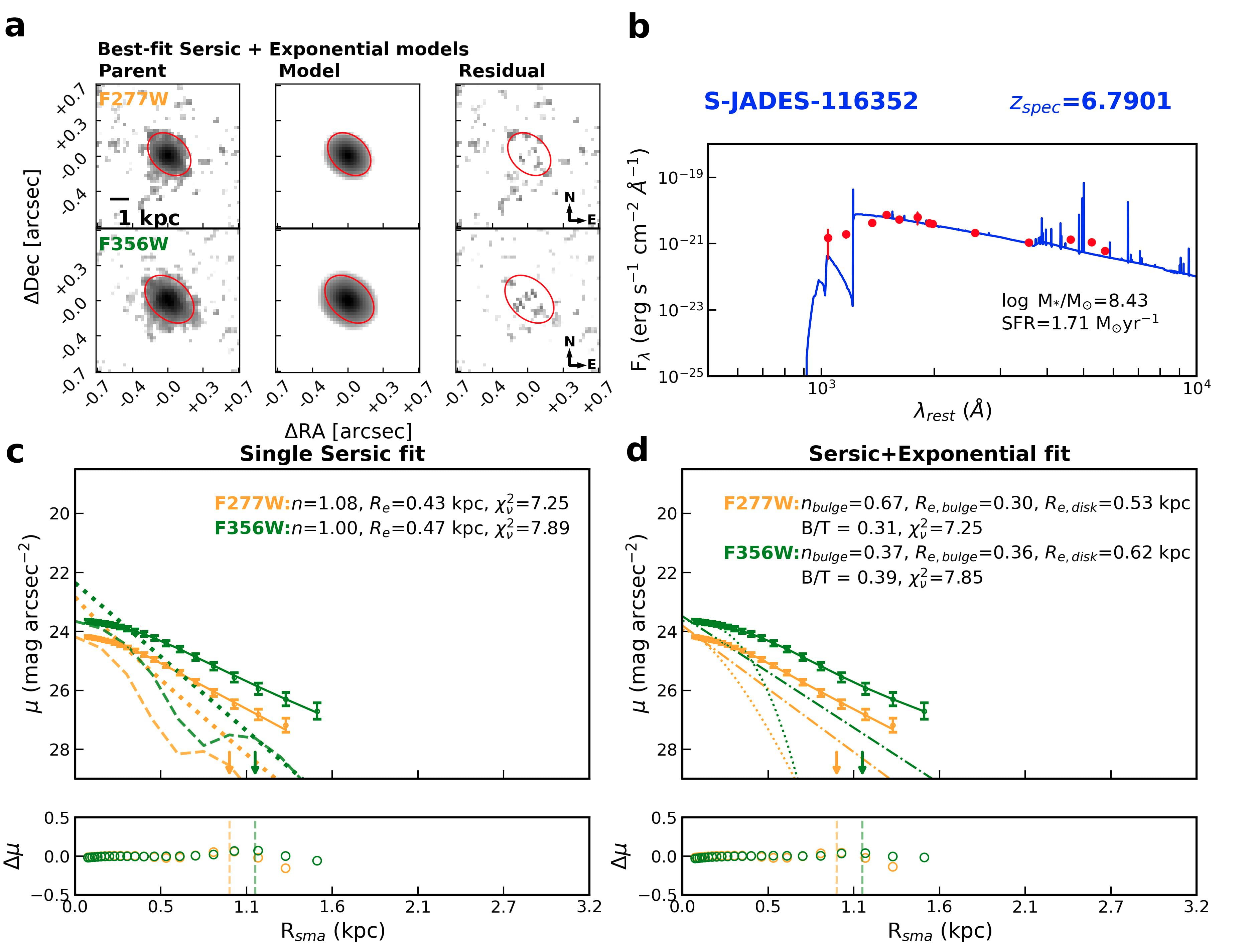}
\caption{Morphological decomposition of S-JADES-116352. See Figure~\ref{fig:representative} for a description of the panels.}
\label{fig:app_S-JADES-116352}
\end{figure*}

\begin{figure*}
\figurenum{A1.15}
\centering
\includegraphics[width=0.7\textwidth]{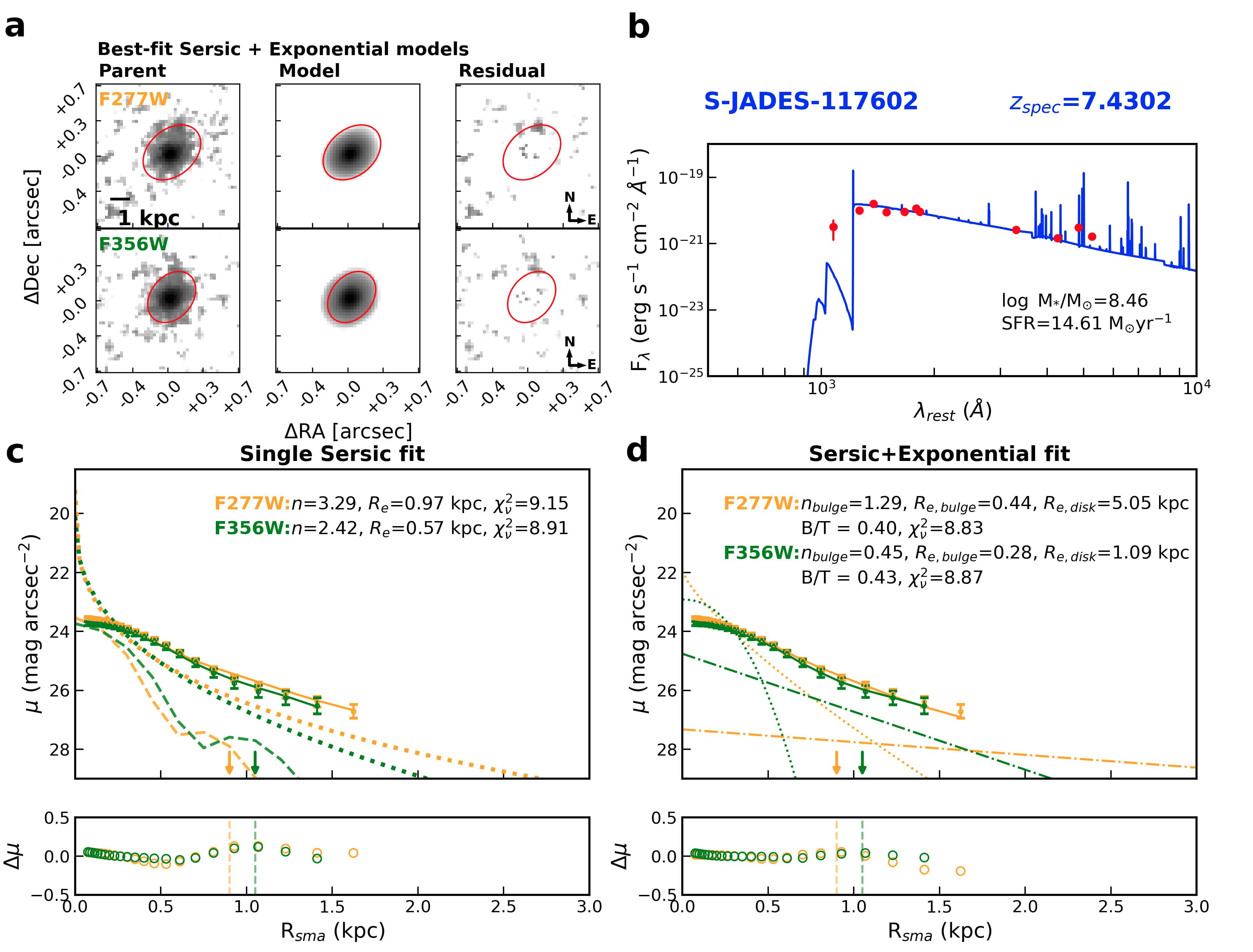}
\caption{Morphological decomposition of S-JADES-117602. See Figure~\ref{fig:representative} for a description of the panels.}
\label{fig:app_S-JADES-117602}
\end{figure*}

\begin{figure*}
\figurenum{A1.16}
\centering
\includegraphics[width=0.7\textwidth]{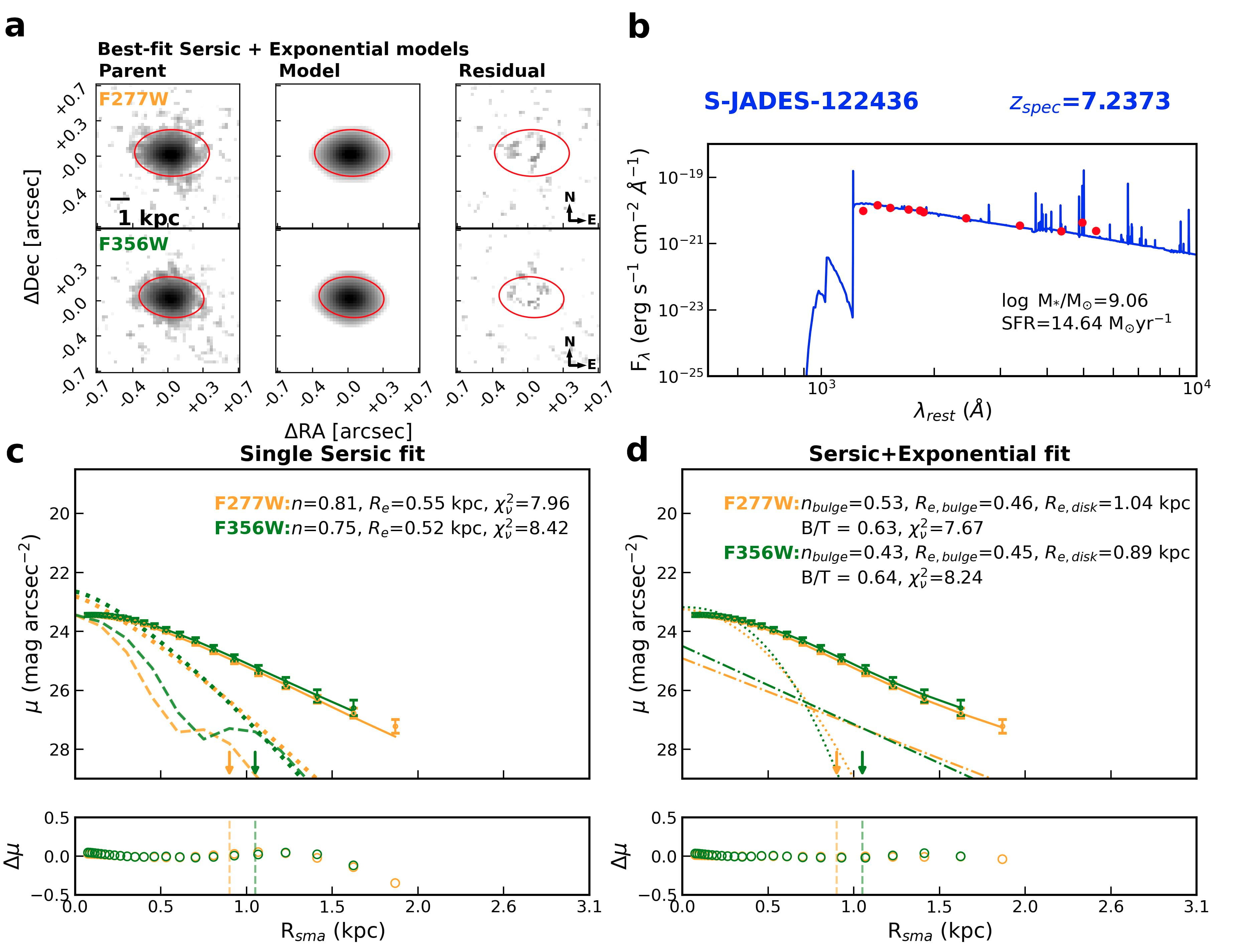}
\caption{Morphological decomposition of S-JADES-122436. See Figure~\ref{fig:representative} for a description of the panels.}
\label{fig:app_S-JADES-122436}
\end{figure*}

\begin{figure*}
\figurenum{A1.17}
\centering
\includegraphics[width=0.7\textwidth]{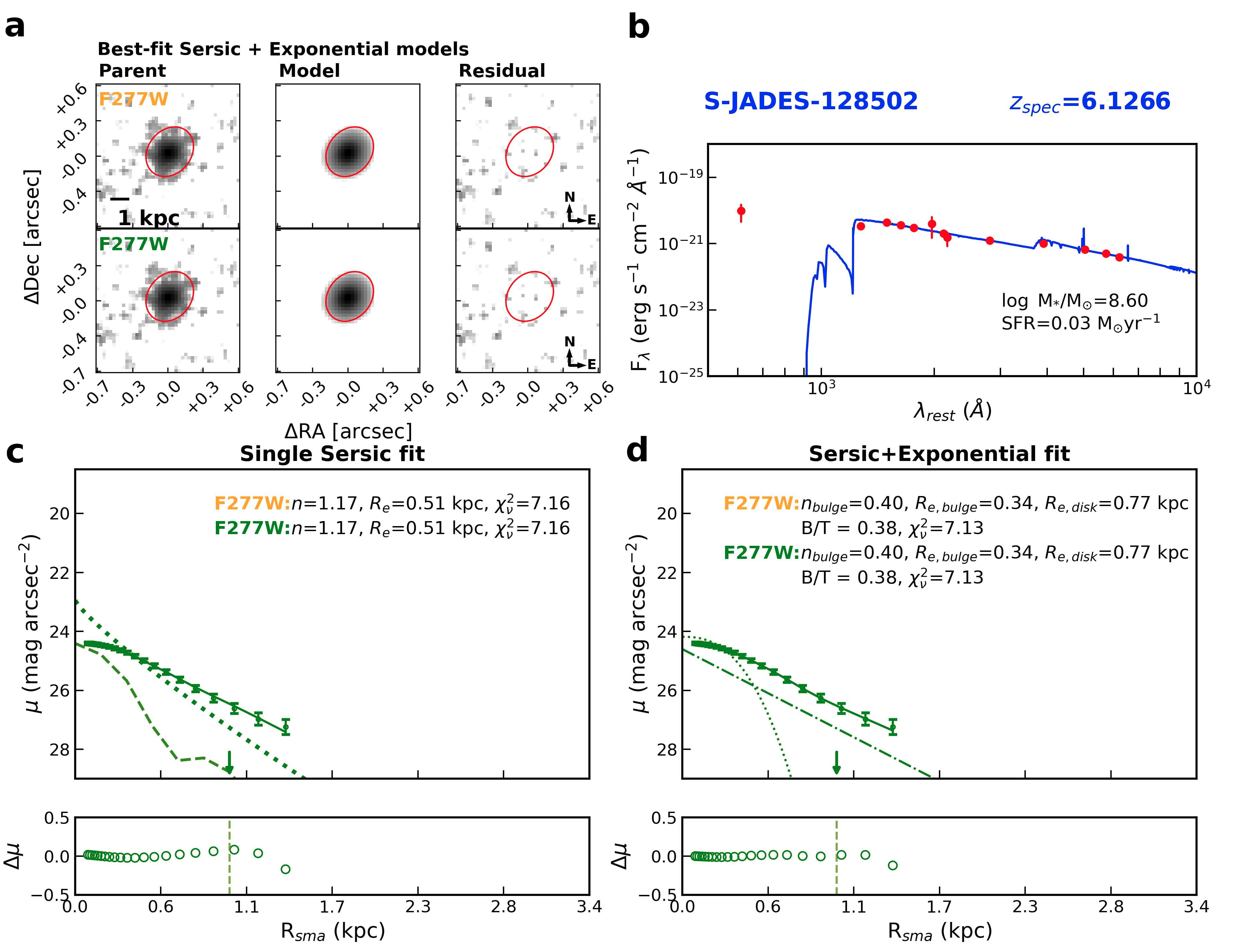}
\caption{Morphological decomposition of S-JADES-128502. See Figure~\ref{fig:representative} for a description of the panels.}
\label{fig:app_S-JADES-128502}
\end{figure*}

\begin{figure*}
\figurenum{A1.18}
\centering
\includegraphics[width=0.7\textwidth]{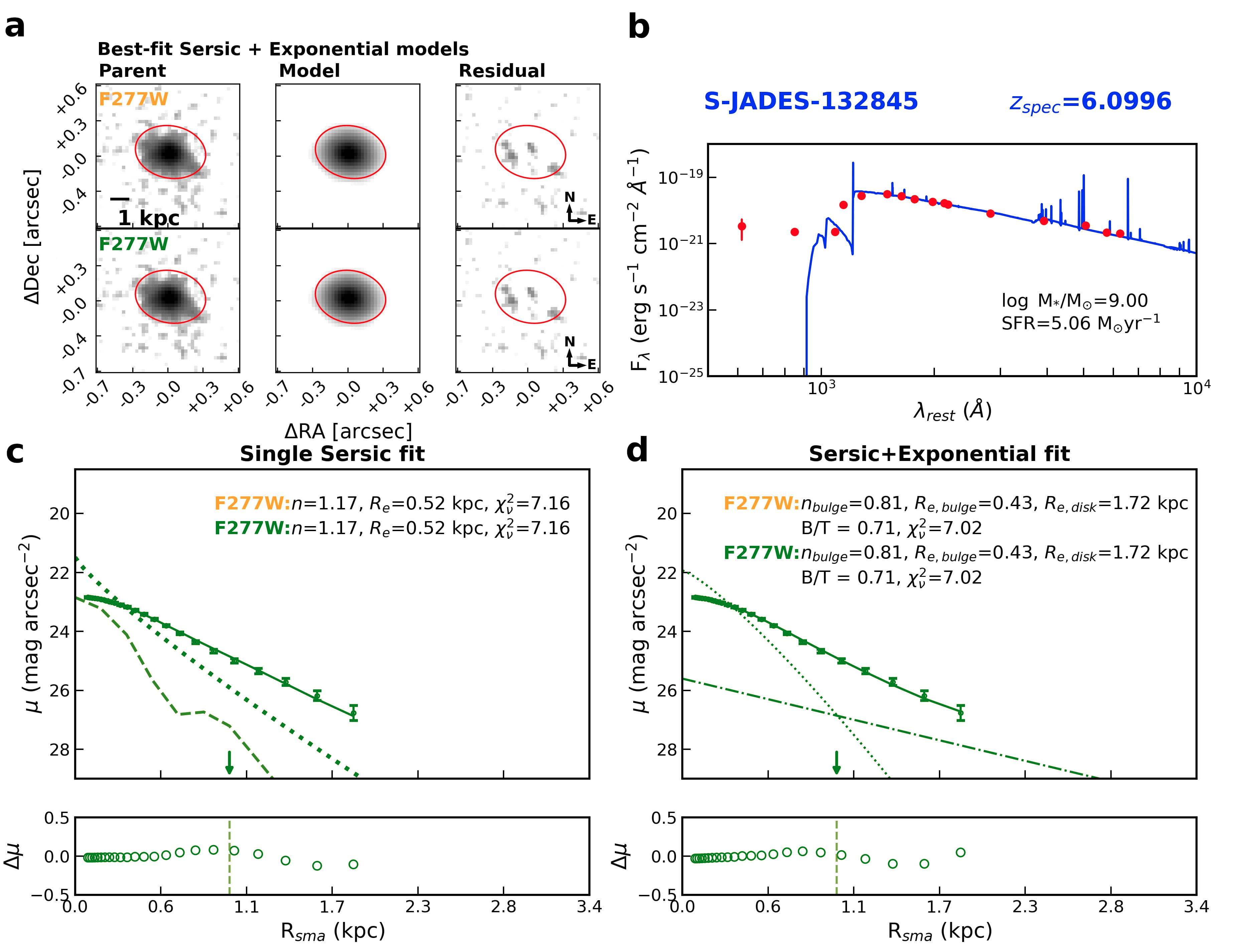}
\caption{Morphological decomposition of S-JADES-132845. See Figure~\ref{fig:representative} for a description of the panels.}
\label{fig:app_S-JADES-132845}
\end{figure*}

\begin{figure*}
\figurenum{A1.19}
\centering
\includegraphics[width=0.7\textwidth]{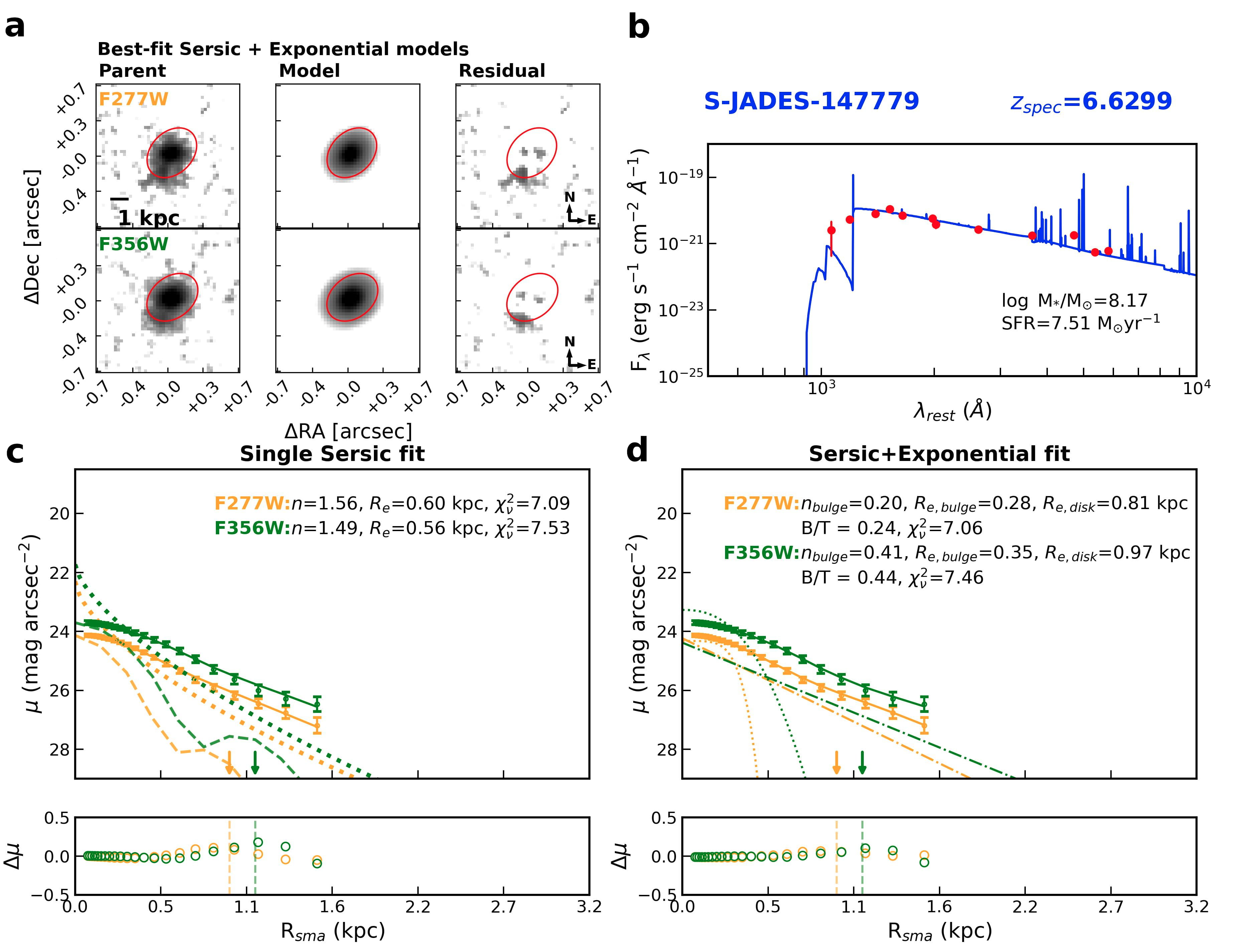}
\caption{Morphological decomposition of S-JADES-147779. See Figure~\ref{fig:representative} for a description of the panels.}
\label{fig:app_S-JADES-147779}
\end{figure*}

\begin{figure*}
\figurenum{A1.20}
\centering
\includegraphics[width=0.7\textwidth]{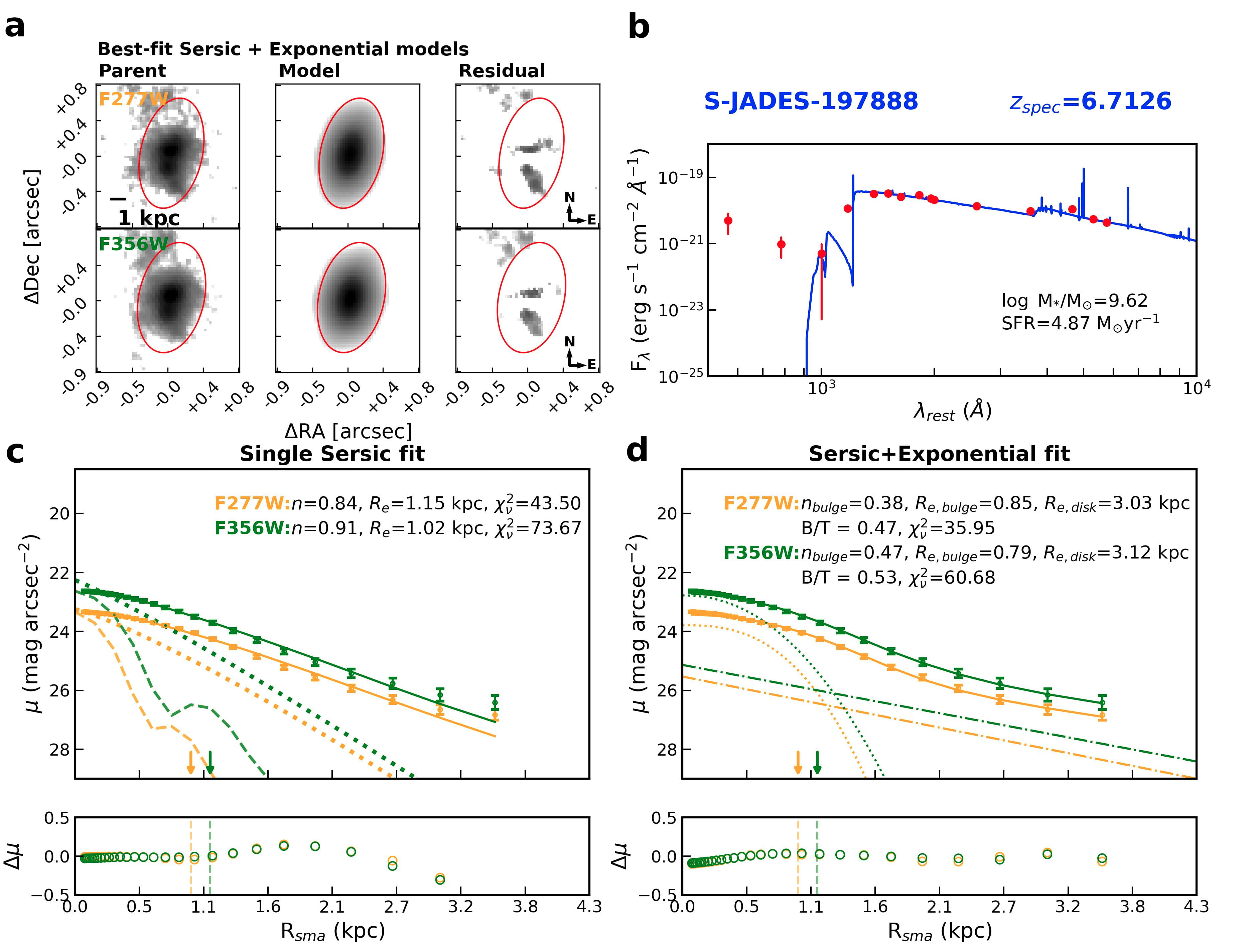}
\caption{Morphological decomposition of S-JADES-197888. See Figure~\ref{fig:representative} for a description of the panels.}
\label{fig:app_S-JADES-197888}
\end{figure*}

\clearpage
\section{Recovery of structural parameters from mock galaxies}

\begin{figure*}[htpb]
\figurenum{A3.1}
\centering
\includegraphics[width=0.7\linewidth]{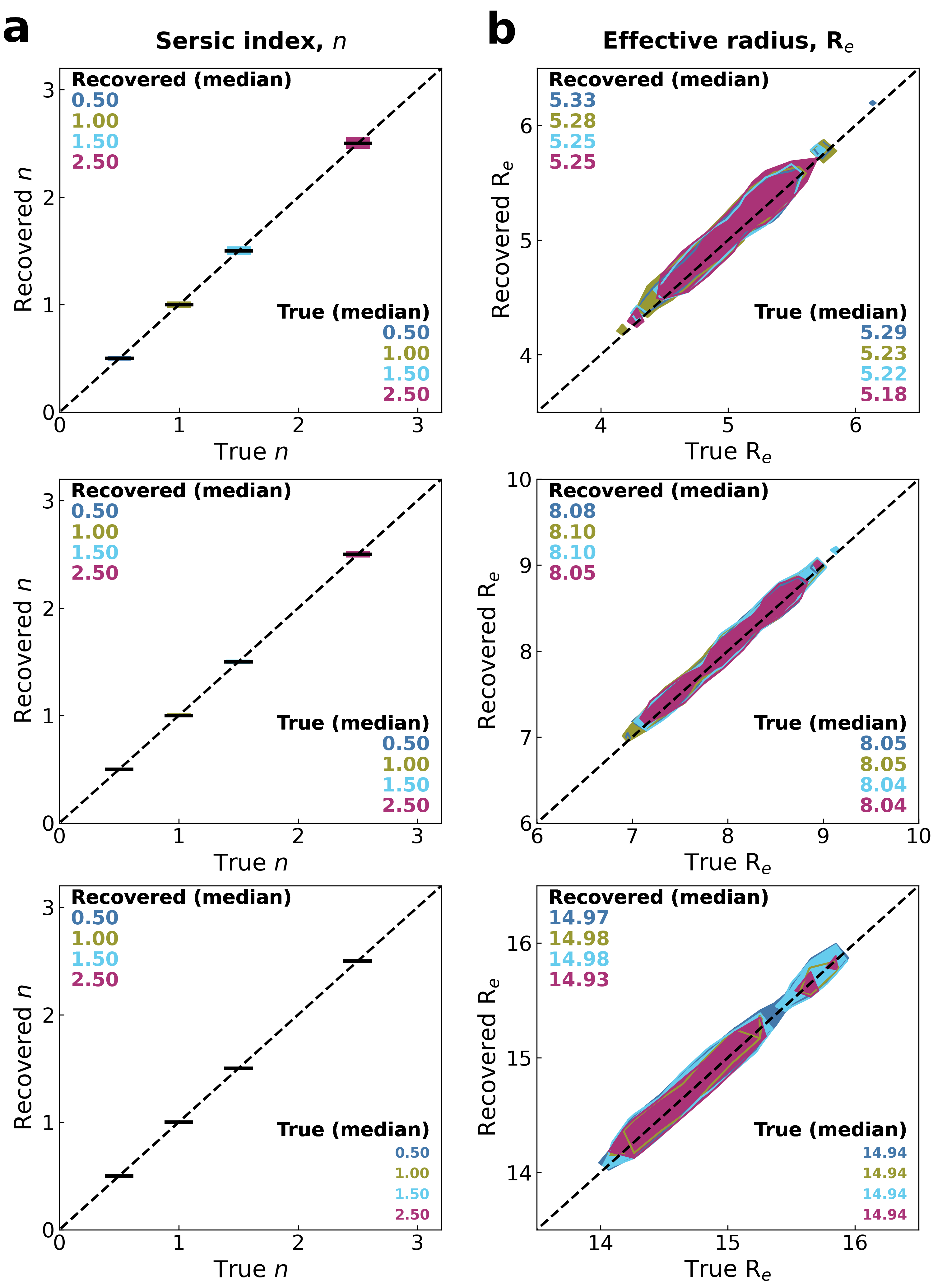}
\caption{\textbf{Comparison between the input and recovered parameters obtained from the \textsc{GALFIT} decomposition of simulated single-Sersic galaxies.} Panels in columns a and b show the recovery of the Sersic index ($n$) and effective radius (R$_e$), respectively. From top to bottom, the rows correspond to progressively larger input galaxy sizes, with mean R$_e$ = 5, 8, and 15 pixels. Panels in column \textbf{a} show the median recovered $n$ value, with the coloured bars spanning the 16$^{\rm th}$--84$^{\rm th}$ percentile range of the recovered distribution. Panels in column \textbf{b} show the coloured contours that represent the respective 1$\sigma$ levels corresponding to a 2D Gaussian distribution for four input Sersic indices: $n=$ 0.5 (blue), 1.0 (olive), 1.5 (cyan), and 2.5 (purple). The dashed line indicates the one-to-one relation. In each panel, the numbers denote the median values of the true and recovered parameters, color-coded according to the value of $n$. Note that, only mock galaxies with sizes larger than the radius enclosing 80\% of the PSF flux are shown. Panel for mocks with R$_e$ = 2 pixels are not shown due to negligible objects being larger than the 80\% PSF flux radius.}
\label{fig:recovery_3_1}
\end{figure*}

\begin{figure*}[htpb]
\figurenum{A3.2}
\centering
\includegraphics[width=\textwidth]{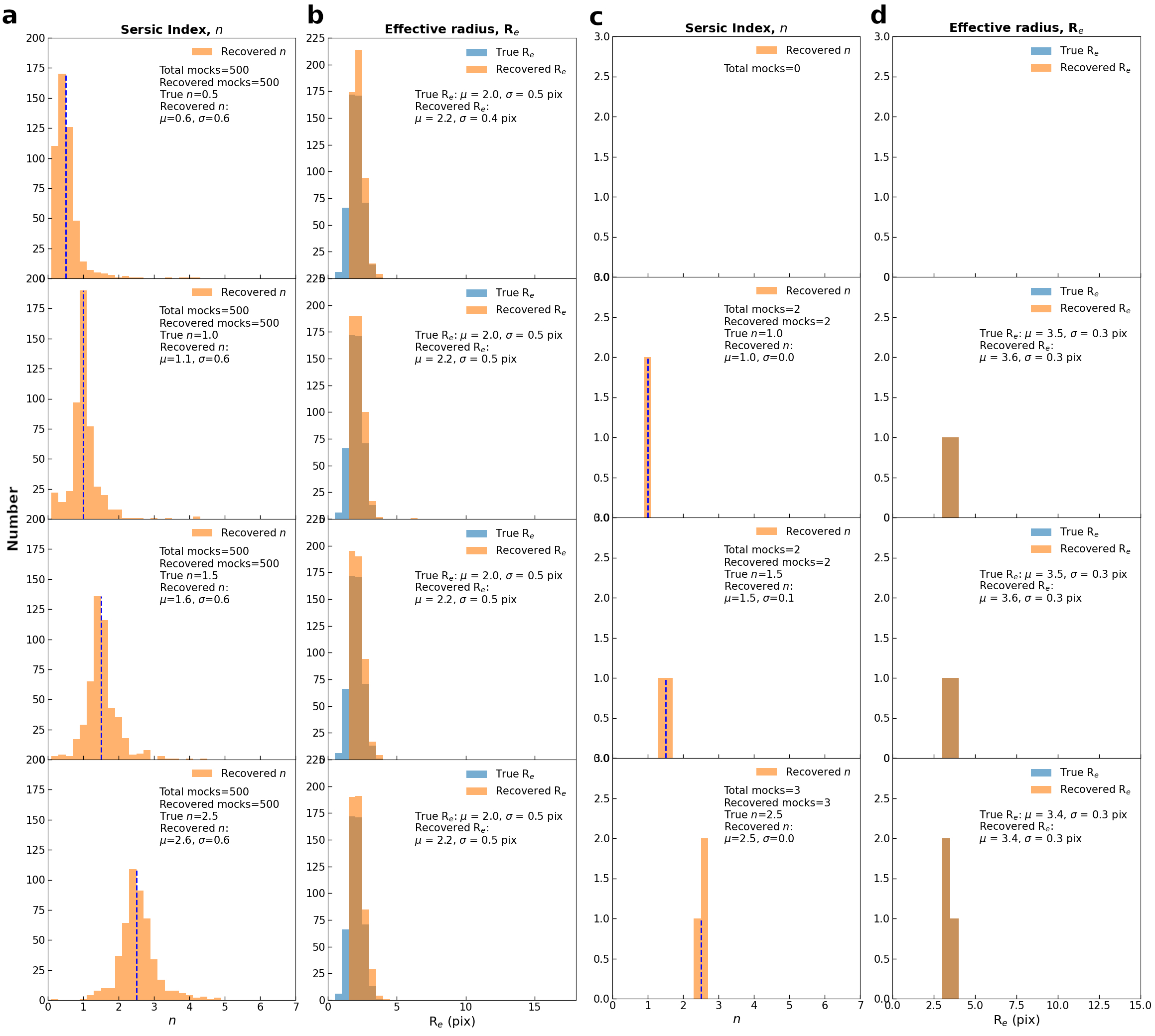}
\caption{\textbf{Comparison of input and recovered parameters from \textsc{GALFIT} modeling of one-component mock galaxies with true R$_{e}$ = 2 pixels.} Each row corresponds to a specific input \textit{n} value of 0.5, 1, 1.5 and 2.5. Columns a and b represent all mocks without taking into consideration the extent of the surface brightness profiles. Columns c and d represent mocks whose sizes (based on threshold of $\mu_{err}<0.3$ mag arcsec$^{-2}$) extended beyond the size that encloses 80\% of the PSF flux. For each panel in column a, the vertical line marks the true $n$ value.}
\label{fig:recovery_3_2}
\end{figure*}

\begin{figure*}
\figurenum{A3.3}
\centering
\includegraphics[width=\textwidth]{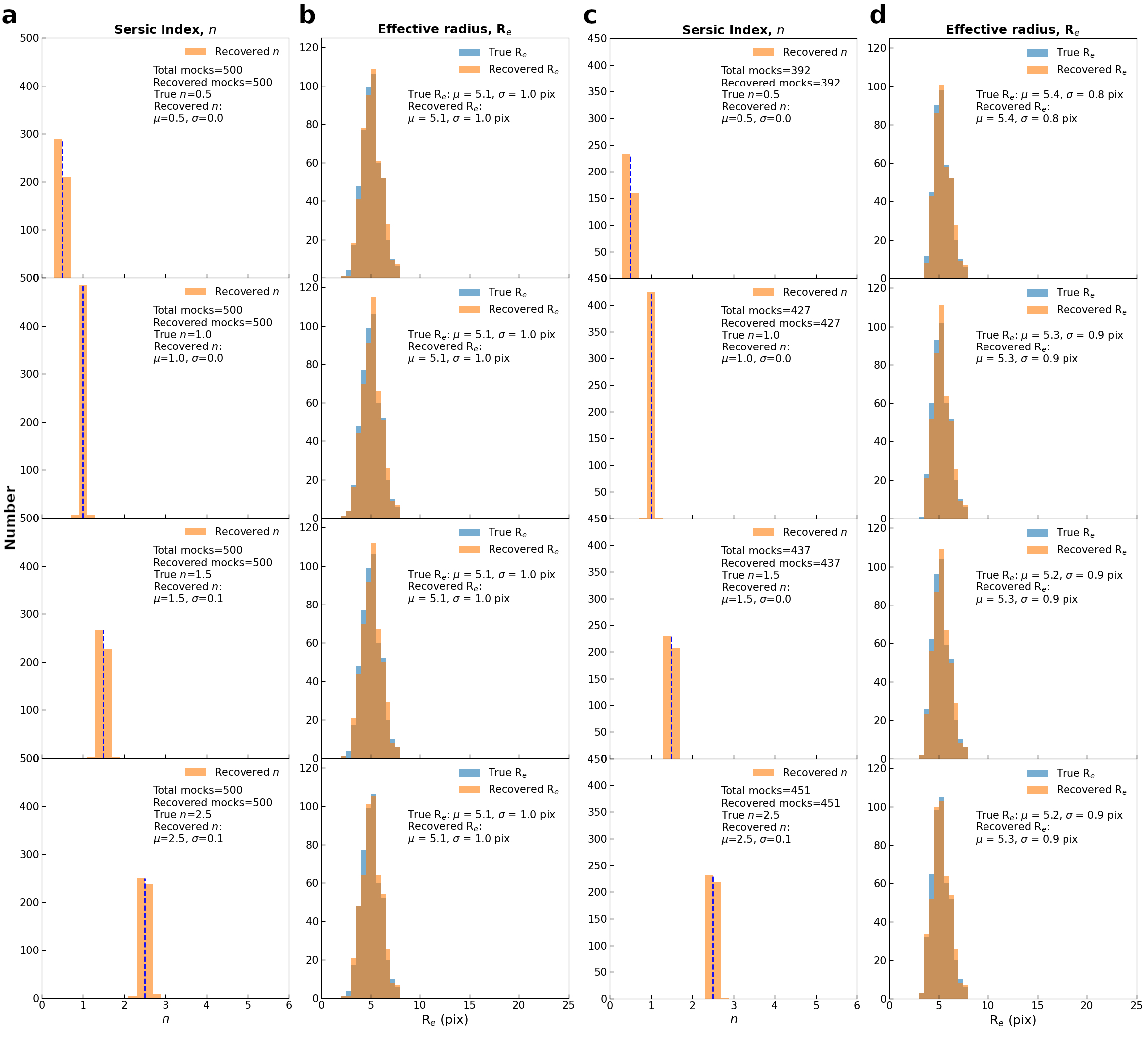}
\caption{Same as Figure A3.2, but for mocks with true R$_{e}$ = 5 pixels.}
\label{fig:recovery_3_2}
\end{figure*}

\begin{figure*}
\figurenum{A3.4}
\centering
\includegraphics[width=\textwidth]{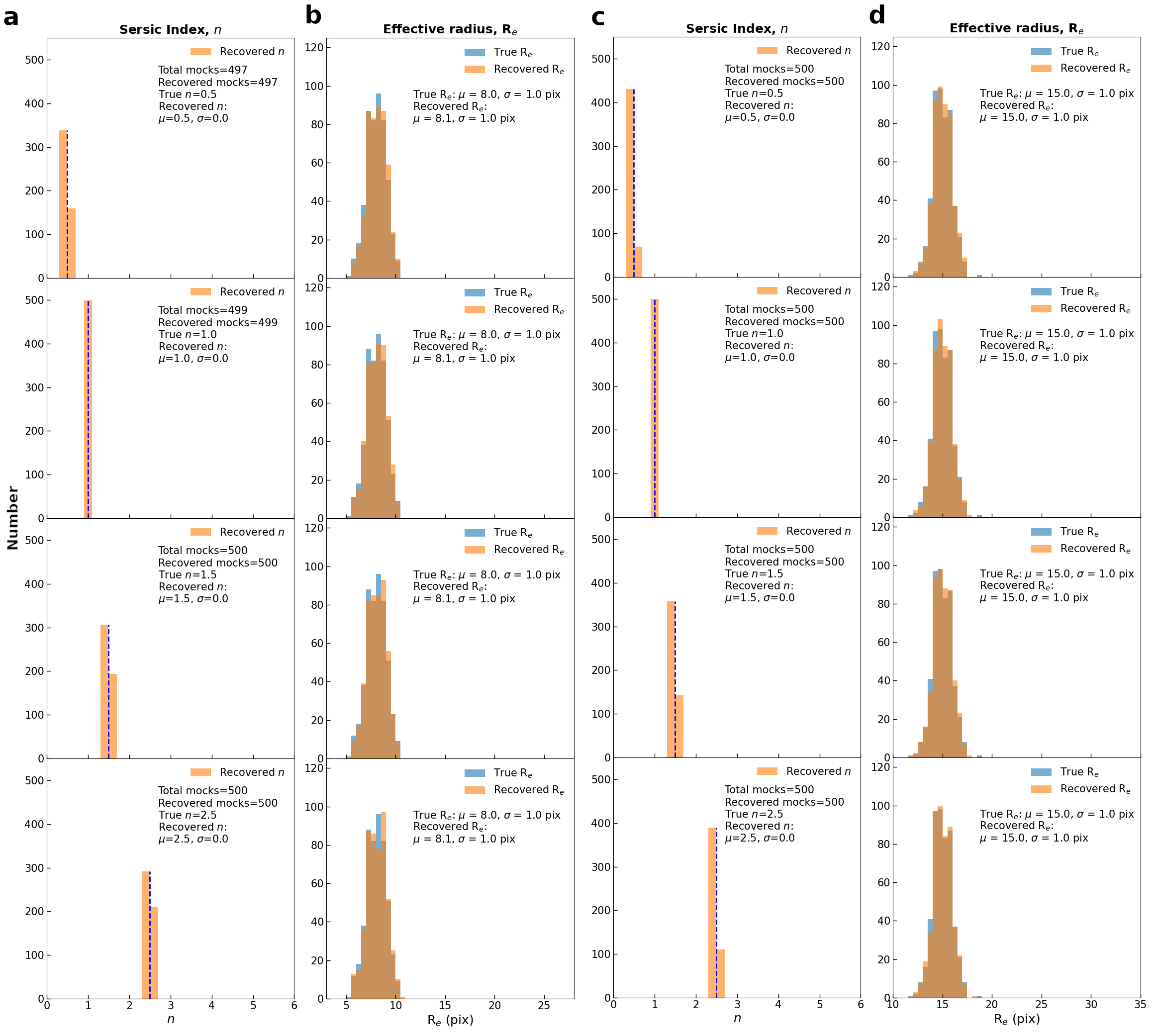}
\caption{\textbf{Comparison of input and recovered parameters from \textsc{GALFIT} modeling of one-component mock galaxies with true R$_{e}$ = 8 pixels (columns a and b) and 15 pixels (columns c and d).} As earlier, each row corresponds to a specific input \textit{n} value of 0.5, 1, 1.5 and 2.5. Here, all mocks with sizes extending beyond the 80\% flux enclosing area of the PSF are shown.}
\label{fig:recovery_3_4}
\end{figure*}

\begin{figure*}
\figurenum{A3.5}
\centering
\includegraphics[width=\textwidth]{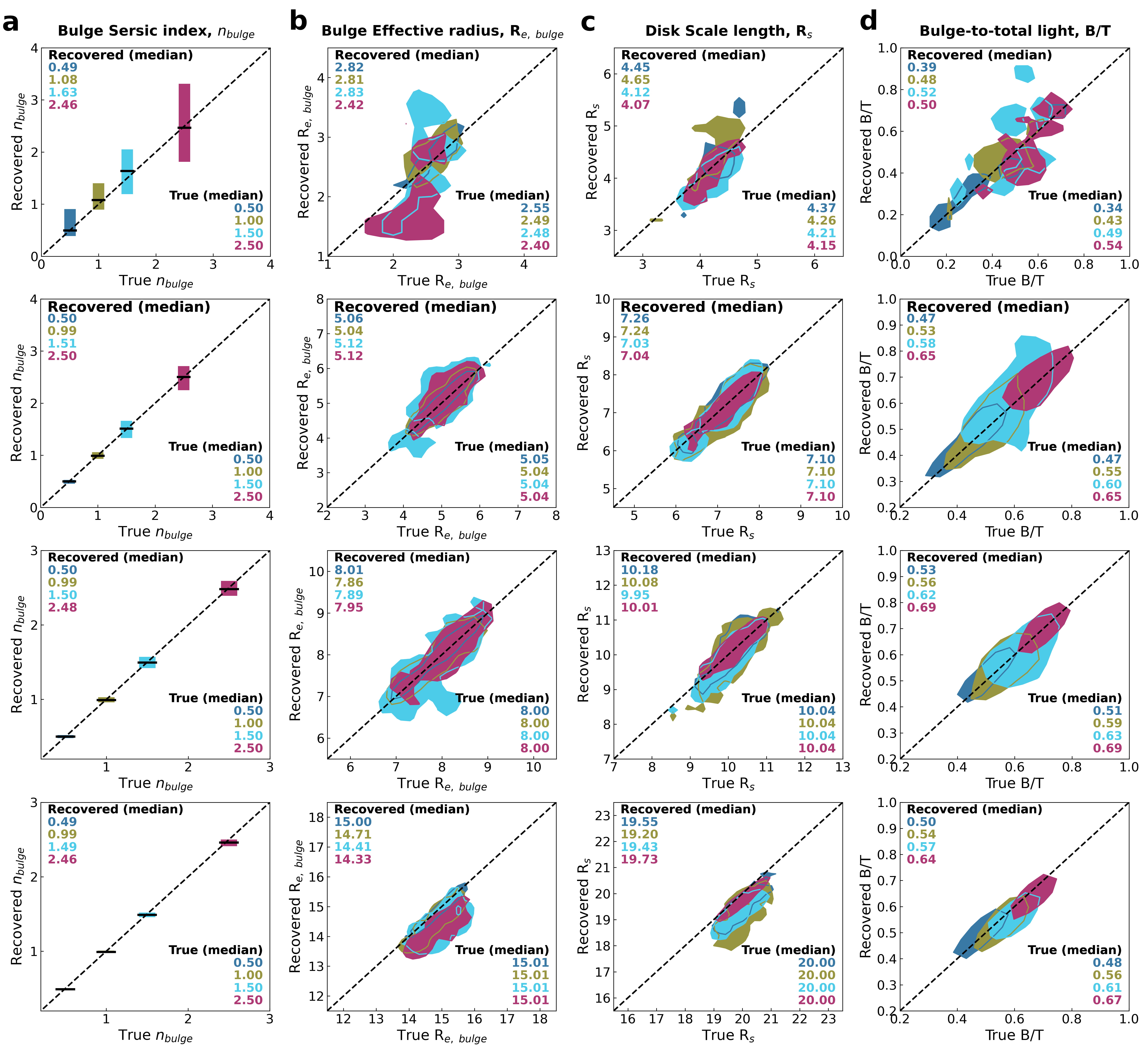}
\caption{\textbf{Comparison between the input and recovered parameters obtained from the \textsc{GALFIT} decomposition of simulated Sersic+exponential disk galaxies (Section~\ref{sec:recovery-tests})}. Panels in columns a to d show the recovery of the bulge Sersic index ($n_{bulge}$), bulge effective radius (R$_{e,~bulge}$), disk scale length (R$_s$), and bulge-to-total light ratio (B/T), respectively. From top to bottom, the rows correspond to progressively larger input galaxy sizes, with mean R$_{e,~bulge}$ = 2, 5, 8, and 15 pixels and corresponding R$_s$ = 4, 7, 10, and 20 pixels, respectively. Panels in column \textbf{a} show the median recovered value of $n_{bulge}$, with the coloured bars spanning the 16$^{\rm th}$--84$^{\rm th}$ percentile range of the recovered distribution. For the remaining panels (columns b to d), the coloured contours represent the respective 1$\sigma$ levels corresponding to a 2D Gaussian distribution for four input bulge Sersic indices: $n_{bulge}=0.5$ (blue), 1.0 (olive), 1.5 (cyan), and 2.5 (purple). The dashed line denotes the one-to-one relation. In each panel, the numbers denote the median values of the true and recovered parameters, color-coded according to the input value of $n_{bulge}$. Note that, only mock galaxies with sizes larger than the radius enclosing 80\% of the PSF flux are shown.}
\label{fig:recovery_3_5}
\end{figure*}

\begin{figure*}
\figurenum{A3.6}
\centering
\includegraphics[width=\textwidth]{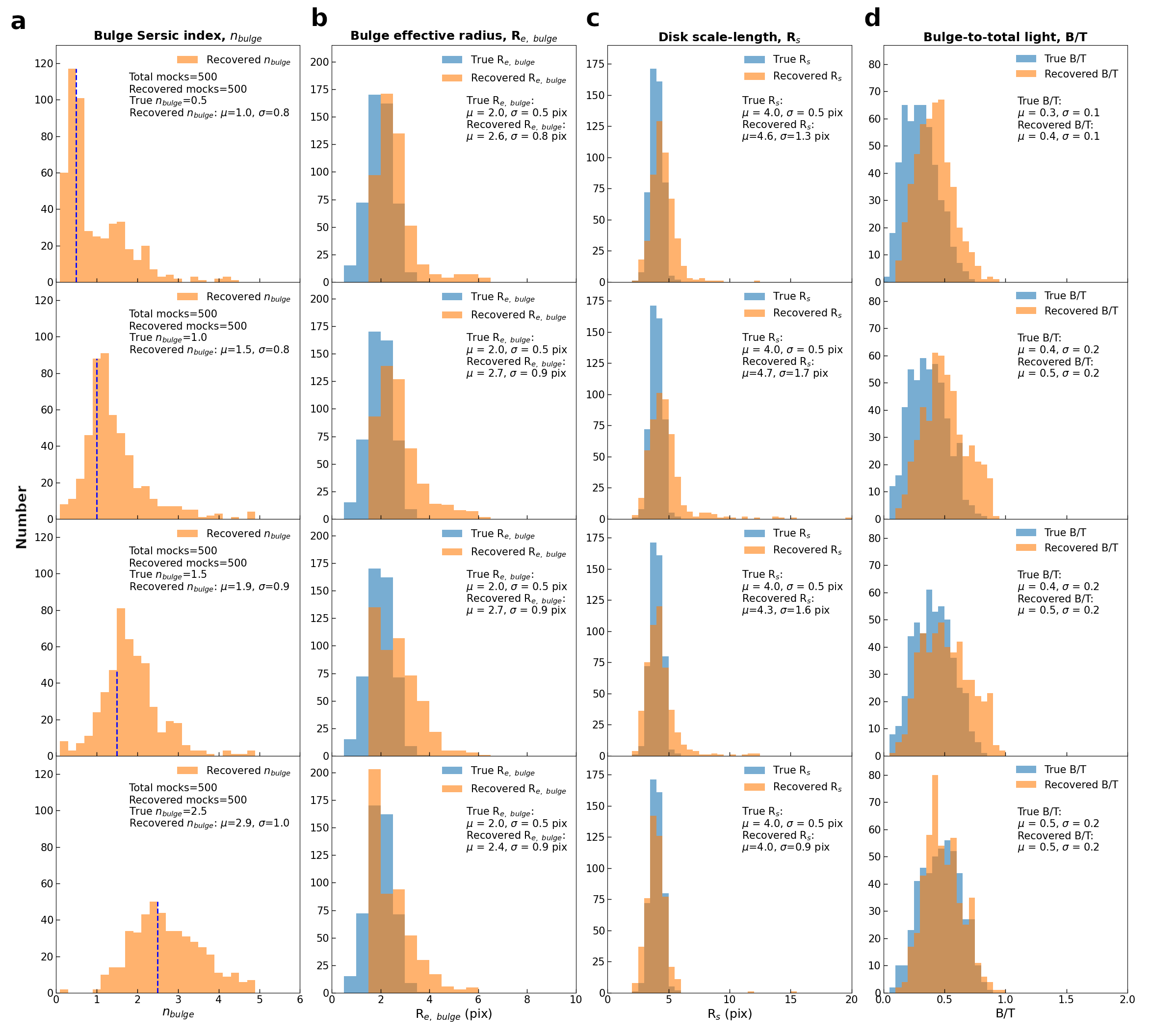}
\caption{Comparison of input and recovered parameters from \textsc{GALFIT} modeling of all \textbf{two-component} mock galaxies with true R$_{e,~bulge}$ = 2 pixels, true R$_{s}$ = 4 pixels. As earlier, each row corresponds to a specific input $n_{bulge}$ value of 0.5, 1, 1.5 and 2.5.}
\label{fig:recovery_3_6}
\end{figure*}

\begin{figure*}
\figurenum{A3.7}
\centering
\includegraphics[width=\textwidth]{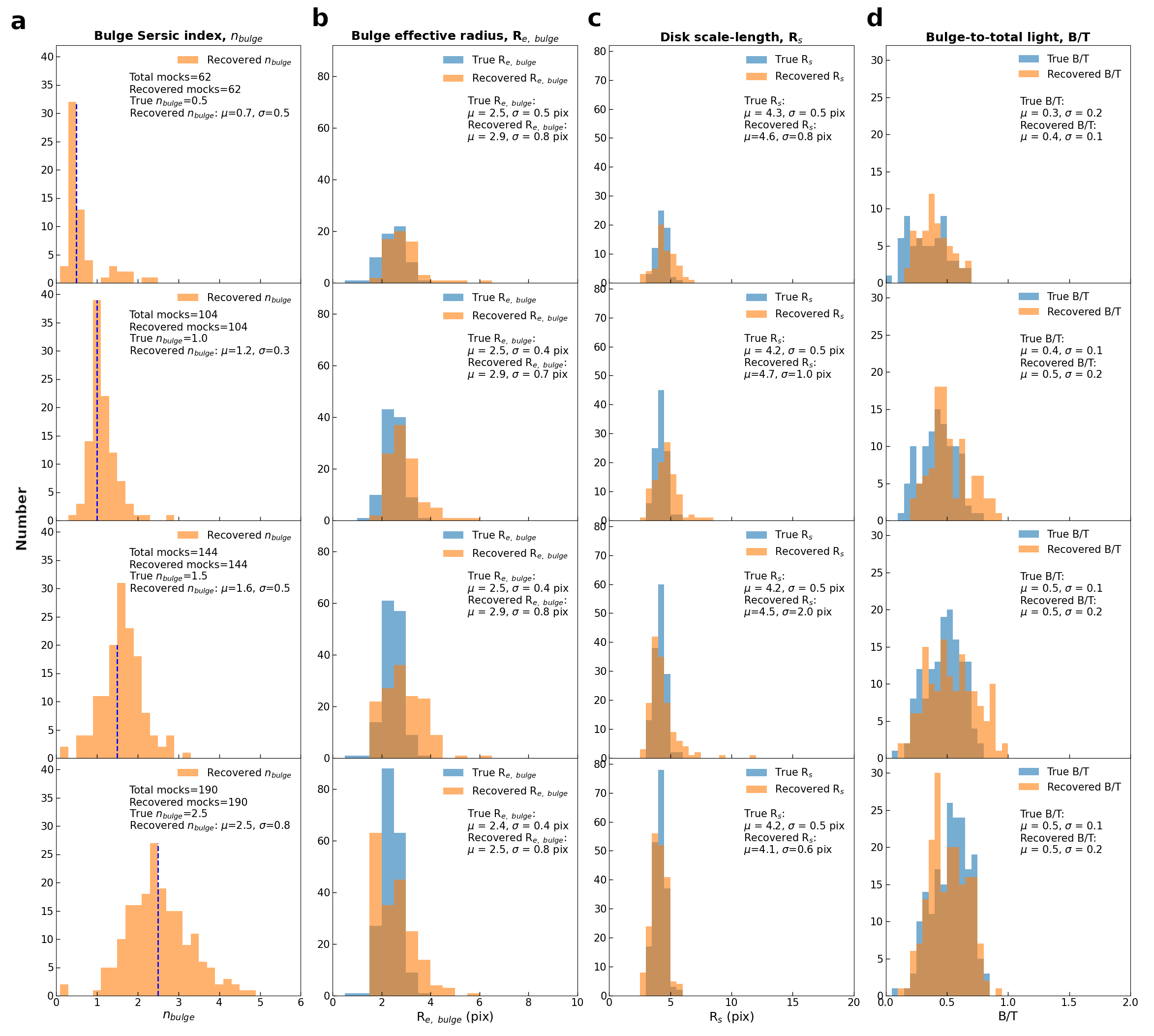}
\caption{Same as Figure A3.6 but using mocks which have sizes larger than the size that encloses 80\% of the PSF flux.}
\label{fig:recovery_3_7}
\end{figure*}

\begin{figure*}
\figurenum{A3.8}
\centering
\includegraphics[width=\textwidth]{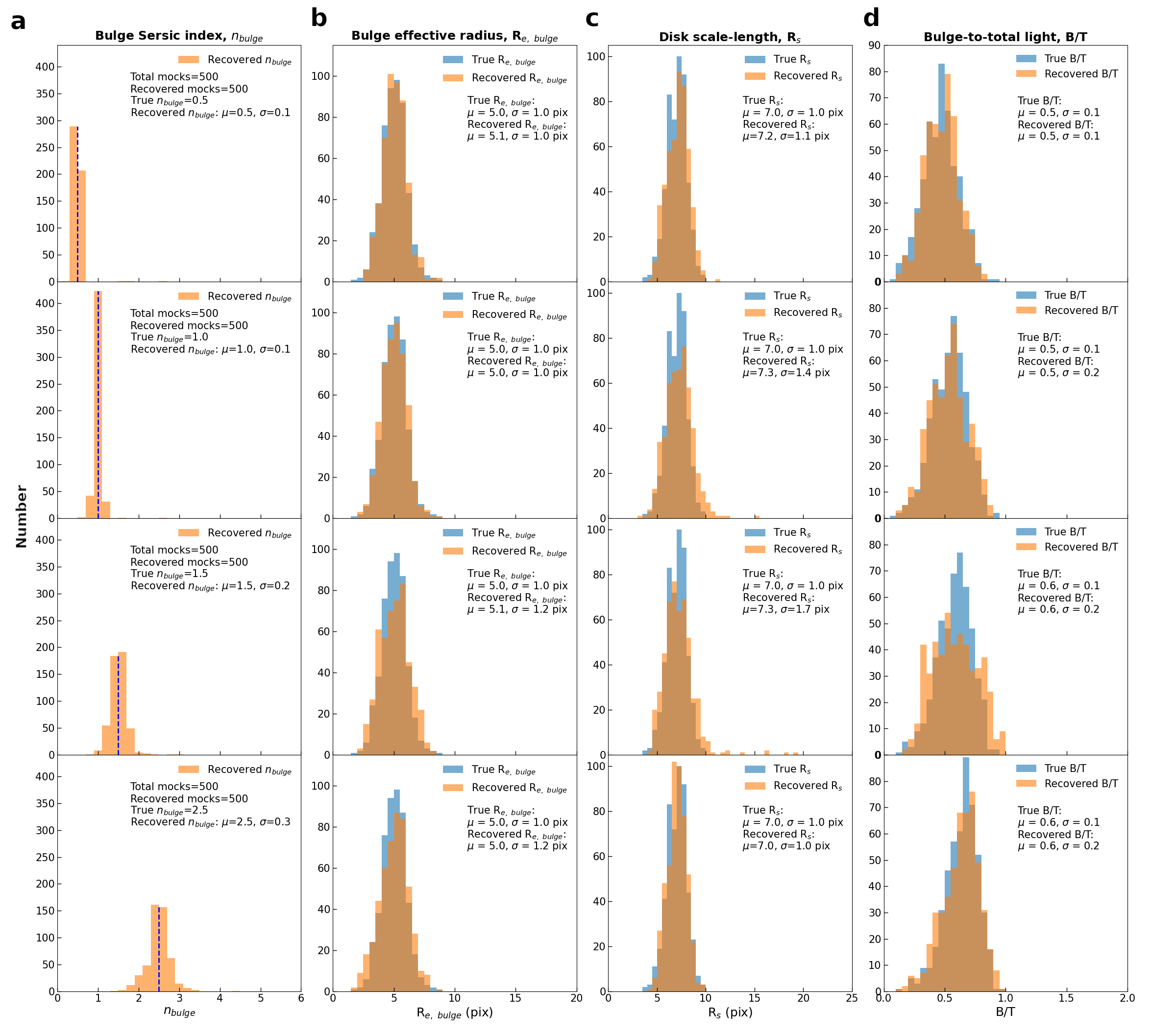}
\caption{Same as Figure A3.6 but true R$_{e,~bulge}$ = 5 pixels and true R$_{s}$ = 7 pixels.}
\label{fig:recovery_3_8}
\end{figure*}

\begin{figure*}
\figurenum{A3.9}
\centering
\includegraphics[width=\textwidth]{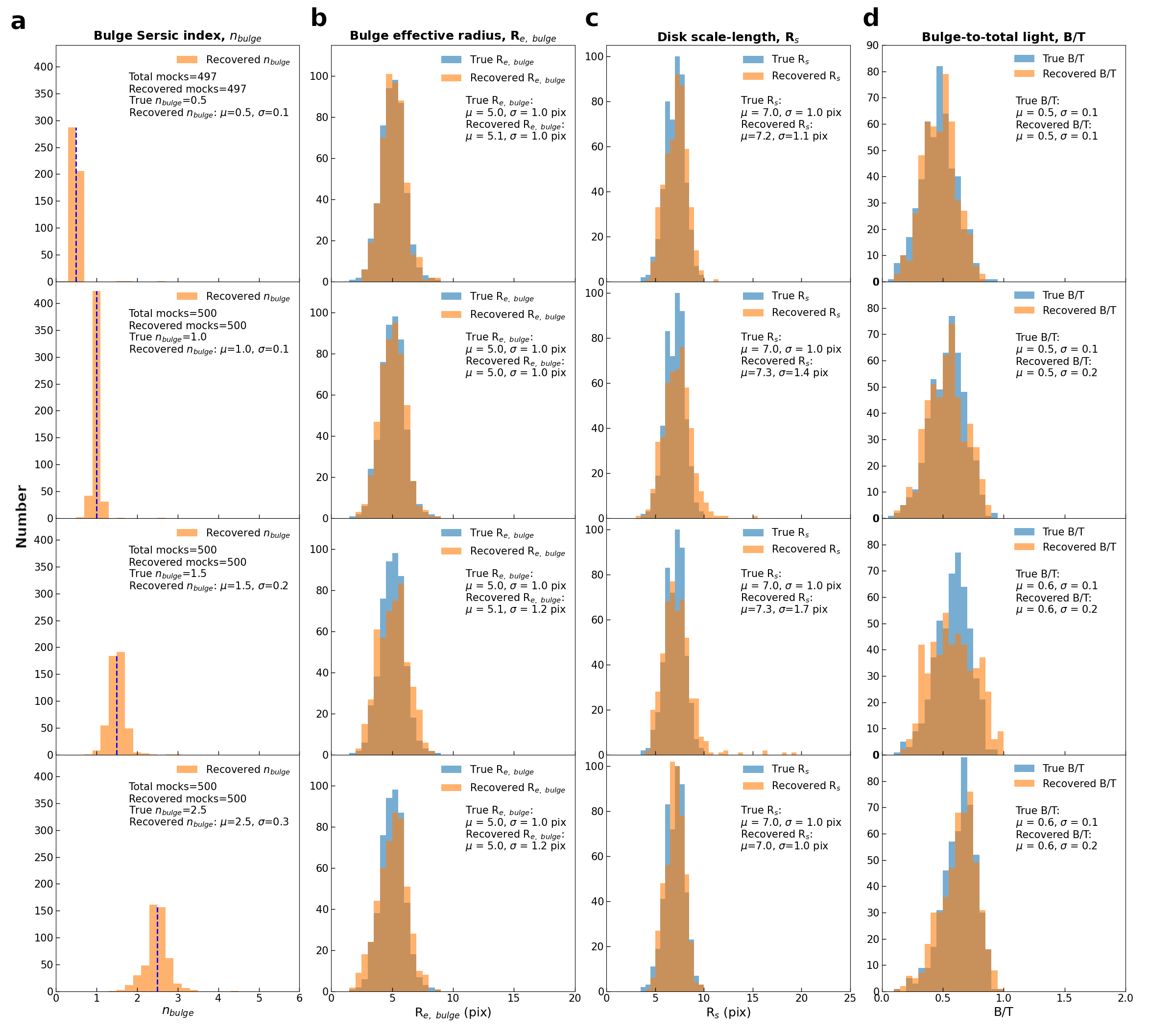}
\caption{Same as Figure A3.8 but using mocks which have sizes larger than the size that encloses 80\% of the PSF flux.}
\label{fig:recovery_3_9}
\end{figure*}

\begin{figure*}
\figurenum{A3.10}
\centering
\includegraphics[width=\textwidth]{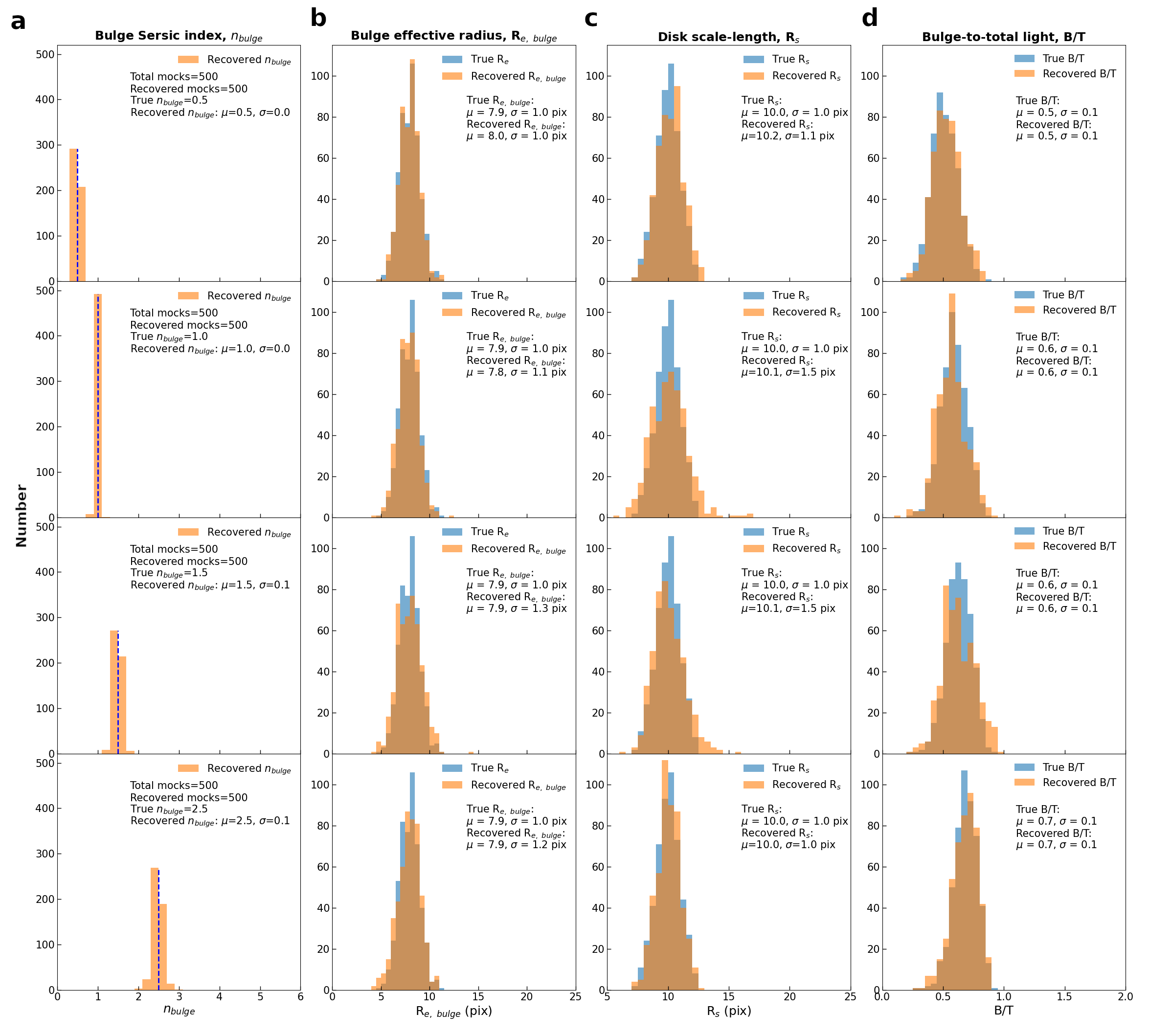}
\caption{Same as Figure A3.6 but true R$_{e,~bulge}$ = 8 pixels and true R$_{s}$ = 10 pixels. Here, all galaxies have sizes sufficiently larger than the PSF.}
\label{fig:recovery_3_10}
\end{figure*}

\begin{figure*}
\figurenum{A3.11}
\centering
\includegraphics[width=\textwidth]{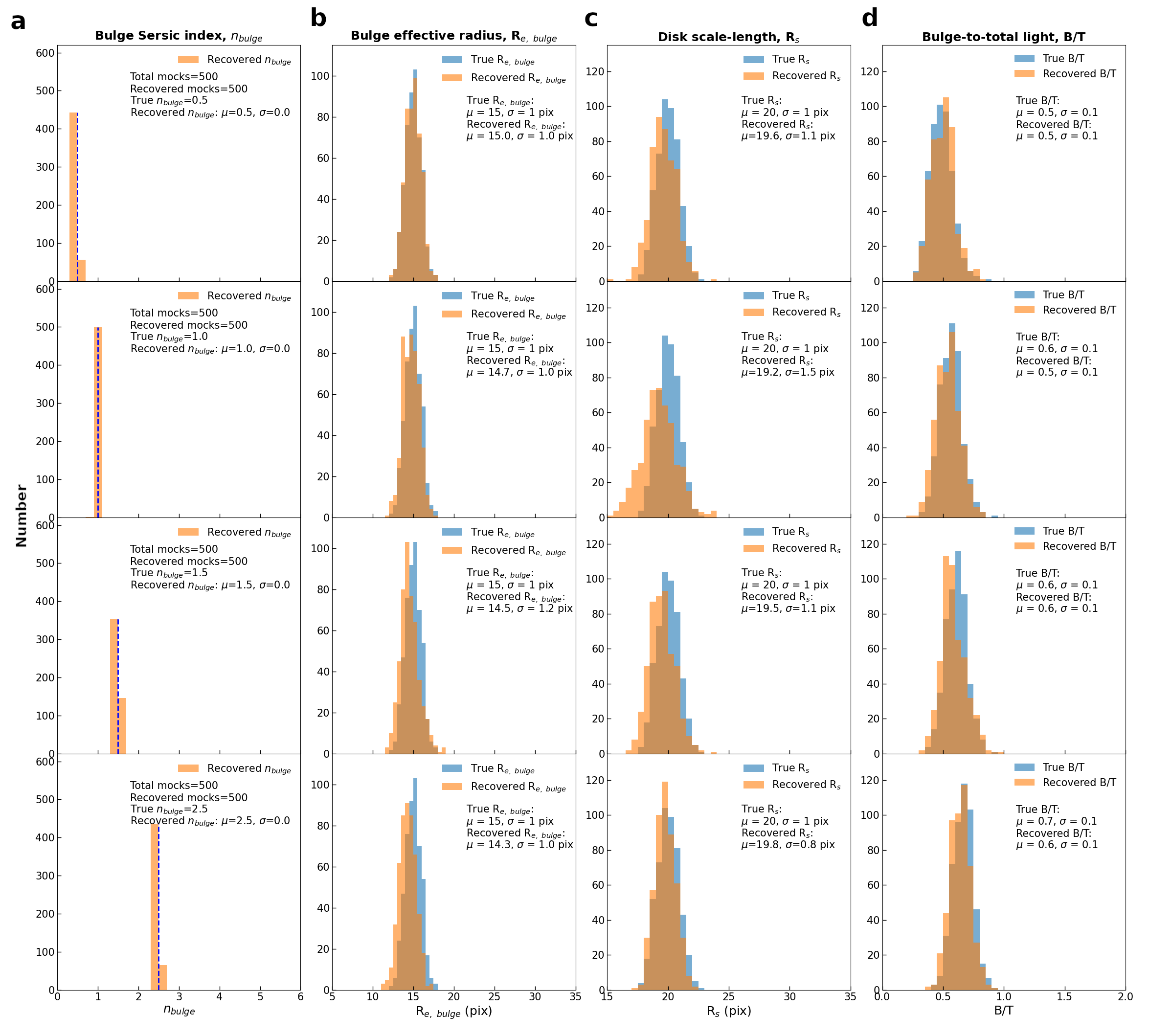}
\caption{Same as Figure A3.6 but true R$_{e,~bulge}$ = 15 pixels and true R$_{s}$ = 20 pixels. Here, all galaxies have sizes sufficiently larger than the PSF.}
\label{fig:recovery_3_11}
\end{figure*}

\end{document}